\documentclass{aa}

\usepackage{graphicx}
\usepackage{txfonts}
\usepackage{color}
\usepackage{enumitem}
\usepackage{subfigure}

\begin{document}

   \title{Joint signal extraction from galaxy clusters in X-ray and SZ surveys: A matched-filter approach}

   \author{P. Tarr\'io
		   	\inst{1}
          \and
          J.-B. Melin
          \inst{2}
          \and
          M. Arnaud
          \inst{1}
          \and
          G. W. Pratt
          \inst{1}
          }

   \institute{Laboratoire AIM, IRFU/Service d'Astrophysique - CEA/DRF - CNRS - Université Paris Diderot, B\^at. 709, CEA-Saclay, 91191, Gif-sur-Yvette Cedex, France\\
   	\email{paula.tarrio-alonso@cea.fr}
   	\and
   	DRF/IRFU/SPP, CEA-Saclay, 91191, Gif-sur-Yvette Cedex, France\\
}

   \date{Submitted 22 February 2016; accepted 13 April 2016}
   %\date{Received }

   \abstract{The hot ionized gas of the intra-cluster medium emits thermal radiation in the X-ray band and also distorts the cosmic microwave radiation through the Sunyaev-Zel'dovich (SZ) effect. Combining these two complementary sources of information through innovative techniques can therefore potentially improve the cluster detection rate when compared to using only one of the probes. Our aim is to build such a joint X-ray-SZ analysis tool, which will allow us to detect fainter or more distant clusters while maintaining high catalogue purity. We present a method based on matched multifrequency filters (MMF) for extracting cluster catalogues from SZ and X-ray surveys. We first designed an X-ray matched-filter method, analogous to the classical MMF developed for SZ observations. Then, we built our joint X-ray-SZ algorithm by combining our X-ray matched filter with the classical SZ-MMF, for which we used the physical relation between SZ and X-ray observations. We show that the proposed X-ray matched filter provides correct photometry results, and that the joint matched filter also provides correct photometry when the $F_{\rm X}/Y_{500}$ relation of the clusters is known. Moreover, the proposed joint algorithm provides a better signal-to-noise ratio than single-map extractions, which improves the detection rate even if we do not  exactly know the $F_{\rm X}/Y_{500}$ relation. The proposed methods were tested using data from the ROSAT all-sky survey and from the \emph{Planck} survey.}


   \authorrunning{P. Tarr\'io et al.}
   \maketitle

\section{Introduction}\label{sec:intro}

Galaxy clusters, composed of dark matter, hot ionized gas and hundreds or thousands of galaxies, are the largest collapsed structures in the Universe. Since the history of cosmic structure formation depends on the cosmology, studying galaxy clusters at different redshifts can help us testing cosmological models and constraining cosmological parameters. Furthermore, galaxy clusters are celestial laboratories in which we can study different astrophysical phenomena. Therefore, galaxy clusters have been studied for decades both as cosmological tools and as astrophysical laboratories. 

To conduct these studies, there has been an increasing need for cluster catalogues with high purity and completeness.  
Since the first cluster catalogue constructed by \citet{Abell1958} by analysing photographic plates, numerous catalogues have been compiled using 
observational data sets at different wavelengths, from microwaves to X-rays. 
The first cluster catalogues were built from optical  
data sets, where clusters are identified as overdensities of galaxies. Clusters can also be detected in X-ray observations, where they appear as bright sources with extended emission from the hot intra-cluster medium (ICM). Finally, over the last decade, clusters have begun to be detected thanks to the characteristic spectral distortion they produce on the cosmic microwave background (CMB), known as the Sunyaev–Zel’dovich (SZ) effect, due to Compton scattering of the CMB photons by the ICM electrons. 

It is natural to think that cluster detection could be improved by combining observations at different wavelengths and from different surveys. Although multiwavelength, multisurvey detection of clusters was theoretically conceived some years ago \citep{Maturi2007,Pace2008}, it is a very complex task and, so far, it has only been attempted in practice in the pilot study of \citet{Schuecker2004} on X-ray data from the ROSAT All-Sky Survey (RASS) \citep{Truemper1993,Voges1999} and optical data from the Sloan Digital Sky Survey (SDSS) \citep{York2000}.  

The goal of this paper is to advance this topic by proposing a novel cluster extraction technique that is based on simultaneous search on SZ and X-ray maps. Combining these two complementary sources of information can improve the cluster detection rate  
when compared to using only one of the probes because both the cluster thermal radiation in the X-ray band and the distortion of the CMB through the SZ effect probe the same medium: the intra-cluster hot ionized gas. However, the combination is not trivial because of the different statistical properties of the signals (nature of noise, astrophysical background, etc.).  

With the motivation of obtaining a tool that is compatible with \emph{Planck}, we decided to use a matched-filter approach to build our joint X-ray-SZ analysis tool. The Matched Multi-Filter (MMF) \citep{Herranz2002,Melin2006,Melin2012} is a popular approach to detect clusters through the SZ effect, and has been extensively used and tested in several SZ surveys such as \textit{Planck} \citep{PlanckEarlyVIII}, the South Pole Telescope (SPT) survey \citep{Bleem2015}, and the Atacama Cosmology Telescope (ACT) survey \citep{Hasselfield2013}. This technique assumes prior knowledge of the cluster shape and the frequency dependence of the SZ signal and leaves the cluster size and the SZ flux amplitude as free parameters. 

X-ray cluster detection techniques usually follow a different approach that is based on maximum likelihood. 
A classical X-ray cluster detection algorithm is the so-called sliding box, in which windows of varying size are moved across the data, marking the positions where the count rate in the central part of the window exceeds the value expected from the background determined in the outermost regions of the window by a certain predetermined factor. This technique is usually followed by a maximum-likelihood routine that evaluates the source position, the detection significance, and the source extent and its likelihood. The flux is determined in a subsequent step, using a growth curve method \citep{Bohringer2000}, for example.

A popular alternative cluster detection technique was designed by \cite{Vikhlinin1998} for the ROSAT Position Sensitive Proportional Counter (PSPC). This technique combines a wavelet decomposition to find candidate extended sources and a maximum-likelihood fitting of the surface brightness distributions to determine the significance of the source extent. The same principles are followed in the XMM-LSS survey \citep{Pacaud2006}. In this case, the images are filtered in  wavelet space with a rigorous treatment of the Poisson noise, and then SExtractor \citep{Bertin1996} is used to find groups of adjacent pixels above a given intensity level in the filtered image. Cluster analysis again follows a maximum-likelihood approach: for each detected candidate, the model that maximizes the probability of generating the observation is determined and, from this, some cluster characteristics are obtained (source counts, extension probability, etc.). 

The Voronoi tessellation and percolation (VTP) method of \cite{Ebeling1993}, used in the WARPS survey \citep{Scharf1997}, has also proved to be well-suited for the detection of extended and low surface brightness emission. The method finds regions of enhanced surface brightness relative to the Poissonian expectation and then derives the source extent from the measured area and flux. 

As a first step to build our X-ray-SZ detector, we developed an X-ray matched-filter detection method, analogous to the MMF developed for SZ observations, which provides X-ray detection results that are readily compatible with those yielded by the SZ detection algorithm. The main difficulty to solve was tuning the filter to take the effects of the Poisson noise in the X-ray signal into account. Although this approach is different from classic X-ray detection techniques, the idea of using a matched filter for X-ray cluster detection was already proposed by \cite{Pace2008}, where a simple filter matched to the X-ray profile was used to detect clusters on synthetic X-ray maps built from hydrodynamical simulations. However, the filter was not designed to extract cluster characteristics, such as the flux and the size,  
and it did not consider Poisson noise. 

In a second step, we built the joint X-ray-SZ algorithm by combining our X-ray matched filter with the classical MMF for SZ, for which we used the physical relation between SZ and X-ray observations. The main idea of our joint detection algorithm is to consider the X-ray map as an additional SZ map at a given frequency and to introduce it, together with other SZ maps, in the classical SZ-MMF. To our knowledge, our proposal is the first complete analysis tool for X-ray clusters based on the matched-filter approach tuned to take the effect of the Poisson noise into account and, furthermore, is the first combined X-ray-SZ extraction technique.

The goal of this paper is to check whether this X-ray-SZ tool can be used to estimate the flux of a cluster accurately, whether the provided signal-to-noise ratio (S/N) is correct, and in particular to analyse whether it represents a gain with respect to SZ-only or X-ray-only cluster detection; in other words: whether the proposed technique improves the completeness. To this end, we focus on the performance of the proposed X-ray-SZ matched filter assuming that we know the position of the cluster and its size. This means that we will not use the filter to \textup{\textup{\emph{\textup{detect}}} } new clusters, but to \emph{\textup{estimate}} some properties of already detected clusters. The performance of the filter as a blind detection tool, which should include an analysis of both the cluster detection rate (completeness) and the false detection rate (or the purity), will be assessed in future work. Although this is a simplification of the complete problem, it is necessary to correctly understand the behaviour of the filter when the statistical properties of the signal are different from those for which the filter was initially designed. In particular, by adding the X-ray information, we must tune the filter to consider Poisson fluctuations on the signal, which are not present in SZ maps. The main goal of this paper is therefore to master this challenge.

Our approach uses all-sky maps from \textit{Planck} and RASS surveys. RASS is the only full-sky X-ray survey conducted with an X-ray telescope \citep{Truemper1993,Voges1999}, which makes it the ideal data set to combine with the all-sky \textit{Planck} survey and compile a joint all-sky cluster catalogue with a large number of clusters. Nevertheless, the proposed technique is general and is also applicable to other surveys, including those from future missions such as e-ROSITA \citep{Merloni2012}, a four-year X-ray survey that is scheduled to start in 2017 and to be much deeper than RASS. 
Using the currently available observations, we are particularly interested in extending the \emph{Planck} catalogue by pushing its detection threshold towards higher redshift, with the specific aim of detecting \emph{\textup{massive and high-redshift clusters}}.

The structure of the paper is as follows. In Sect. \ref{sec:xraydetection} we present the X-ray matched filter and evaluate its performance by injecting simulated clusters on RASS maps and by extracting known clusters on RASS maps. In Sect. \ref{sec:szdetection} we briefly revise the MMF for SZ maps. Section \ref{sec:jointdetection} describes the joint X-ray-SZ MMF and evaluates its performance using RASS and \textit{Planck} maps. Finally, we 
conclude the paper and discuss ongoing and future research directions in Sect. \ref{sec:conclusions}.  

Throughout, we adopt a flat $\Lambda$CDM cosmological model with $H_0 = 70$ km s$^{-1}$ Mpc$^{-1}$ and $\Omega_{\rm M} = 1-\Omega_{\Lambda} = 0.3$. We define $R_{500}$ as the radius at which the average density of the cluster is 500 times the critical density of the Universe, $\theta_{500}$ as the corresponding angular radius, $M_{500}$ as the mass enclosed within $R_{500}$, and $L_{500}$ as the X-ray luminosity within $R_{500}$ in the [0.1-2.4] KeV band.

\section{Extraction of galaxy clusters on X-ray maps}\label{sec:xraydetection}

In this section, we describe and evaluate the proposed algorithm for extracting galaxy clusters on X-ray maps. The algorithm is based on the matched-filter approach and was designed to be compatible with the SZ MMF known as MMF3, described by \cite{Melin2012} and used by the \cite{Planck2013ResXXIX, Planck2015ResXXVII} to construct their SZ cluster catalogues. This compatibility motivates some of the details of the algorithm and its practical implementation, such as the assumption of a generalized Navarro-Frenk-White (GNFW) profile \citep{Nagai2007} to describe the squared density of the gas and the use of HEALPix maps \citep{Gorski2005}.

\subsection{X-ray matched filter: Description of the algorithm}\label{ssec:xrayalgorithm}

Galaxy clusters are powerful and spatially extended X-ray sources. This X-ray emission is due to the very hot, low-density gas of the ICM. 
The X-ray emission of the ICM is that of a coronal plasma at ionization equilibrium. All emission processes (like the Bremsstrahlung emission) result from collisions between electrons and ions \citep{Arnaud2005}, thus 
the emission scales as $ n_{\rm e}^{2} \varepsilon(T_{\rm e})$, where $n_{\rm e}$ is the electron density and  $T_{\rm e}$ is the temperature. The observed surface brightness scales as the integral of $ n_{\rm e}^{2}$ along the line of sight multiplied by the emissivity $\Lambda(T_{\rm e},z)$, which takes $\varepsilon(T_{\rm e})$, the absorption, the cluster redshift, and the instrumental response into account.

The X-ray brightness profile of a cluster at a given energy band can be written as
\begin{equation}
S(\mathbf{x}) =  s_{0}\tilde{T}^{\rm{x}}_{\theta_{\rm s}}(\mathbf{x}),
\end{equation}
where $\mathbf{x}$ indicates the 2D position on the sky (with $\mathbf{x} = 0$ corresponding to the center of the cluster), $ \tilde{T}^{\rm{x}}_{\theta_{\rm s}}(\mathbf{x}) $ is a normalized cluster spatial profile (normalized so that its central value is 1), and $s_{0}$ is the cluster central surface brightness. The notation $ \tilde{T}^{\rm{x}}_{\theta_{\rm s}}(\mathbf{x}) $ indicates that the cluster profile depends on the apparent size of the cluster through the characteristic cluster scale $\theta_{\rm s}$.

Let us imagine that we have an X-ray map of a certain region of the sky in which a cluster is located at position $\mathbf{x}_0$, characterized by a profile $ \tilde{T}^{\rm{x}}_{\theta_{\rm s}}(\mathbf{x}) $ and a central surface brightness $s_0$. If we denote the convolution of the cluster profile with the point spread function (PSF) of the X-ray instrument $B_{\rm xray}(\mathbf{x})$ by $T^{\rm{x}}_{\theta_{\rm s}}(\mathbf{x}) = \tilde{T}^{\rm{x}}_{\theta_{\rm s}}(\mathbf{x}) \ast B_{\rm xray}(\mathbf{x})$, we can express this X-ray map as
\begin{equation}\label{eq:Xray_map}
M(\mathbf{x}) = s_{0} j_{\rm x} T^{\rm{x}}_{\theta_{\rm s}}(\mathbf{x}-\mathbf{x}_0) + N(\mathbf{x}),
\end{equation}
where $j_{\rm x}$ is simply a conversion factor to express the map in any desired units ($j_{\rm x}=1$ if the map is expressed in surface brightness units), and $N(\textbf{x})$ is the noise map, which includes instrumental noise and astrophysical X-ray background. The nature of the X-ray signal means that the map is affected by Poisson noise.
Consequently, each pixel of our X-ray map can be characterized as a Poisson random variable with variance equal to the expected value of counts. Let us represent the cluster signal, in counts, by $T_{\rm c}(\mathbf{x}) = aT^{\rm{x}}_{\theta_{\rm s}}(\mathbf{x}) + N_{\rm a}(\mathbf{x})$, where $ aT^{\rm{x}}_{\theta_{\rm s}}(\mathbf{x}) $ is the expected number of counts at position $\mathbf{x}$ and  $N_{\rm a}(\mathbf{x})$ is the random noise in addition to this, with zero-mean ($\left\langle N_{\rm a}(\mathbf{x}) \right\rangle =0$) and variance $ \left\langle (N_{\rm a}(\mathbf{x}))^2 \right\rangle = aT^{\rm{x}}_{\theta_{\rm s}}(\mathbf{x})$.
Then, we can rewrite Eq. \ref{eq:Xray_map} as
\begin{equation}\label{eq:Xray_map_2}
{M}(\mathbf{x}) = s_{0} j_{\rm x} T^{\rm{x}}_{\theta_{\rm s}}(\mathbf{x}-\mathbf{x}_0) + N_{\rm sig}(\mathbf{x}) + N_{\rm bk}(\mathbf{x}),
\end{equation}
where $N_{\rm sig}(\mathbf{x})$ is the additional random noise in the cluster signal that is due to Poisson fluctuations, and ${N}_{\rm bk}(\mathbf{x})$ is the background noise. If we define $u=s_{0} j_{\rm x}/a$ as the unit conversion factor from counts to the units of the X-ray map, then we have that 
$N_{\rm sig}(\mathbf{x}) = u N_{\rm a}(\mathbf{x})$ is a random variable with zero-mean ($\left\langle N_{\rm sig}(\mathbf{x}) \right\rangle =0$) and variance equal to 
\begin{equation}\label{eq:Nsig_variance}
\left\langle ({N_{\rm sig}}(\mathbf{x}))^2 \right\rangle = u^2 aT^{\rm{x}}_{\theta_{\rm s}}(\mathbf{x}). 
\end{equation}
It is important to keep in mind that, in general, $u=u(\mathbf{x})$ is not constant across the map, but depends on the position. This is because the conversion from counts to surface brightness depends on the exposure time and on the $N_{\rm H}$ column density, which in turn depend on the position. However, for small scales (a few arcminutes), it can be approximated as constant. 

If the cluster profile is known, the problem in Eqs. \ref{eq:Xray_map} or \ref{eq:Xray_map_2} reduces to  estimating the amplitude $s_0$ of a known signal from an observed signal that is contaminated by noise. Generally speaking, if we do not know the probability density function  
of the noise, we cannot calculate the optimal estimator, that is, the minimum variance unbiased (MVU) estimator. However, we can restrict the estimator to be linear and then find the linear estimator that is unbiased and has minimum variance, that is, the best linear unbiased estimator (BLUE). 

We can construct a linear estimator, $\hat{s}_0$, of the central brightness $s_0$ as a linear combination of the observed data,
\begin{equation}\label{eq:s0_estim}
\hat{s}_{0} = \sum_{\mathbf{x}} \Psi_{\theta_{\rm s}}(\mathbf{x}-\mathbf{x}_0) M(\mathbf{x}),
\end{equation}
where $\Psi_{\theta_{\rm s}}$ can be interpreted as a filter to be applied to the X-ray map. We note that Eq. \ref{eq:s0_estim} yields a scalar value if we know the position $\mathbf{x}_0$ of the cluster, whereas if it is unknown, we can apply the equation for every possible value of $\mathbf{x}_0$ to obtain a $\hat{s}_{0}$-map with the same size as the observed map. 

If we restrict this linear estimator to be unbiased and to have minimum variance, we obtain the following expression for the filter in Fourier space (the derivation is analogous to that in \citet{Haehnelt1996, Herranz2002, Melin2006, Melin2012}): 
\begin{equation}\label{eq:Xray_MF}
\Psi_{\theta_{\rm s}}(\mathbf{k}) = \sigma_{\theta_{\rm s}}^2   j_{\rm x} \frac{T^{\rm{x}}_{\theta_{\rm s}}(\mathbf{k})}{{P}(\mathbf{k})} ,
\end{equation}
where
\begin{equation}\label{eq:Xray_sigmaMF}
\sigma_{\theta_{\rm s}}^2 = \left[ j_{\rm x}^2 \sum_{\mathbf{k}} \frac{\left|   T^{\rm{x}}_{\theta_{\rm s}}(\mathbf{k})\right|^2}{{P}(\mathbf{k})}   \right] ^{-1}
\end{equation}
is, approximately, the background noise variance after filtering (see Appendix \ref{app:sigma_poisson} for derivation) and ${P}(\mathbf{k})$ is the noise power spectrum, given by $\left\langle N(\mathbf{k})N^\ast(\mathbf{k}')\right\rangle  = P(\mathbf{k}) \delta(\mathbf{k}-\mathbf{k}')$. Here and in the remainder of the paper, we use $\mathbf{k}$ to denote the two-dimensional spatial frequency, corresponding to $\mathbf{x}$ in the Fourier space. All the variables expressed as a function of $\mathbf{k}$ are then to be understood as variables in the Fourier space. The filter is determined by the shape of the cluster X-ray signal and by the power spectrum of the noise, hence, the name of X-ray matched filter.

Taking the Fourier transform of Eq. \ref{eq:Xray_map_2}, we have
\begin{equation}\label{eq:XraymapFT}
{M}(\mathbf{k}) =  s_{0} j_{\rm x} T^{\rm{x}}_{\theta_{\rm s}}(\mathbf{k}) +N_{\rm sig}(\mathbf{k}) + N_{\rm bk}(\mathbf{k}).
\end{equation}
When we apply the matched filter given by Eq. \ref{eq:Xray_MF}, the filtered map in Fourier space is
\begin{align}\label{eq:filteredmapFT}
	\sum_{\mathbf{k}}^{}\Psi_{\theta_{\rm s}}^\ast(\mathbf{k}) {M}(\mathbf{k}) = & s_{0} j_{\rm x}^2 \sigma_{\theta_{\rm s}}^2 \sum_{\mathbf{k}}^{} \frac{\left| T^{\rm{x}}_{\theta_{\rm s}}(\mathbf{k}) \right|^2 }{P(\mathbf{k})} + j_{\rm x} \sigma_{\theta_{\rm s}}^2 \sum_{\mathbf{k}}^{} \frac{ T^{\rm{x}\ast}_{\theta_{\rm s}}(\mathbf{k})  }{P(\mathbf{k})} {N_{\rm sig}}(\mathbf{k}) \nonumber\\
	+ & j_{\rm x} \sigma_{\theta_{\rm s}}^2 \sum_{\mathbf{k}}^{} \frac{ T^{\rm{x}\ast}_{\theta_{\rm s}}(\mathbf{k})  }{P(\mathbf{k})} N_{\rm bk}(\mathbf{k}),
\end{align}
where the first term on the right-hand side of the equation is equal to $s_{0}$ (the amplitude of the cluster profile), the third term is the filtered background noise, whose variance is given by Eq. \ref{eq:Xray_sigmaMF} (approximately), and the second term is the filtered Poisson fluctuations on the signal. The variance due to the Poisson fluctuations on the signal, after passing through the filter, can be written as (see derivation in Appendix \ref{app:sigma_poisson})
\begin{align}\label{eq:poissonvariance}
	\sigma_{\rm Poisson}^2 = \frac{u s_{0} j_{\rm x}^3 \sigma_{\theta_{\rm s}}^4}{n^2}\sum_{\mathbf{k}}\sum_{\mathbf{k}'}\frac{T^{\rm{x}\ast}_{\theta_{\rm s}}(\mathbf{k})T^{\rm{x}}_{\theta_{\rm s}}(\mathbf{k}')}{P(\mathbf{k}) P(\mathbf{k}')} T^{\rm{x}}_{\theta_{\rm s}}\left( \mathbf{k}-\mathbf{k}'\right),
\end{align}
where $n^2$ is the total number of pixels in the map and the double sum can be computed efficiently by making use of the cross-correlation theorem, as explained in Appendix \ref{app:sigma_poisson}.  We note that this variance depends on $s_0$, the real value of the central surface brightness. As this is not known in practice, it is necessary to approximate it by its estimated value $\hat{s}_0$. 

Therefore, we can characterize our central brightness estimator $\hat{s}_{0}$ as a random variable with mean equal to the true central brightness $s_0$ and variance given by the sum of the variances of the filtered background noise and the filtered Poisson fluctuations on the signal. That is,
\begin{equation}\label{eq:totalvariance}
\sigma_{\hat{s}_0}^2 =  \sigma_{\theta_{\rm s}}^2 + \sigma_{\rm Poisson}^2,
\end{equation}
where we have assumed that the Poisson fluctuations on the signal are independent of the background noise, as expected, given their independent origin. Considering this Poisson term in the final variance is essential to correctly characterize the errors on our flux estimation, and it is specially important for bright clusters, where the Poisson noise is dominant over the background noise. 

As previously said, if the exact position of the cluster is unknown (or if its accuracy is not high enough), we can apply Eq. \ref{eq:s0_estim} for every possible value of $\mathbf{x}_0$ and obtain a $\hat{s}_{0}$-map with the same size as the observed map. In this case, we also obtain a $\sigma_{\rm Poisson} $-map that represents the variance due to the Poisson fluctuations on the signal at every position of the map (because $u$ and $\hat{s}_{0}$ depend on the position)  
and a S/N map 
($\hat{s}_{0}$/$\sigma_{\theta_{\rm s}}$) 
with the same size. The cluster is then detected as a peak in this S/N map, down to a given threshold. In this way, the proposed filter could also be used as a blind detection tool and not only as an estimator of the cluster properties. In this paper we focus on the performance of the filter as an extraction tool (once we know that there is a cluster at a given position); the assessment of its performance as a detector is beyond the scope of this paper and will be undertaken in future work. 

It is important to remark that this X-ray matched-filter approach relies on the knowledge of the normalized cluster brightness profile $T^{\rm{x}}_{\theta_{\rm s}}(\mathbf{x})$. In practice this profile is not known, therefore we need to use a theoretical profile that represents the average brightness profile of the clusters we wish to detect as well as possible. Here we assume the average gas density profile from \citet{Piffaretti2011}:
\begin{equation}\label{eq:densprof}
\rho_{\rm gas} \propto \left( \frac{x}{x_{\rm c}}\right)^{-\alpha '}  \times \left[ 1+\left( \frac{x}{x_{\rm c}}\right)^{2}\right] ^{-3\beta'/2+\alpha'/2},
\end{equation}
where $ x_{\rm c} $ = 0.303, $ \alpha' $ = 0.525, $ \beta' $ = 0.768, $x=\theta/\theta_{500}$ is the distance to the center of the cluster in $\theta_{500}$ units, and the free parameter $\theta_{500}$ relates to the characteristic cluster scale $\theta_{\rm s}$ through the concentration parameter $c_{500}$ ($\theta_{\rm s} = \theta_{500}/c_{500}$). It can be seen that $\rho_{\rm gas}^2$ can be written as a GNFW profile given by
\begin{equation}\label{eq:pressure_prof}
p(x) \propto \frac{1}{\left( c_{500}x\right) ^{\gamma} \left[ 1+\left( c_{500}x\right) ^{\alpha}\right]^{(\beta-\gamma)/\alpha}  }
\end{equation}
with $ \alpha = 2 $, $ \beta=6\beta ' $, $ \gamma=2\alpha ' $  and $ c_{500}=1/x_{\rm c} $. We therefore describe the squared average density profile using Eq. \ref{eq:pressure_prof} with the parameters
\begin{equation}\label{eq:xray_param}
\left[ \alpha, \beta, \gamma, c_{500}\right] = \left[ 2.0, 4.608, 1.05, 1/0.303\right].
\end{equation}
This expression is convenient to facilitate compatibility with the classical SZ MMF, which uses a GNFW model to describe the cluster profile (see Sect. \ref{sec:szdetection}). 
As previously said, the observed intensity at a given energy in an X-ray map depends on $ n_{\rm e}^{2}$ and 
$\Lambda(T_{\rm e},z)$. In the soft X-ray band (below 2 keV), the emissivity $\Lambda(T_{\rm e},z)$ is approximately independent of the temperature, so
the X-ray emission is approximately proportional to the square of the gas density $ \rho_{\rm gas}^{2} $.
Thus, the cluster profile $\tilde{T}^{\rm{x}}_{\theta_{\rm s}}(\mathbf{x})$ can be obtained by numerically integrating the cluster 3D squared-density profile (Eq. \ref{eq:pressure_prof}) along the line of sight.

Finally, to take the effect of beam smoothing affecting the observed signal into account, the cluster profile $\tilde{T}^{\rm{x}}_{\theta_{\rm s}}(\mathbf{x})$ has to be convolved by the PSF of the instrument, namely $B_{\rm xray}(\mathbf{x})$. Here we used the X-ray maps of the ROSAT All-Sky Survey, whose PSF is not Gaussian \citep[see][]{Boese2000, Bohringer2013}. In absence of an analytical expression for the RASS PSF, we have estimated it numerically by stacking observations of X-ray point sources. In particular, we stacked all the point sources in the Bright Source Catalogue \citep{Voges1999} with Galactic latitude $\left| l \right| > 30\degr$ and computed the azimuthal average of the stack.

On the other hand, the noise power spectrum ${P}(\mathbf{k})$ can be estimated in practice from the X-ray image itself, after masking and inpainting the cluster region.

\subsection{Performance evaluation: Simulation results}\label{ssec:xray_simu}

\subsubsection{Description of the simulations}\label{sssec:XraySimuDescription}
To assess the performance of the proposed X-ray matched filter, we carried out an experiment in which we injected simulated clusters into real X-ray maps and extracted them using the proposed filter, assuming their positions and their sizes are known. The objective of this experiment was to check whether the flux estimated with the matched filter and its associated error bar are consistent with the flux of the injected clusters because a correct photometry at this point will be important for the joint X-ray-SZ algorithm, which relies on the $F_{\rm X}/Y_{500}$ relation (i.e., the ratio between the X-ray flux of the cluster within $R_{500}$ in the [0.1-2.4] KeV band and the SZ flux of the cluster within $R_{500}$). The injection into real maps provides a very realistic environment, with real background contributions and perfectly known cluster characteristics against which to compare the results. 

We simulated the clusters as follows. Given the redshift $z$, the mass $M_{500}$, and the luminosity $L_{500}$ of a cluster, we first created a map containing the cluster profile corresponding to the size $\theta_{500}$ of the cluster (calculated from $z$ and $M_{500}$). This was done by integrating the average profile defined by Eq. \ref{eq:pressure_prof} with the parameters given in Eq. \ref{eq:xray_param}. Then we normalized this map so that its total flux coincided with the flux of the cluster within $5R_{500}$ 
(we assumed that the total flux of the cluster is contained within this radius and calculated it by extrapolating $L_{500}$ up to $5R_{500}$ using the shape of the cluster profile) and convolved this map with the instrument beam. Finally, to obtain a simulated image of the cluster that reproduces the observational noise, we added Poisson fluctuations according to the local photon flux. To this end, we converted the flux at each pixel into counts by using the $N_{\rm H}$ value and the exposure time corresponding to the position where the simulated cluster was going to be injected.

\begin{figure*}[]
	\centering
	\subfigure[]{\includegraphics[width=.95\columnwidth]{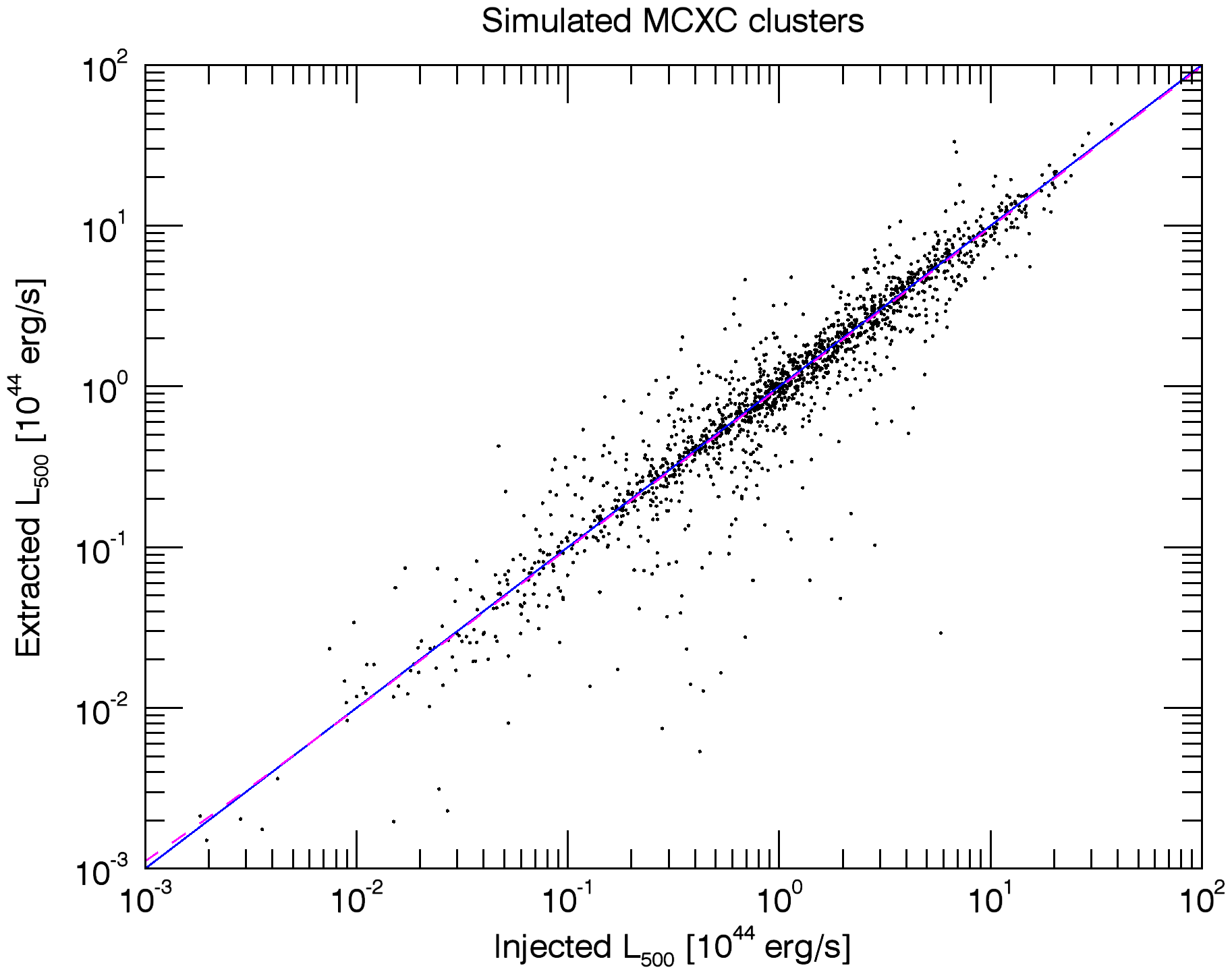}\label{fig:extraction_simupxcc}}
	\subfigure[]{\includegraphics[width=.95\columnwidth]{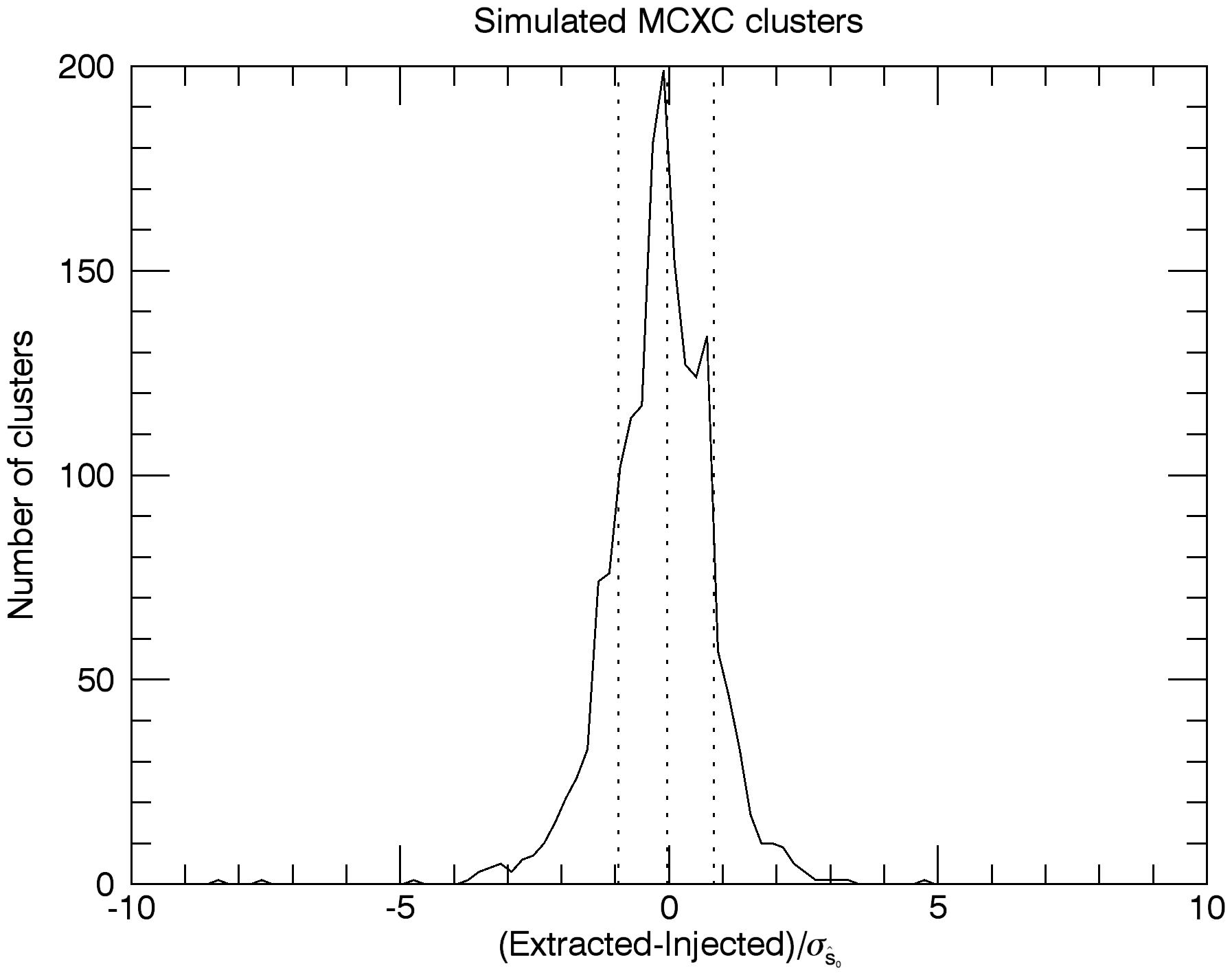}\label{fig:hist_simupxcc}}
	\subfigure[]{\includegraphics[width=.95\columnwidth]{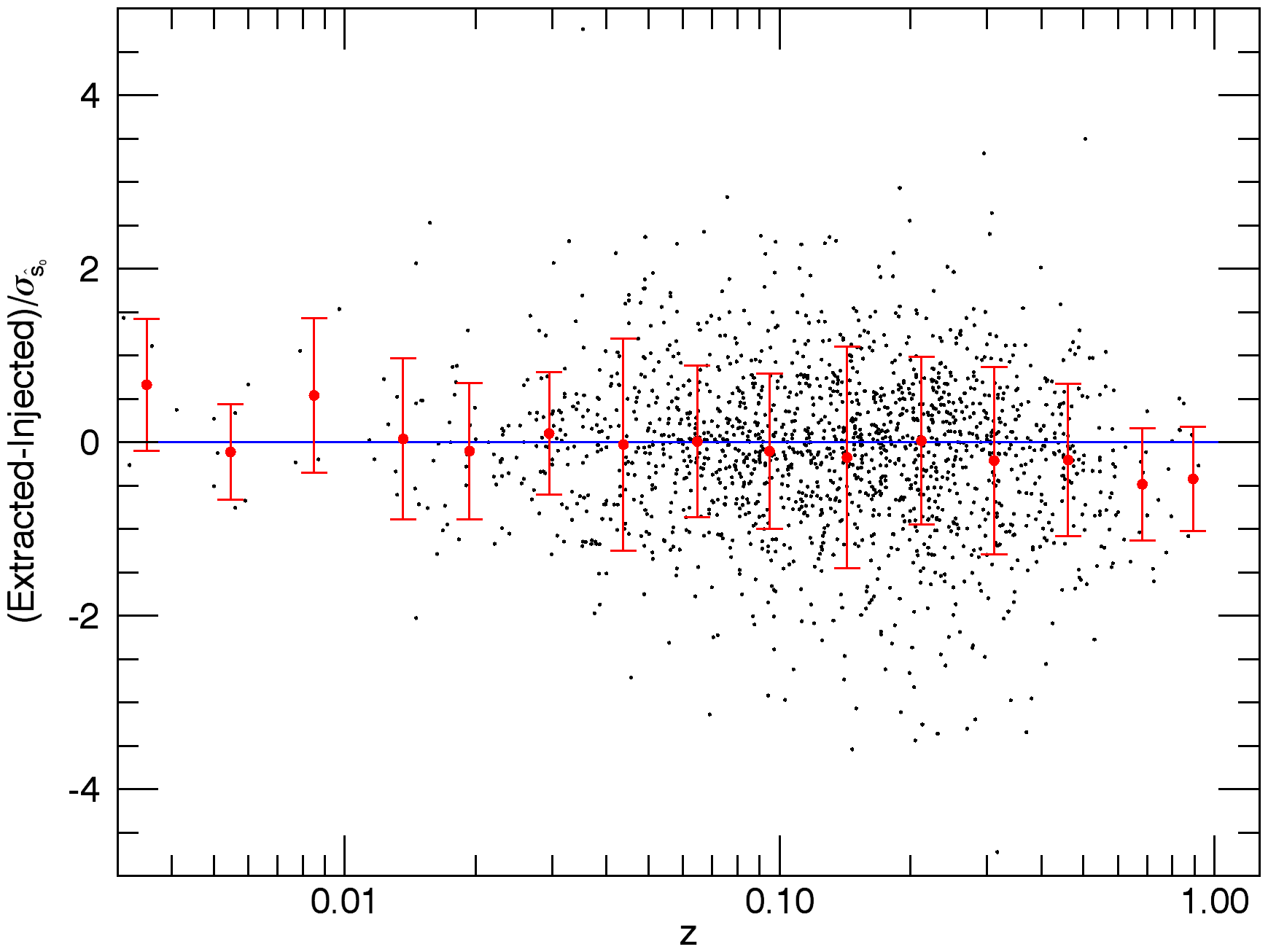}\label{fig:simupxcc_vs_z}}
	\subfigure[]{\includegraphics[width=.95\columnwidth]{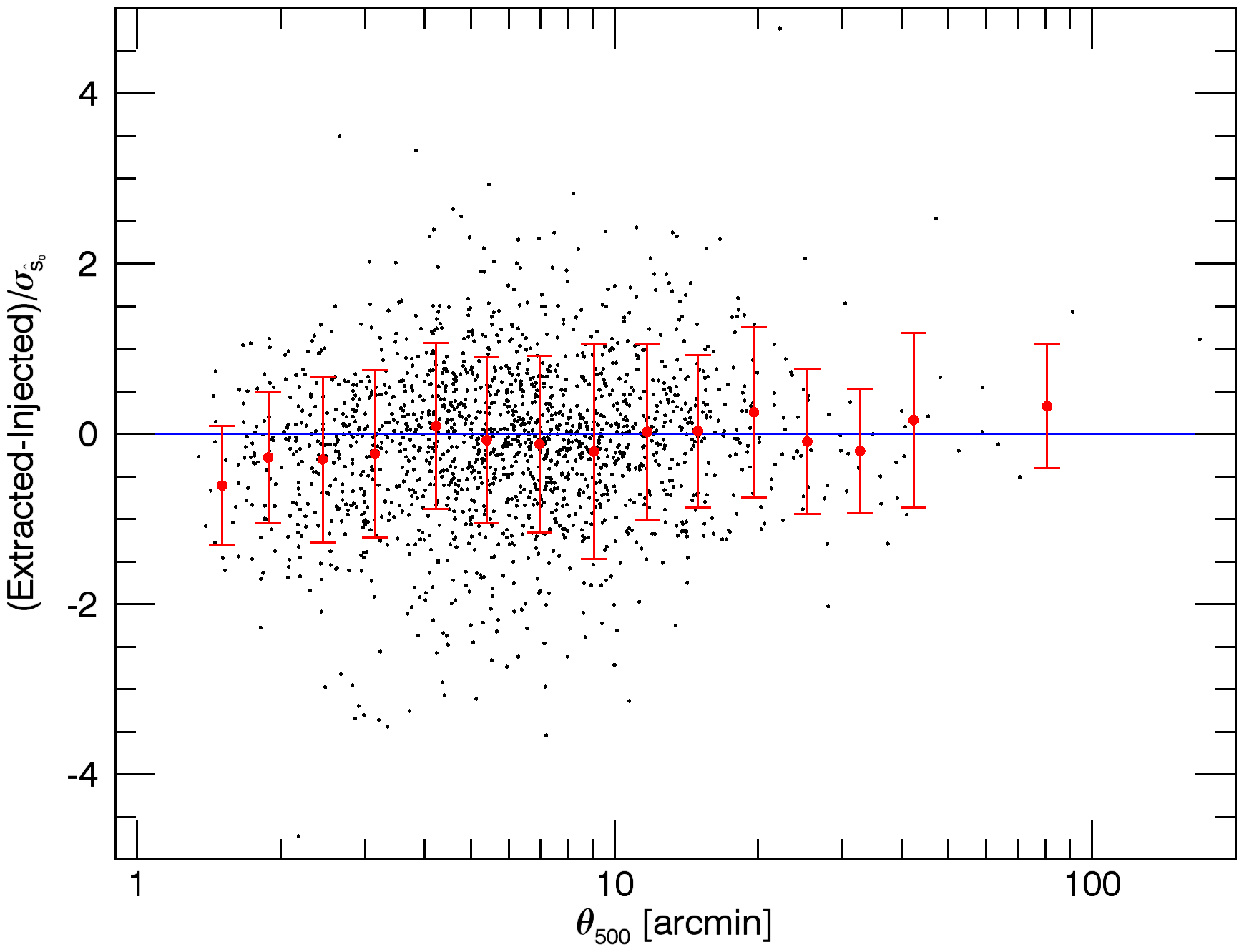}\label{fig:simupxcc_vs_theta}}
	\subfigure[]{\includegraphics[width=.95\columnwidth]{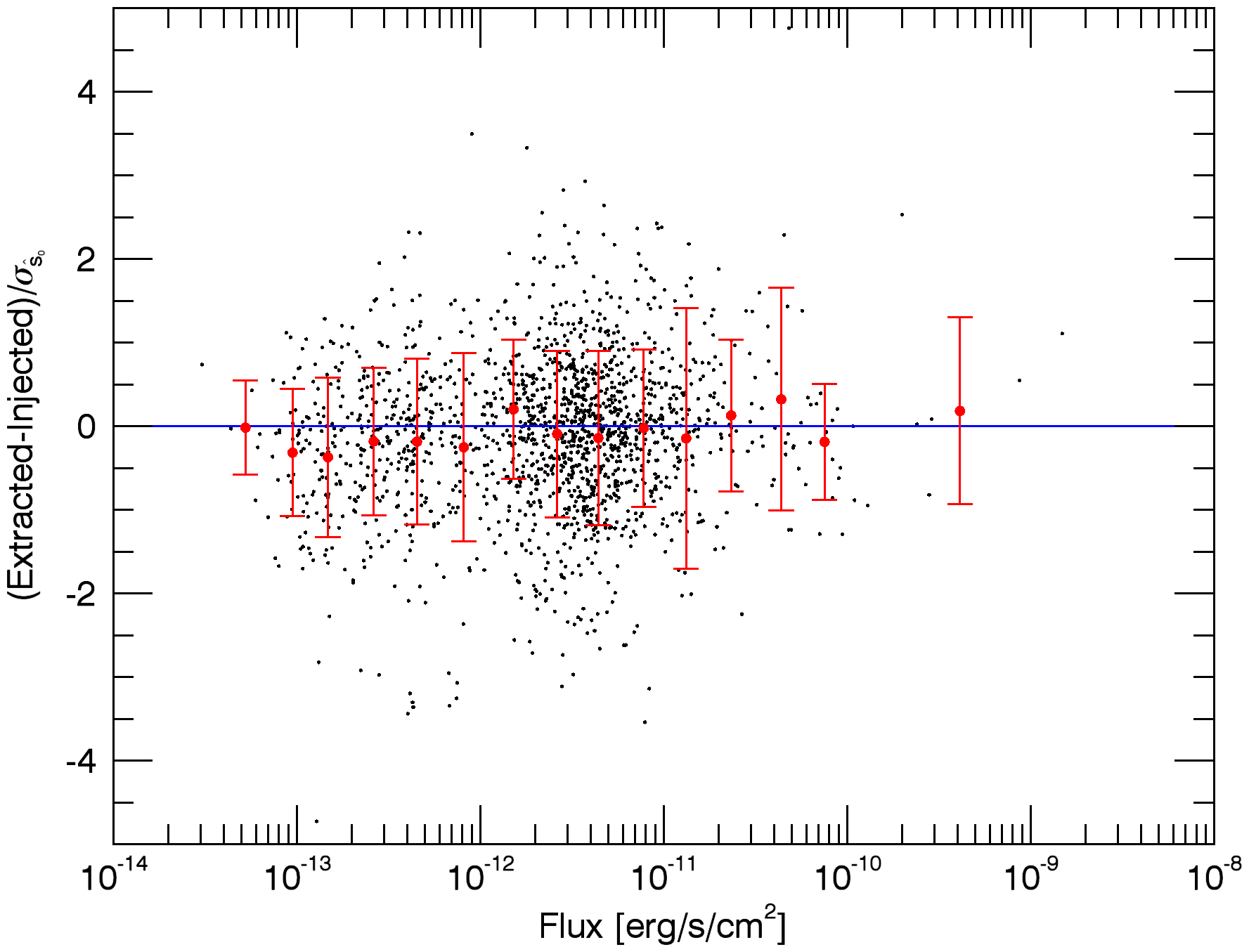}\label{fig:simupxcc_vs_flux}}
	\subfigure[]{\includegraphics[width=.95\columnwidth]{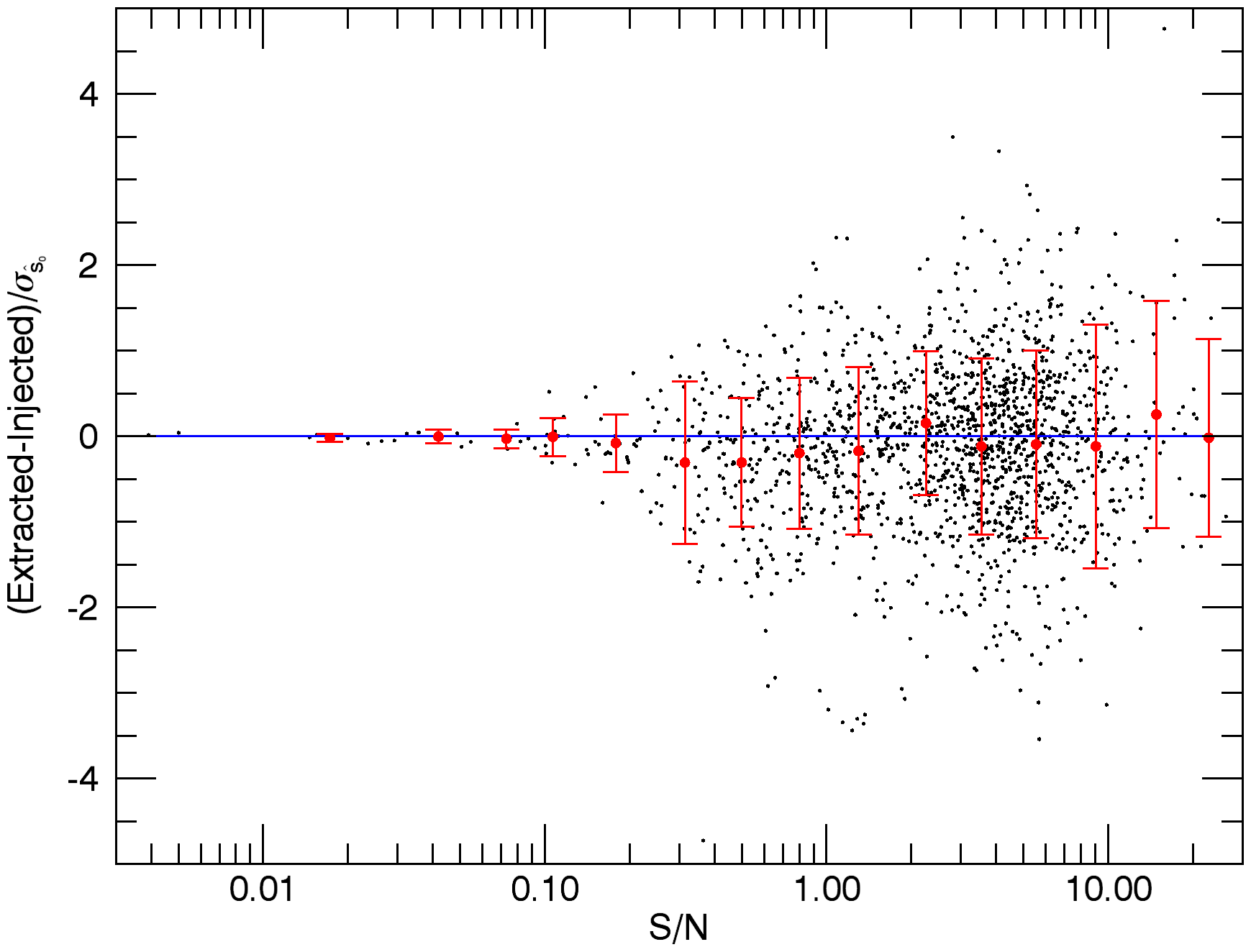}\label{fig:simupxcc_vs_snr}}
	\caption{Photometry results of the extraction of simulated MCXC clusters (as described in Sect. \ref{sssec:XraySimuDescription}) using the proposed X-ray matched filter and assuming the position and size of the clusters are known. Top left panel: Extracted versus injected $L_{500}$. Individual measurements are shown as black dots. 
		The solid blue line shows the line of zero intercept and unity slope and the dashed magenta line the best linear fit to the data. 
		Top right panel: Histogram of the difference between the extracted and the injected $L_{500}$, divided by the estimated $\sigma_{\hat{s}_0}$ (scaled to $L_{500}$ units). The central vertical line shows the median value; the other two vertical lines indicate the region inside which 68$\% $ of the clusters are located. 
		Middle and bottom panels: Difference between the extracted and the injected $L_{500}$, divided by the estimated  $\sigma_{\hat{s}_0}$, as a function of (c) the redshift, (d) the size, (e) the flux, and (f) the S/N of each cluster (S/N is defined here as the injected signal divided by the theoretical $\sigma_{\hat{s}_0}$). Individual measurements are shown as black dots. The red filled circles represent the corresponding averaged values in different bins, calculated as described in the text (Sect. \ref{sssec:XraySimuResults}). The 
			error bars represent the standard deviation of the values in the bin.}
	\label{fig:simupxcc}
\end{figure*}

\begin{figure*}[]
	\centering
	\subfigure[]{\includegraphics[width=.99\columnwidth]{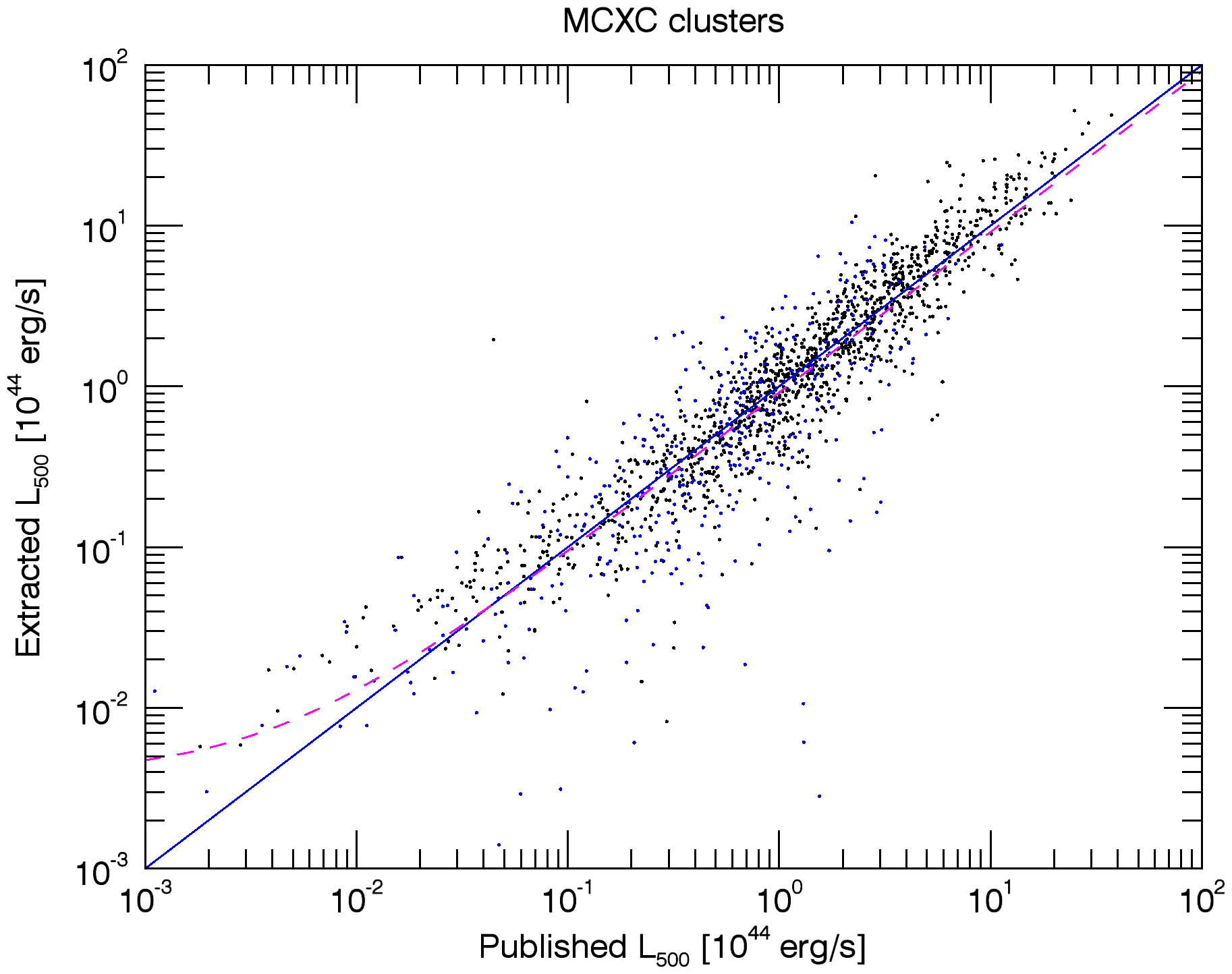}\label{fig:extraction_pxcc}}
	\subfigure[]{\includegraphics[width=.99\columnwidth]{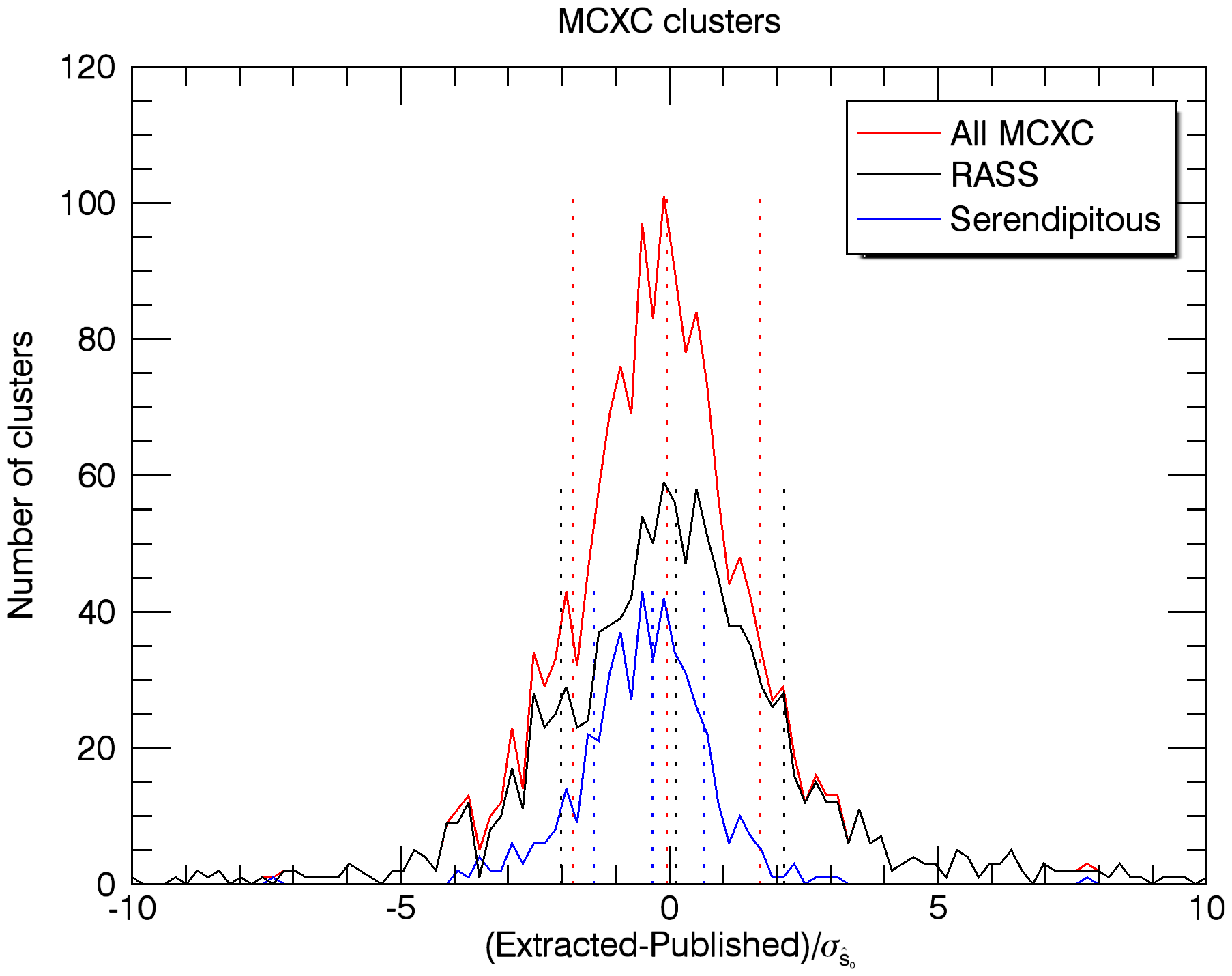}\label{fig:hist_pxcc}}
	\subfigure[]{\includegraphics[width=.99\columnwidth]{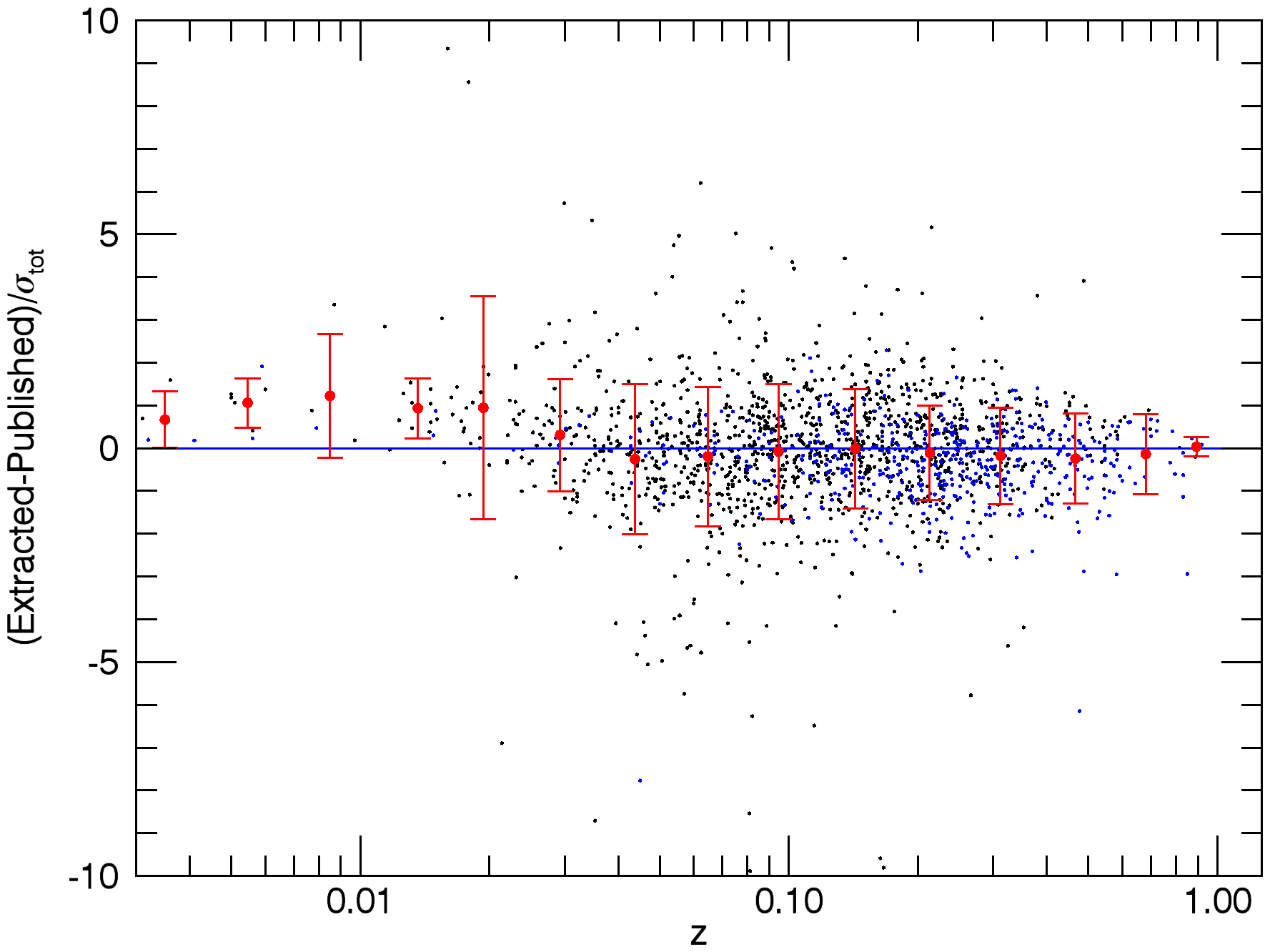}\label{fig:realpxcc_vs_z}}
	\subfigure[]{\includegraphics[width=.99\columnwidth]{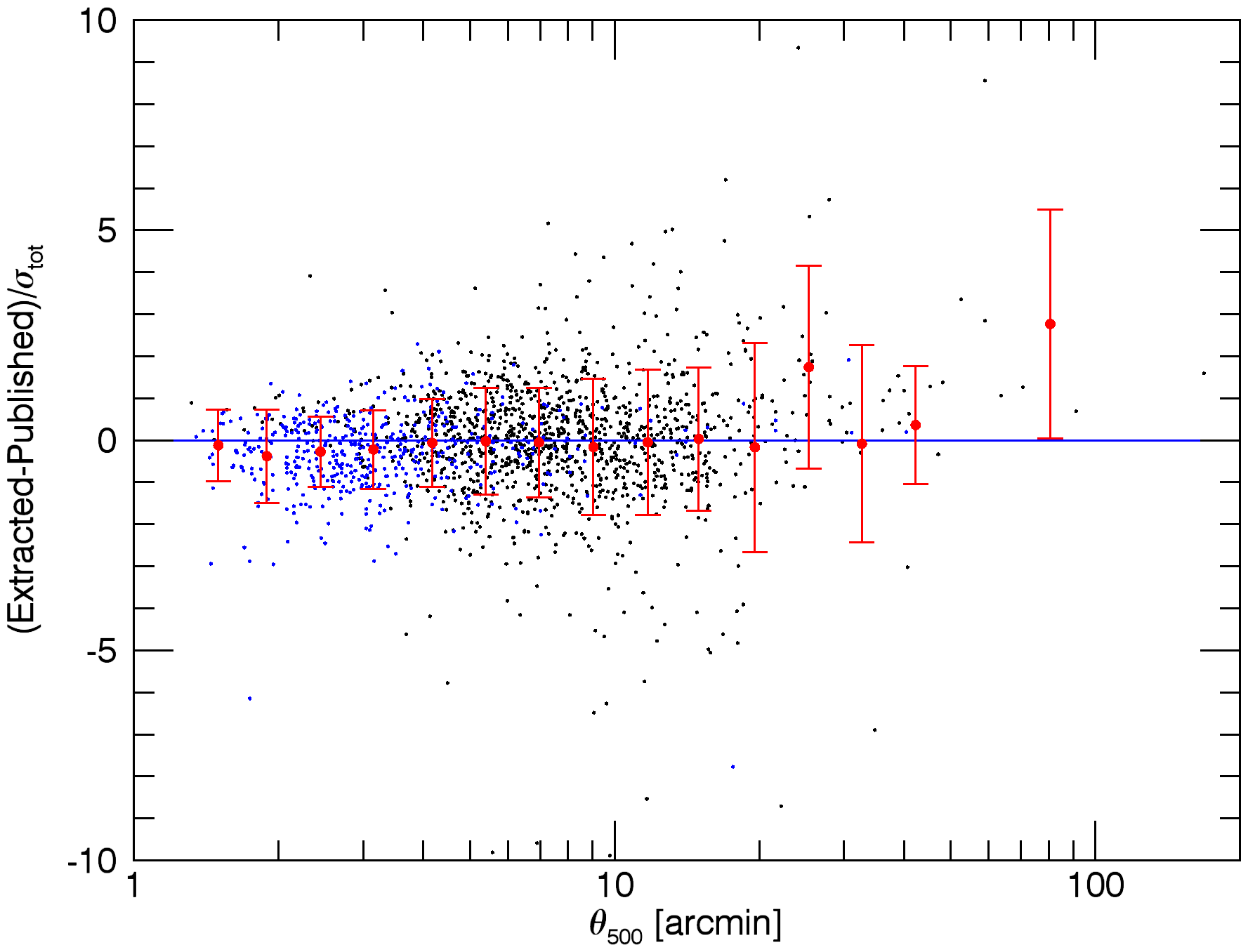}\label{fig:realpxcc_vs_theta}}
	\subfigure[]{\includegraphics[width=.99\columnwidth]{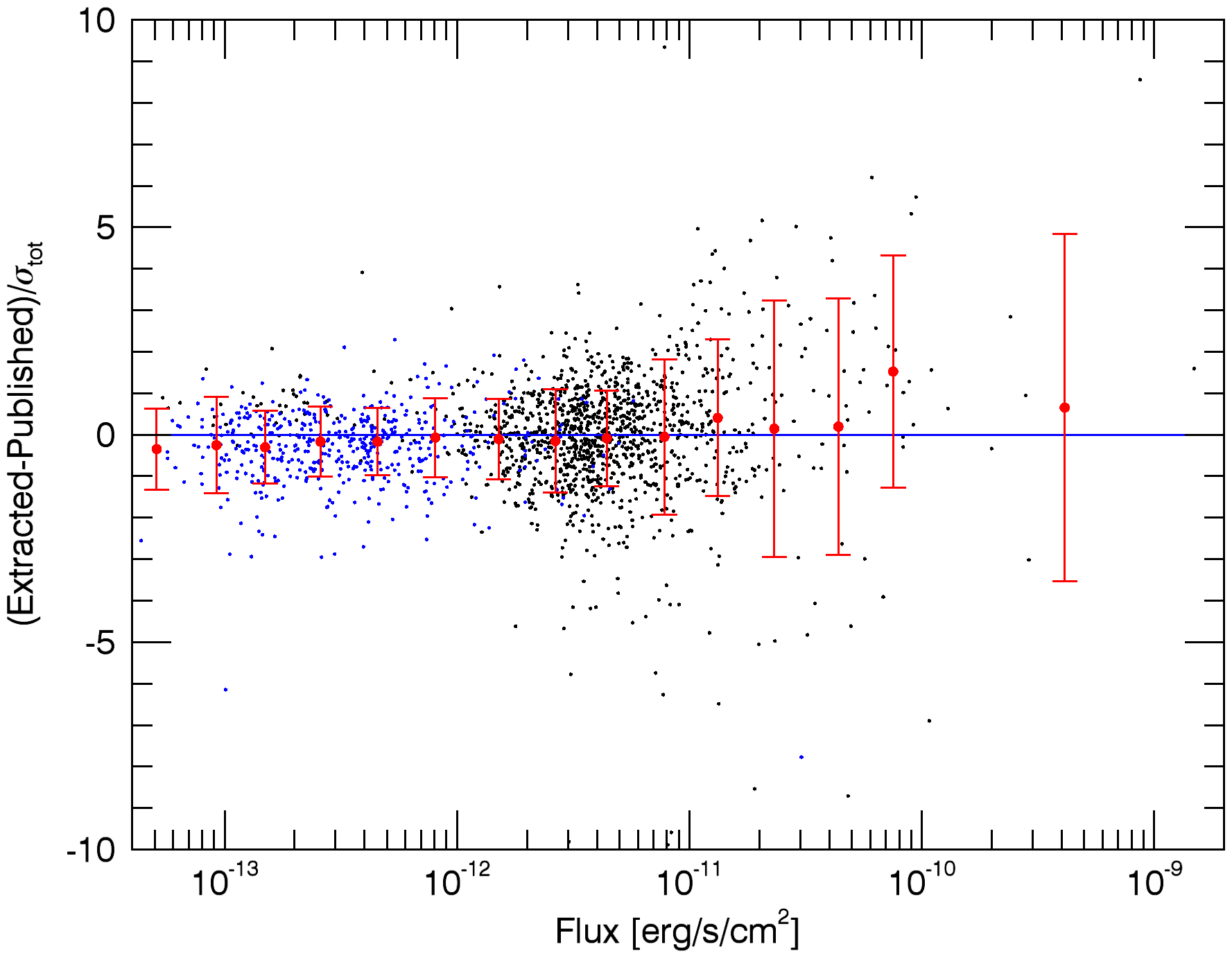}\label{fig:realpxcc_vs_flux}}
	\subfigure[]{\includegraphics[width=.99\columnwidth]{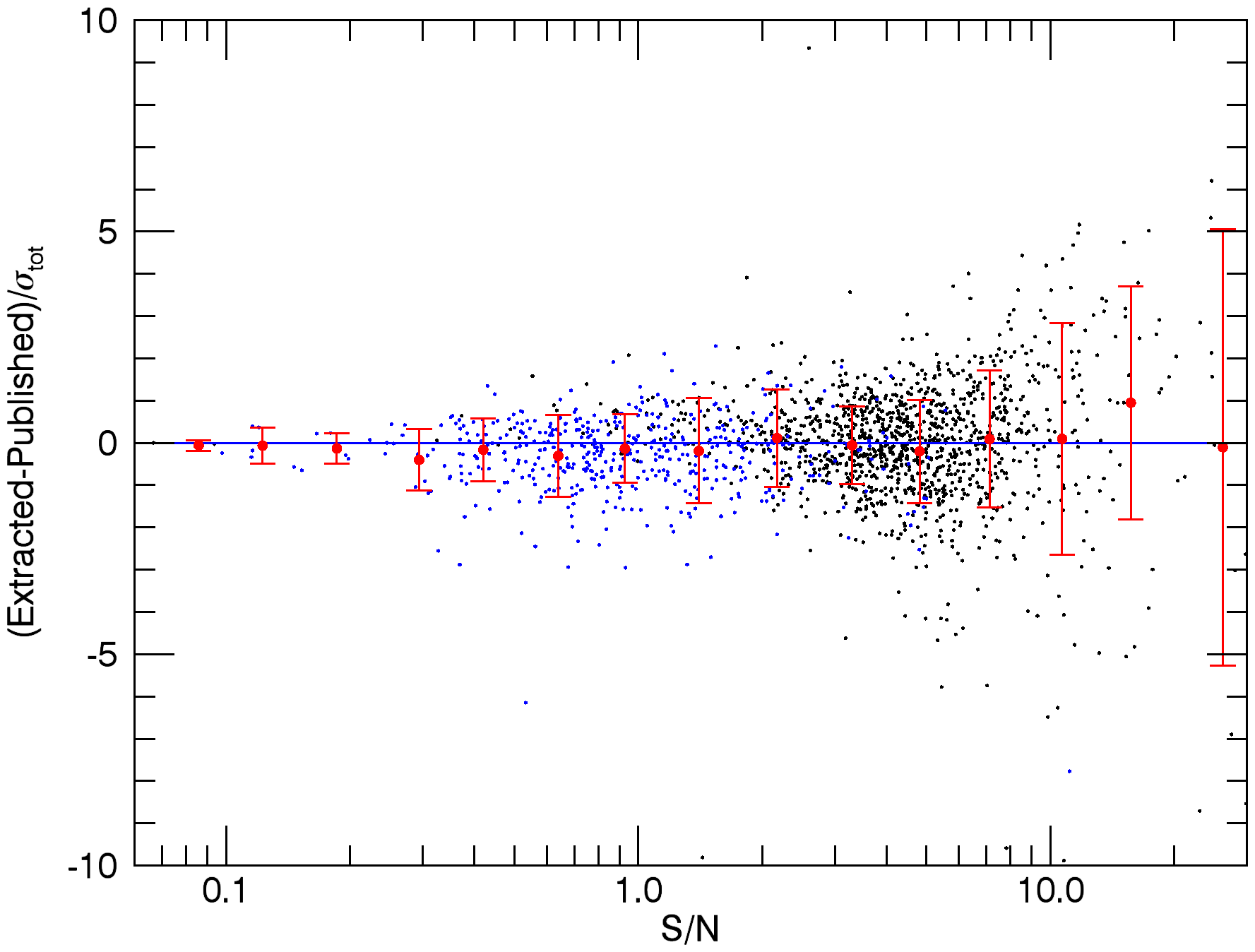}\label{fig:realpxcc_vs_snr}}
	\caption{Photometry results of the extraction of the real MCXC clusters using the proposed X-ray matched filter and assuming the position and size of the clusters are known. The six panels are analogous to those in Fig. \ref{fig:simupxcc}, but comparing the extracted $L_{500}$ with the published $L_{500}$. Individual measurements in panels a, c, d, e, and f are shown as black and blue dots: the black dots correspond to clusters originally detected in RASS, the blue dots to serendipitous clusters. Panel b shows three histograms that correspond to the RASS clusters (black), the serendipitous clusters (blue), and the complete MCXC sample (red).}
	\label{fig:realpxcc}
\end{figure*}

These simulated maps were then added to real X-ray patches centered on random positions of the sky, corresponding to the positions of the simulated clusters. These X-ray patches were constructed in two steps. First, we created an all-sky HEALPix map \citep{Gorski2005} using all the fields of the ROSAT all-sky survey, as described in Appendix \ref{app:HealpixMap}. Second, we projected this HEALPix map onto small $10\degr \times 10\degr$ flat patches centered on the cluster position. This way of creating the X-ray patches is, of course, not the only alternative, but we chose it to guarantee compatibility with the SZ MMF that we have used, so that it will be useful for the joint X-ray-SZ algorithm.

In particular, we injected 1743 clusters at random positions of the sky, with characteristics ($z$, $M_{500}$, and $L_{500}$) taken from the 1743 clusters in the MCXC catalogue \citep{Piffaretti2011}. 
Then, we applied the X-ray matched filter described in Sect. \ref{ssec:xrayalgorithm} at each cluster position, fixing the cluster size to the true value. 
We chose to simulate this sample because it is an X-ray selected sample, and thus it is appropriate for evaluating the performance of our X-ray filter.

\subsubsection{Results of the simulations}\label{sssec:XraySimuResults}

Figure \ref{fig:simupxcc} shows the extraction results for the experiment described above. Figure \ref{fig:extraction_simupxcc} shows the extracted value of $L_{500}$ for each cluster as a function of the injected value. We used the redshift in the catalogue to convert from flux to $L_{500}$. The black dots show the individual measurements, and the red filled circles represent the corresponding averaged values in several luminosity bins. These bin-averaged values were calculated as $\sum(y_i/\sigma_i^2)/\sum(1/\sigma_i^2)$, where $y_i$ is the estimated flux of cluster $i$ scaled to $L_{500}$ units and $\sigma_i$ is the estimated $\sigma_{\hat{s}_0}$ (also scaled to $L_{500}$ units) for cluster $i$. The position of these bin-averaged values in the x-axis was calculated by averaging the injected flux of the clusters in the bin with the same weights that were used to average the extracted flux. 
The extracted flux follows the injected flux very well, with some dispersion. The best linear fit to these data is given by 
$y=0.978(\pm0.005)x+1.2(\pm2.7)\cdot10^{-4}$, which is very close to the unity-slope line, as shown in the figure. 

Figure \ref{fig:hist_simupxcc} shows the histogram of the difference between the extracted and the injected value, divided by the estimated standard deviation $\sigma_{\hat{s}_0}$. Some of the properties of this histogram are summarized in Table \ref{table:simupxcc}. In this histogram there is no bias, and the estimated error bars describe the dispersion on the results well (as 68$\%$ of the extractions fall in an interval that is close to $\pm1\sigma_{\hat{s}_0}$). The small asymmetry is produced by the fact that we are approximating $s_0$ in Eq. \ref{eq:poissonvariance} by its estimated value $\hat{s}_0$, which yields larger error bars for the clusters with overestimated flux and smaller error bars for the clusters with underestimated flux.

Figures \ref{fig:simupxcc_vs_z} to \ref{fig:simupxcc_vs_snr} show the difference between the extracted and the injected value, divided by the estimated standard deviation $\sigma_{\hat{s}_0}$ as a function of the redshift, the size, the flux, and the S/N of each cluster. The extraction behaves correctly for all the values of these parameters, and they do not introduce any systematic error or bias in the results.

\subsection{Performance evaluation: Extraction of real clusters}\label{ssec:xray_real}
Given that the X-ray matched filter performs well when applied to simulated clusters, the next step is to check its performance on real clusters. This section presents the results of the extraction of known clusters from real X-ray data, assuming that we know their position and size. The objective is to check whether the photometry is still correct for real clusters, and also to quantify the detection probability in this case. 

\subsubsection{Extraction of MCXC clusters} 
\begin{figure}[]
   		\centering
   		\includegraphics[width=\columnwidth]{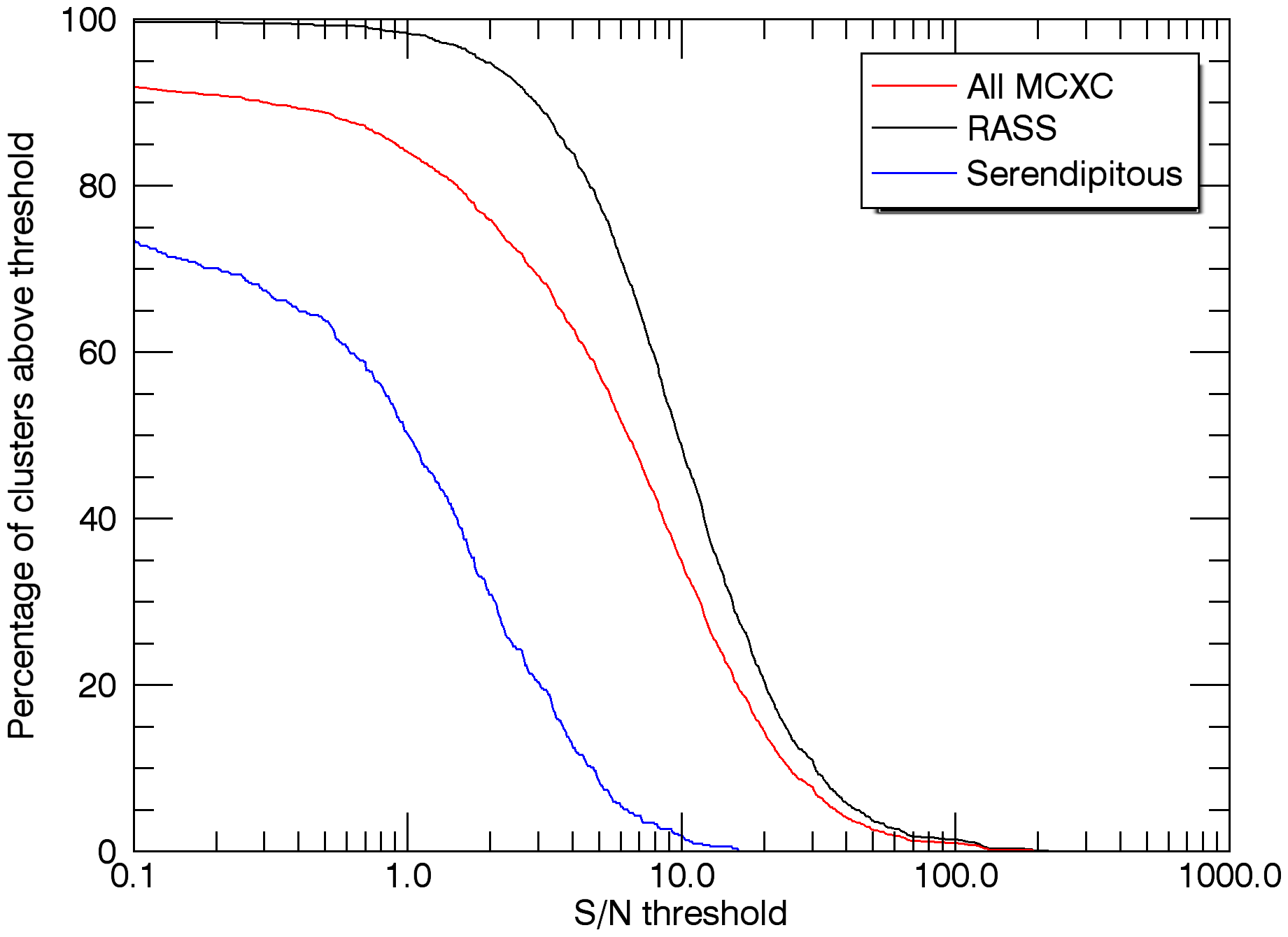}
   		\caption{Percentage of MCXC clusters whose extracted S/N, using the proposed X-ray matched filter and assuming the position and size of the clusters are known, is above a given S/N threshold. Red corresponds to the complete MCXC sample, while black and blue correspond to the RASS and serendipitous subsamples, respectively. S/N is defined here as the estimated signal $\hat{s}_0$ divided by the estimated background noise $\sigma_{\theta_{\rm s}}$.}
   		\label{fig:detections_vs_snr_realmcxc_rosatonly}
   	\end{figure}

We started the analysis with the 1743 clusters of the MCXC sample \citep{Piffaretti2011}. We extracted these clusters on $10\degr \times 10\degr$ patches centered on the cluster position, following the same procedure as in the previous section. 
Figures \ref{fig:realpxcc} and \ref{fig:detections_vs_snr_realmcxc_rosatonly} show the extraction results for this experiment, where we divided the MCXC sample into two subsamples: clusters that were originally detected in RASS and clusters detected in ROSAT serendipitous (deeper) observations. 

Figure \ref{fig:extraction_pxcc} shows the extracted value of $L_{500}$ for each cluster as a function of the published value. As in the simulations, we used the redshift in the catalogue to convert from flux to $L_{500}$. 
The extracted flux follows the published flux quite well, but the dispersion is larger than in the simulations (Fig. \ref{fig:extraction_simupxcc}). The best linear fit to these data is given by $y=0.913(\pm0.004)x-3.76(\pm0.29)\cdot10^{-3}$.

Figure \ref{fig:hist_pxcc} shows the histogram of the difference between the extracted and the published value, divided by the estimated standard deviation $\sigma_{\hat{s}_0}$. Some of the properties of this histogram are summarized in Table \ref{table:simupxcc}. In this histogram we show again that there is no bias for the complete sample, but this time the estimated error bars do not describe the dispersion of the results (as the 68$\%$ of the extractions fall in an interval that is almost $\pm2\sigma_{\hat{s}_0}$). It is reasonable to assume that this additional dispersion comes from the difference between the profile used for extraction and the real profile of each particular cluster (see Sect. \ref{sssec:prof_mismatch}). We also need to take into account that the published flux has some uncertainties It is difficult to characterize how these affect the dispersion in our histogram because some of the clusters were observed in RASS, that is, using the same data as we used, and others were serendipitous (deeper) observations, for which the published flux is expected to have a lower uncertainty.

\begin{figure*}[]
	\centering
	\subfigure[]{\includegraphics[width=.98\columnwidth]{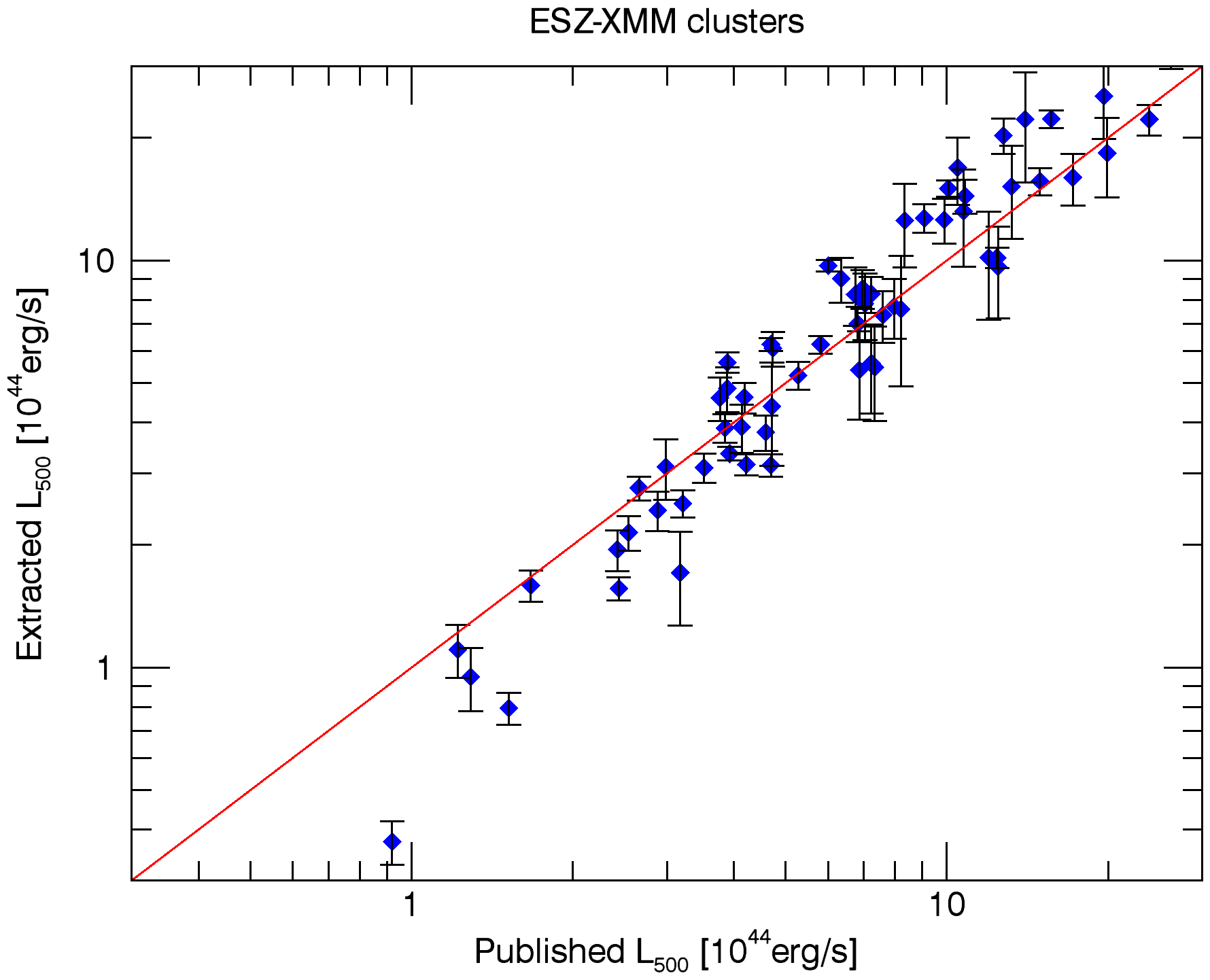}\label{fig:extraction_esz_aver}}
	\subfigure[]{\includegraphics[width=.98\columnwidth]{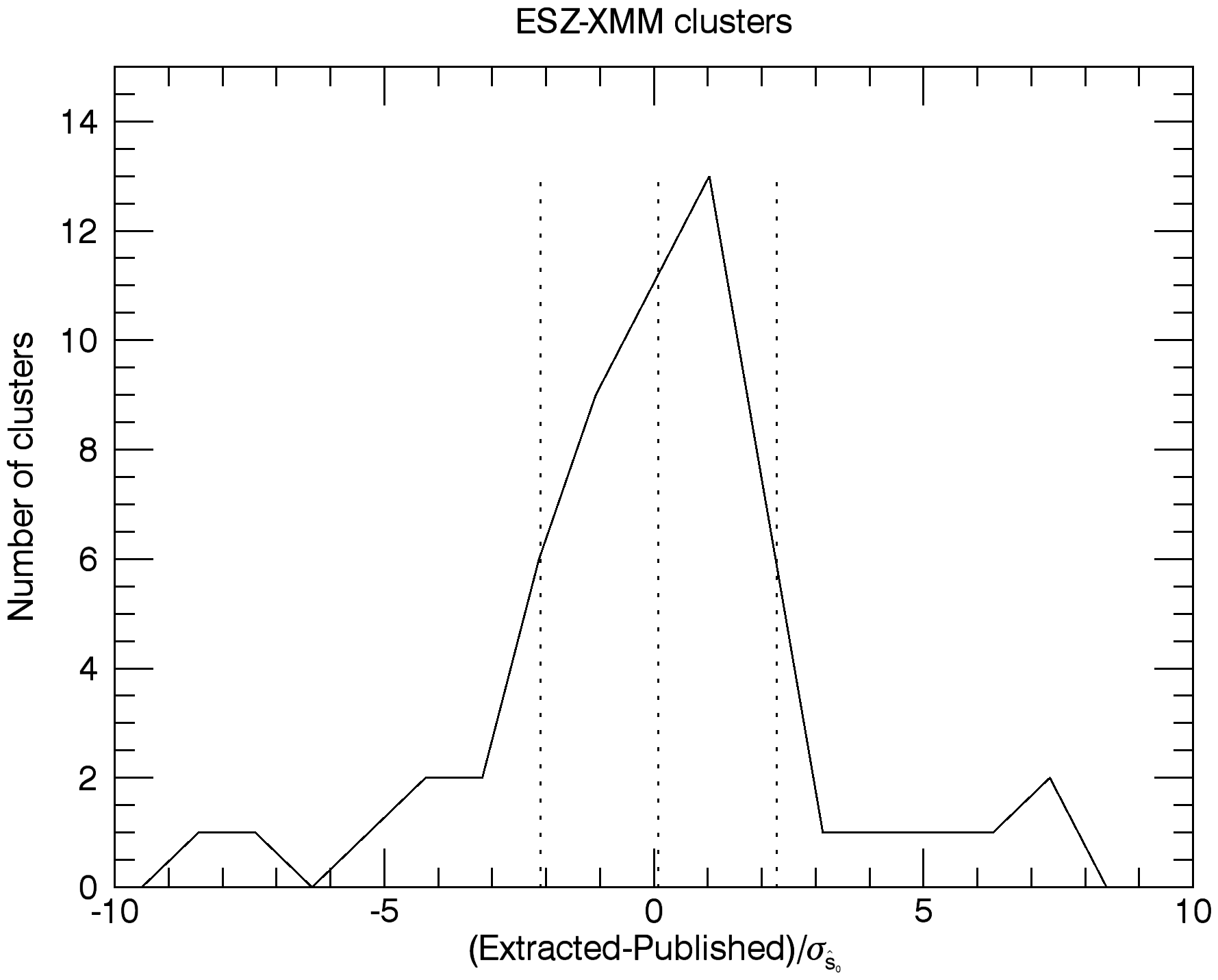}\label{fig:hist_esz_aver}}
	\subfigure[]{\includegraphics[width=.98\columnwidth]{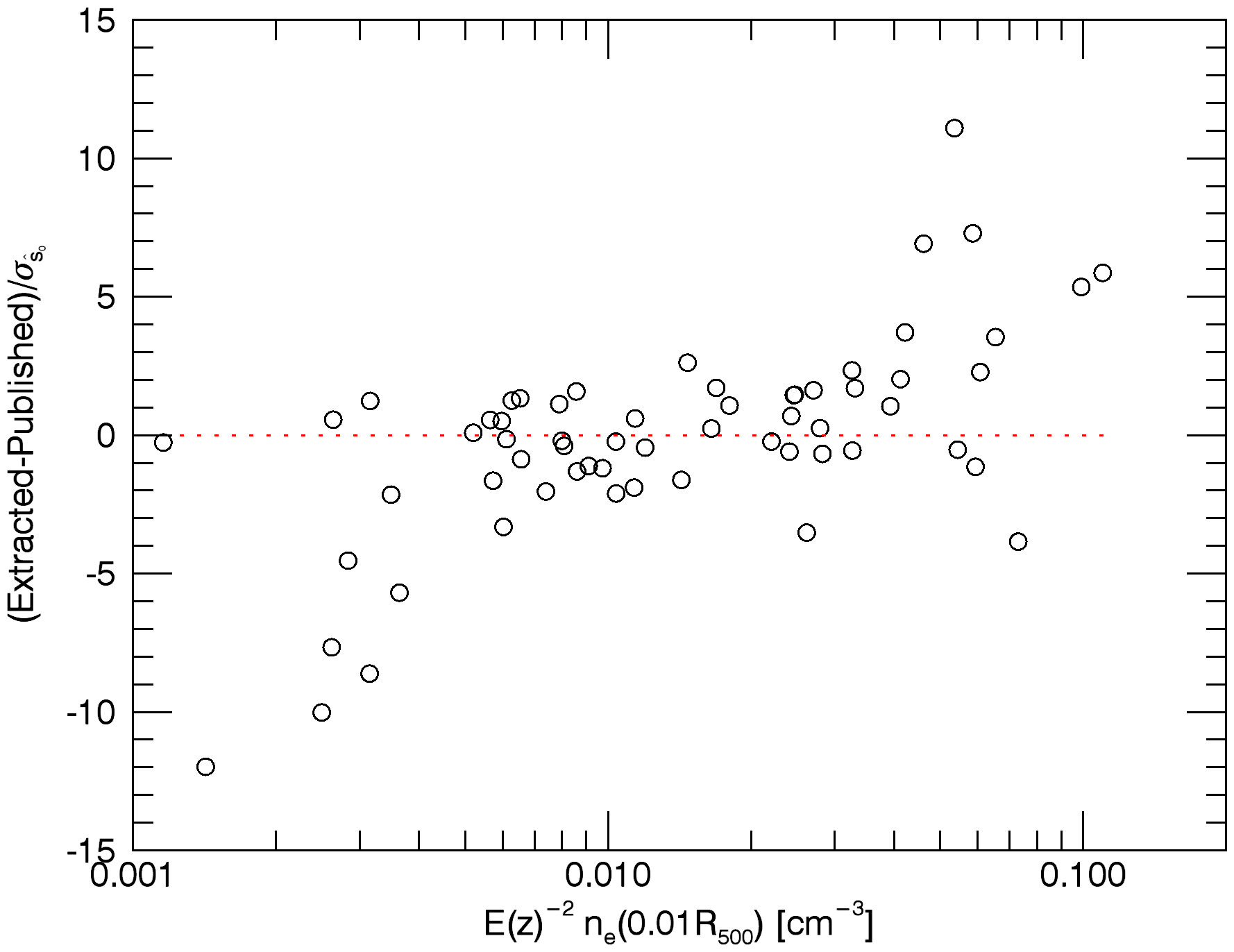}\label{fig:dispdens_esz_aver}}
	\subfigure[]{\includegraphics[width=.98\columnwidth]{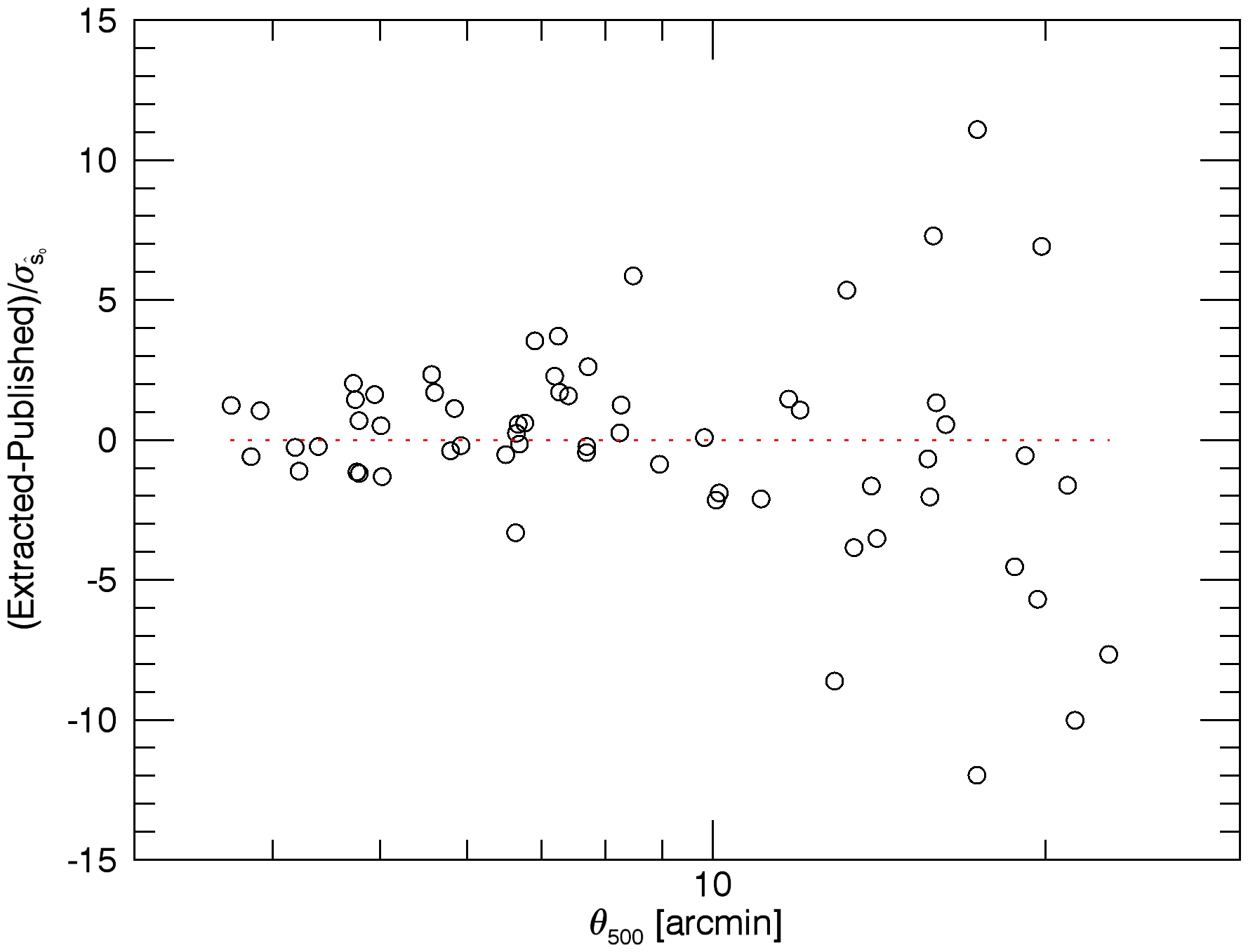}\label{fig:dispsize_esz_aver}}
	\caption{Photometry results of the extraction of the ESZ-XMM clusters using the proposed X-ray matched filter with the average cluster profile and assuming the position and size of the clusters are known. Top left panel: Extracted versus published $L_{500}$. 
		The error bars correspond to the estimated $\sigma_{\hat{s}_0}$ (scaled to $L_{500}$ units). Top right panel: Histogram of the difference between the extracted and the published $L_{500}$, divided by the estimated $\sigma_{\hat{s}_0}$. The central vertical line shows the median value, whereas the other two vertical lines indicate the region inside which 68$\% $ of the clusters lie. Bottom panels: Difference between the extracted and the published $L_{500}$, divided by the estimated $\sigma_{\hat{s}_0}$ as a function of (c) the scaled central density of the cluster (at 0.01$R_{500}$), which is an indicator of the shape of the cluster profile, and (d) the size $\theta_{500}$ of the cluster.}
	\label{fig:esz_aver}
\end{figure*}

\begin{figure*}[]
	\centering
	\subfigure[]{\includegraphics[width=.98\columnwidth]{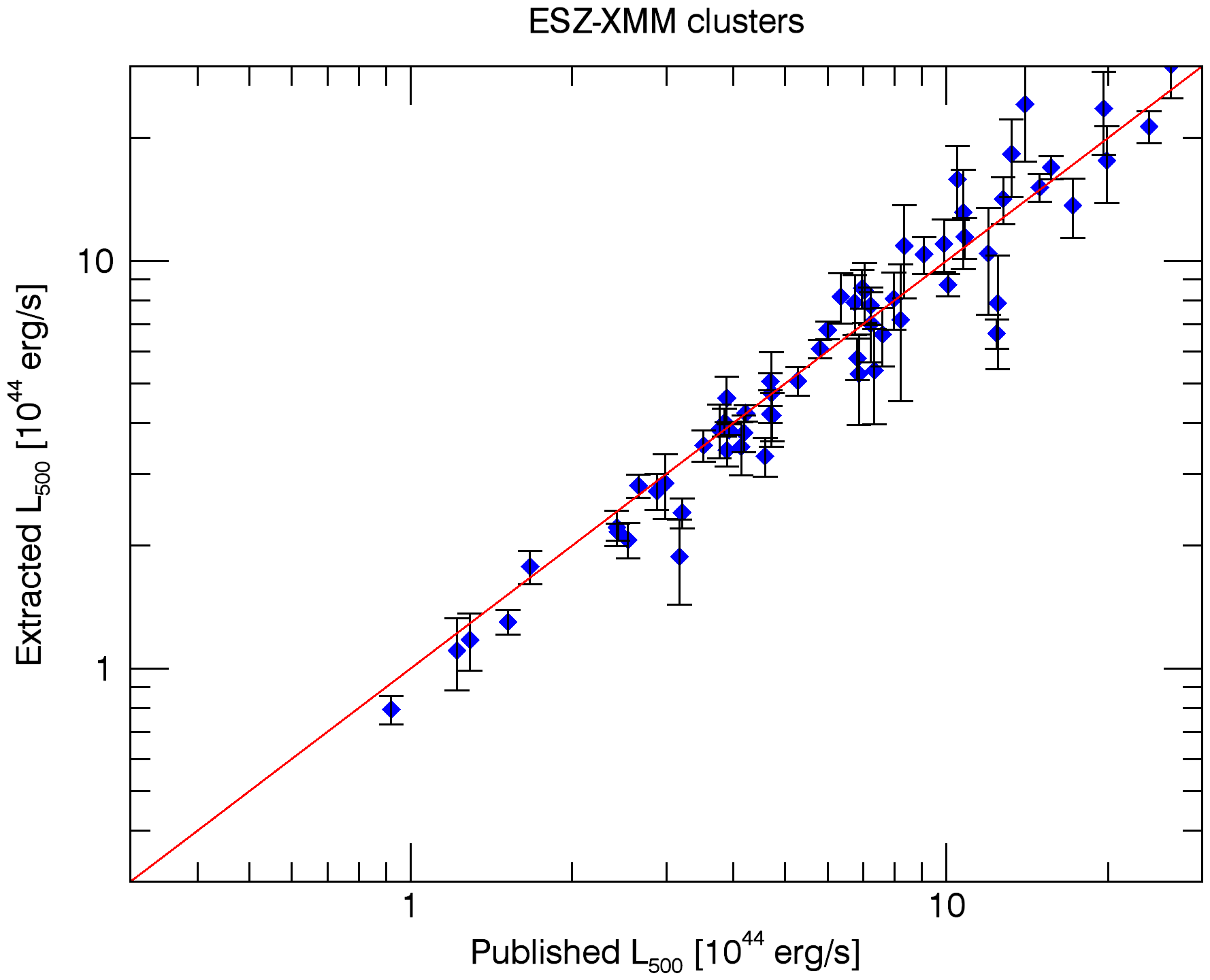}\label{fig:extraction_esz_indiv}}
	\subfigure[]{\includegraphics[width=.98\columnwidth]{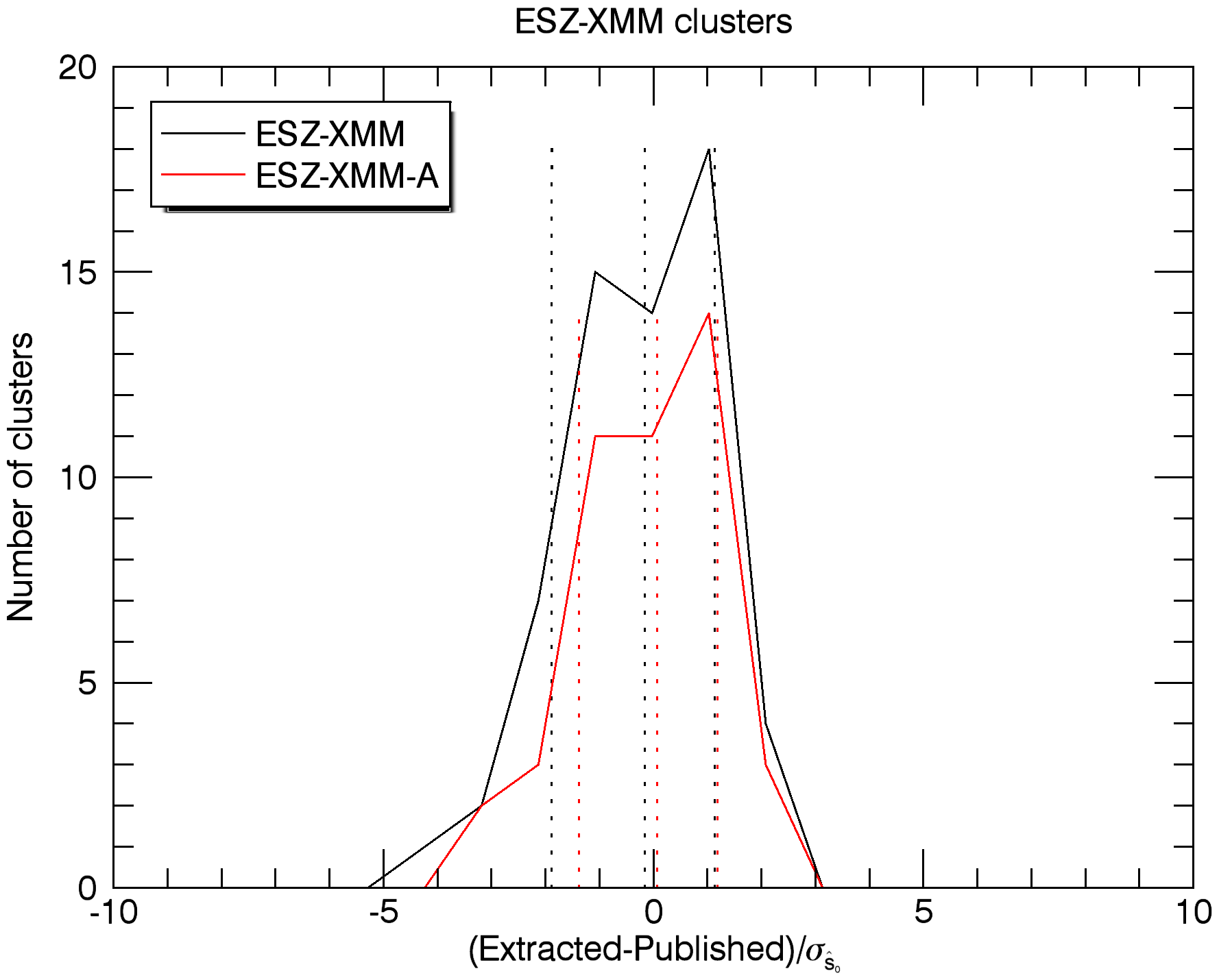}\label{fig:hist_esz_indiv}}
	\subfigure[]{\includegraphics[width=.98\columnwidth]{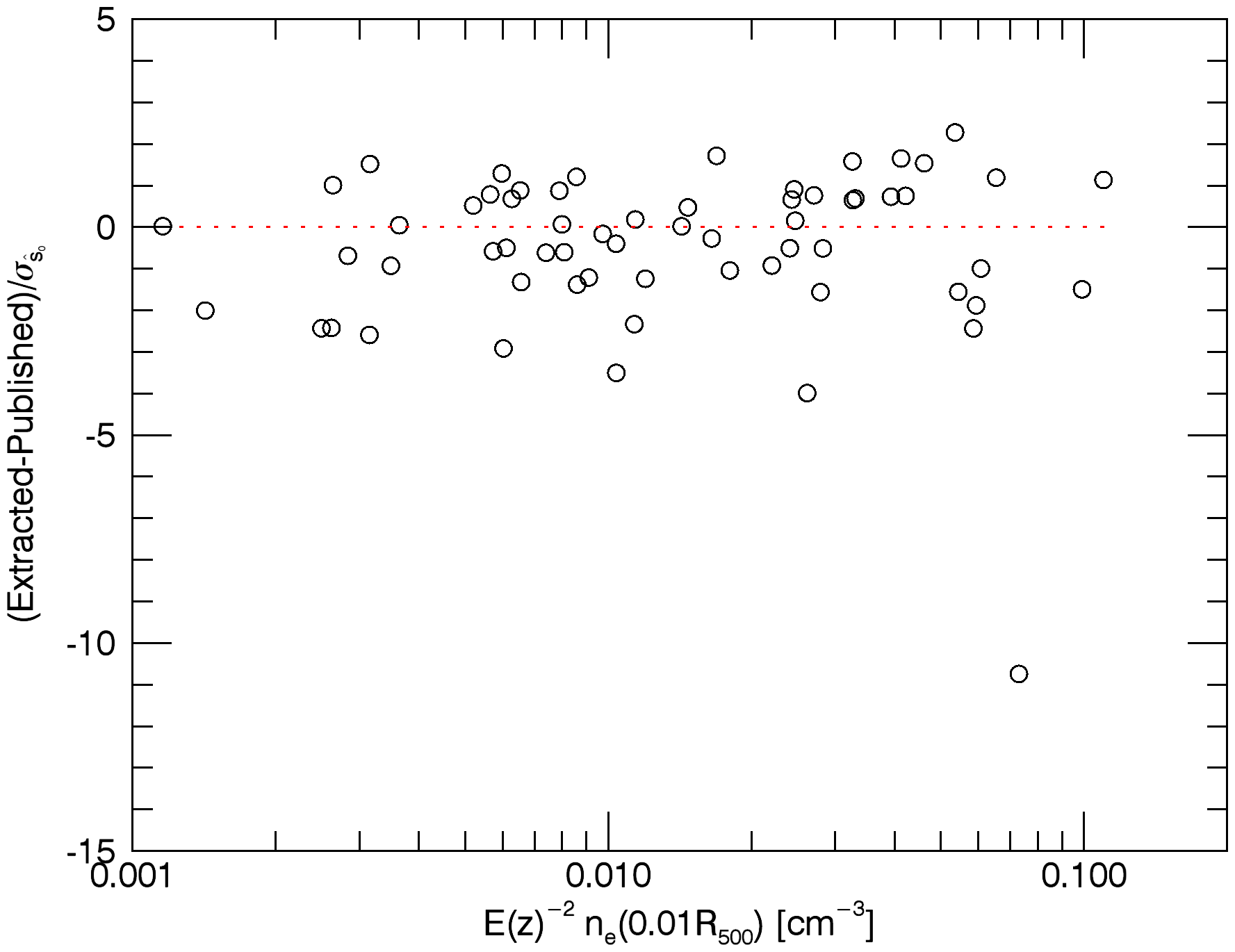}\label{fig:dispdens_esz_indiv}}
	\subfigure[]{\includegraphics[width=.98\columnwidth]{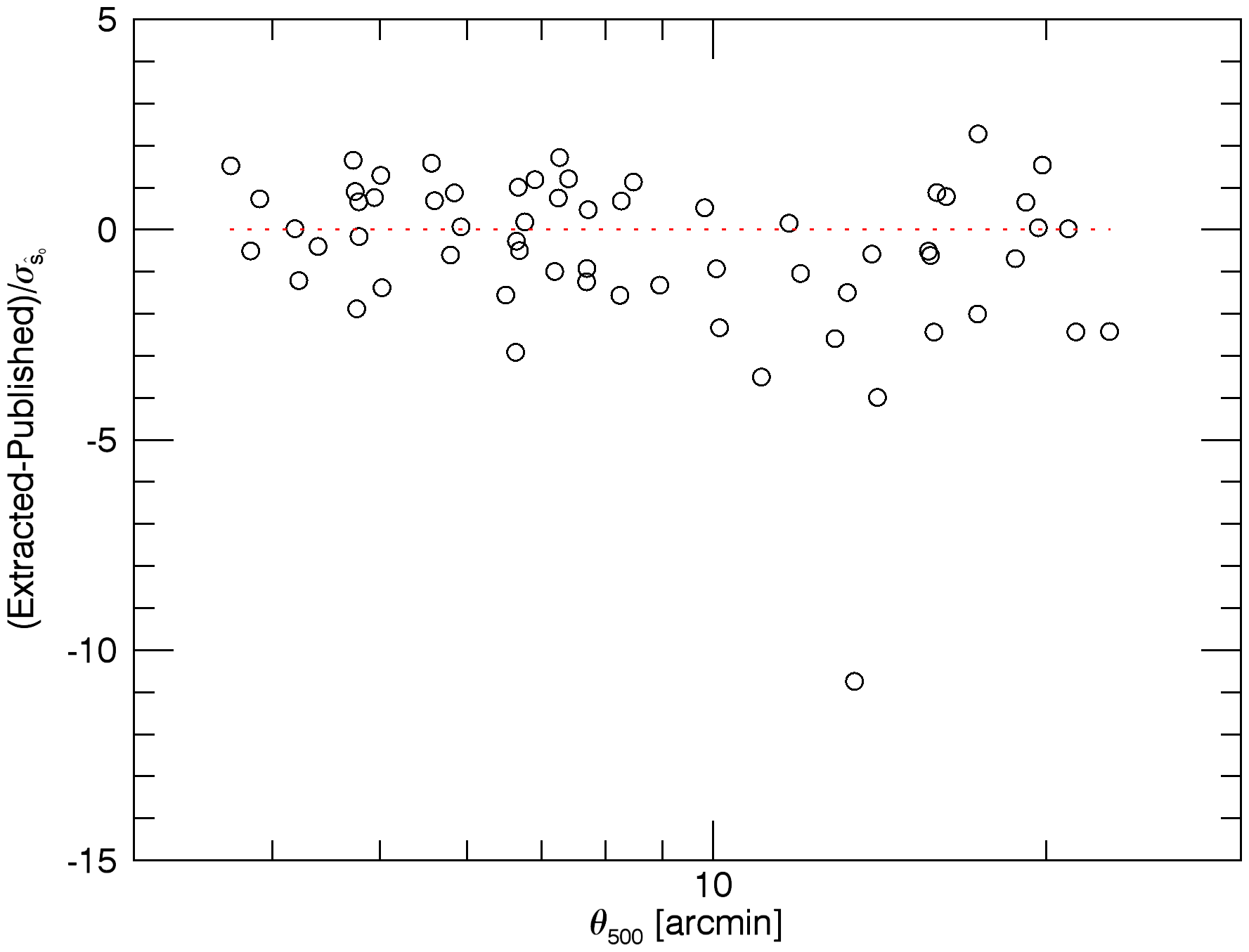}\label{fig:dispsize_esz_indiv}}
	\caption{Photometry results of the extraction of the ESZ-XMM clusters using the proposed X-ray matched filter with the individual cluster profiles and assuming the position and size of the clusters are known. The four panels are analogous to those in Fig. \ref{fig:esz_aver}. The top right panel includes in this case two histograms, one corresponding to the whole ESZ-XMM sample (black) and another corresponding to the ESZ-XMM-A subsample (red), which includes the clusters well within the XMM-\emph{Newton} field of view.}
	\label{fig:esz_indiv}
\end{figure*}

Figures \ref{fig:realpxcc_vs_z} to \ref{fig:realpxcc_vs_snr} show the difference between the extracted and the published value, divided by the estimated standard deviation $\sigma_{\hat{s}_0}$ as a function of the redshift, the size, the flux, and the S/N of each cluster. The extraction behaves correctly for almost all the values of these parameters (except for clusters with a very large apparent size, usually very nearby in redshift, which are not the main objects of our interest), and no systematic error is introduced in our region of interest (more distant clusters).

Figure \ref{fig:detections_vs_snr_realmcxc_rosatonly} shows the percentage of clusters whose extracted S/N is above a given S/N threshold, which is an indicator of the detection probability of our method. We defined this S/N threshold in terms of the estimated signal $\hat{s}_0$ divided by the estimated \emph{\textup{background}} noise $\sigma_{\theta_{\rm s}}$ (and not by the estimated total noise $\sigma_{\hat{s}_0}$), as done in the classical SZ MMF. For the clusters that were originally detected with RASS observations (the same as we used here), our method finds 95\% of them above a S/N threshold of 2, 90\% above a S/N threshold of 3, and 84\% above a S/N threshold of 4. We recall that the proposed method was designed to be compatible with the MMF used for SZ cluster detection, meaning that it is not specifically optimized for the detection of X-ray clusters. Nevertheless, its performance in this sense is satisfactory. Obviously, our method is not able to detect many of the serendipitous clusters because they were originally detected using deeper observations, but still, some of them are detected: 31\%, 20\%, and 13\% above S/N thresholds of 2, 3, and 4, respectively.

\subsubsection{Effect of profile mismatch}\label{sssec:prof_mismatch}

As we mentioned above, the additional dispersion we found in the extraction of real clusters may come from the mismatch between the cluster profile and the profile used for the extraction. Since we do not know the real profiles of all the MCXC clusters, we checked the effect of the profile mismatch with the ESZ-XMM sample \citep{PlanckEarlyXI}, a well-studied cluster sample composed of 62 clusters detected at high S/N in the first \textit{Planck} data set and present in \textit{XMM-Newton} archival data. We chose this sample because its good quality data allows accurately computing the individual cluster profiles.

These 62 clusters were extracted from $10\degr \times 10\degr$ patches centered at the cluster positions. 
We repeated the extraction of these clusters twice. First, we performed the extraction using the average profile defined by Eq. \ref{eq:pressure_prof} with parameters given in Eq. \ref{eq:xray_param}, as previously. Second, we also performed the extraction using the individual profile of each cluster, which was obtained by fitting an AB model (Eq. \ref{eq:densprof}) to the density profile data used to derive the pressure profiles presented by the \cite{PlanckIntV2013}, obtained from \textit{XMM-Newton} observations.

Figure \ref{fig:esz_aver} shows the results of extracting the 62 ESZ-XMM clusters using the average profile, and Fig. \ref{fig:esz_indiv} shows the results using the individual profile of each cluster. 
Figures \ref{fig:extraction_esz_aver} and \ref{fig:extraction_esz_indiv} show that the extracted flux is consistent with the published flux, as there is no bias (see also Figs. \ref{fig:hist_esz_aver} and \ref{fig:hist_esz_indiv}). The dispersion in Fig. \ref{fig:hist_esz_aver} shows the same behaviour as for the MCXC clusters (Fig. \ref{fig:hist_pxcc}): when the average profile is used, the estimated error bars are not enough to describe the dispersion of the results (68$\%$ of the extractions fall in the interval $\pm2\sigma_{\hat{s}_0}$). Figure \ref{fig:dispdens_esz_aver} shows that in this case the extracted flux value depends on the shape of the "real" profile of the cluster: if the cluster is very peaked, we tend to overestimate the flux, whereas if the cluster has a flat profile, we tend to underestimate its flux. This effect is especially strong in clusters with larger apparent size $\theta_{500}$, as shown in Fig. \ref{fig:dispsize_esz_aver}. However, when individual profiles, which are better matched to the "real" profiles of the cluster, are used, this dependency disappears, as shown in Figs. \ref{fig:dispdens_esz_indiv} and \ref{fig:dispsize_esz_indiv}. The dispersion of the results when we used the individual profiles for the extraction (Fig. \ref{fig:hist_esz_indiv}) is smaller than when we used the average profile, but it is not completely well characterized by the estimated standard deviation. However, if we focus on subsample ESZ-XMM-A, which includes the clusters with $\theta_{500} < 12$ arcmin (well within the \textit{XMM-Newton} field of view), the result is much better. 
We therefore conclude that the additional dispersion when we use the average profile for extraction comes mainly from the profile mismatch.

Since the profile mismatch produces an additional scatter in the estimated flux, it will also produce an additional scatter in the estimated S/N of the clusters. The average S/N, and consequently the global detection probability, will not be affected, but clusters with a peaked profile will be more easily detected than clusters with a flat profile, as in standard detection techniques.

   	\begin{table}
   		\caption{Main properties of the histograms in Figs. \ref{fig:hist_simupxcc}, \ref{fig:hist_pxcc}, and \ref{fig:hist_pxcc_corr}, corresponding to the extraction of clusters using the proposed X-ray matched filter. The first column corresponds to the histogram of the extraction results for the simulated MCXC clusters (Fig. \ref{fig:hist_simupxcc}). The second and third columns correspond to the histograms of the extraction results for the real MCXC clusters, before (Fig. \ref{fig:hist_pxcc}) and after (Fig. \ref{fig:hist_pxcc_corr}) correction for the profile mismatch effect.}
   		\label{table:simupxcc}
   		\centering 
   		\begin{tabular}{c c c c}
   			\hline
   			\noalign{\smallskip}
   			&  Fig. \ref{fig:hist_simupxcc} & Fig. \ref{fig:hist_pxcc} & Fig. \ref{fig:hist_pxcc_corr}\\
   			\noalign{\smallskip}
   			\hline
   			\noalign{\smallskip}
   			Median 				& -0.032 & -0.042 & -0.023\\
   			Mean        		& -0.086 & +0.018 & -0.071\\
   			Skewness    		& -1.799 & +2.744 & -0.759\\
   			Kurtosis    		& +18.280 & +61.560 & +9.325\\
   			Standard deviation 	& 1.020  & 3.539  & 1.422\\
   			68\% lower limit 	& -0.946 & -1.791 & -1.090\\
   			68\% upper limit 	& +0.839 & +1.684 & +0.976\\
   			\noalign{\smallskip}
   			\hline
   		\end{tabular}
   	\end{table}

\subsection{Practical form of the algorithm}\label{ssec:finalxrayerror}
In practice, we do not know the exact profile of the clusters we will detect, so in any case we need to use the average profile. If we assume that the additional dispersion we will have in this case is due to the profile mismatch, we could correct for it using a simple expression depending only on the apparent size of the cluster. Thus, we propose to correct the estimated standard deviation $\sigma_{\hat{s}_0}$, given in Eq. \ref{eq:totalvariance}, in the following way:
\begin{equation}\label{eq:sigma_corrected}
\sigma_{\rm corr}^2 =  \sigma_{\hat{s}_0}^2  \cdot \left( 1+0.1 \frac{\theta_{500}}{1 \rm arcmin}\right) ^2.
\end{equation}
The correction factor was obtained by calculating for the MCXC clusters in several $\theta_{500}$ bins the standard deviation of the difference between the extracted and the published value, divided by the estimated $\sigma_{\hat{s}_0}$, and checking that it increased roughly linearly with $\theta_{500}$ and tended to 1 when $\theta_{500}\rightarrow 0$. It is important to remark that this correction factor is not universal: it depends on the beam, on the cluster sample we are working with (meaning that it depends on the selection function), and on the evolution of the sample. However, this correction is stronger in the regime we are not interested in (clusters with large apparent sizes, hence at low redshift), therefore we consider it as a good approximation for our purposes.

\begin{figure}[]
	\centering
	\includegraphics[width=\columnwidth]{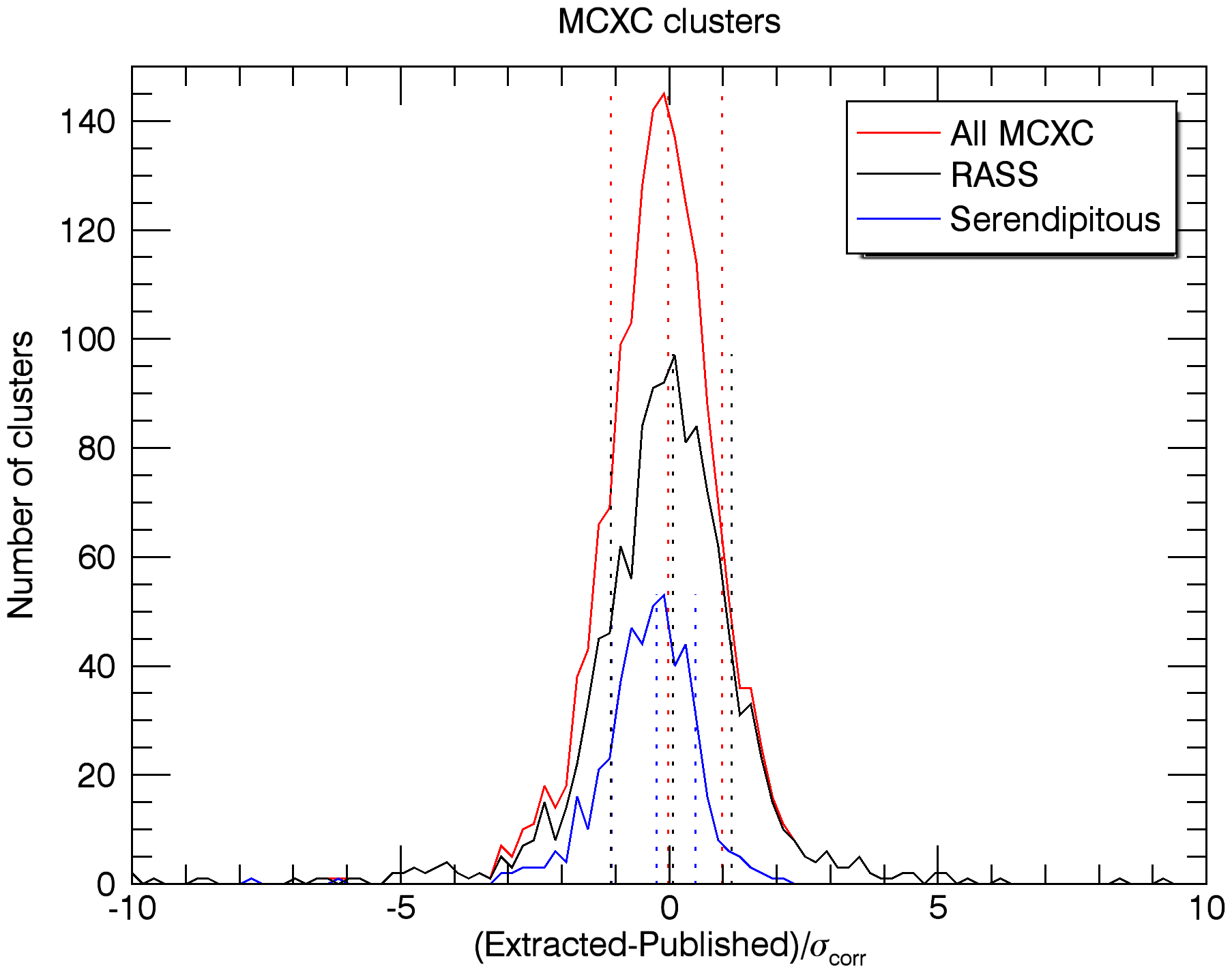}
	\caption{Histogram of the difference between the extracted and the published $L_{500}$, divided by the standard deviation after correcting for the effect of the profile mismatch ($\sigma_{\rm corr}$, scaled to $L_{500}$ units) for the MCXC clusters extracted with the proposed X-ray matched filter using the average profile and assuming the position and size of the clusters are known. Red corresponds to the complete MCXC sample, while black and blue correspond to the RASS and serendipitous subsamples, respectively. For each color, the central vertical line shows the median value, whereas the other two vertical lines indicate the region inside which 68$\% $ of the clusters lie.}
	\label{fig:hist_pxcc_corr}
\end{figure}

When we apply this empirical correction to the estimated $\sigma_{\hat{s}_0}$ in the MCXC cluster sample, we obtain the histogram in Fig. \ref{fig:hist_pxcc_corr} (see main properties summarized in Table \ref{table:simupxcc}), which again shows that there is no bias and that the corrected error bars now describe  the dispersion on the results well (as the 68$\%$ of the extractions fall in the interval $\pm1\sigma$).

\section{Matched multifilter (MMF) for SZ cluster detection}\label{sec:szdetection}

The matched multifilter (MMF) is a well-studied approach that was developed for SZ detection within the \textit{Planck} mission \citep{PlanckEarlyVIII}. It has also been used to detect clusters in other SZ surveys, such as the South Pole Telescope (SPT) survey \citep{Bleem2015} and the Atacama Cosmology Telescope (ACT) survey \citep{Hasselfield2013}. In this section we recall how it works, since its formulation is used for the joint X-ray-SZ extraction technique.

When CMB photons pass through a galaxy cluster, they can interact with the high-energy electrons in the ICM, gaining energy in the process. This effect, known as thermal Sunyaev-Zel'dovich (SZ) effect, produces a small distortion in the CMB spectrum, which can be observed as a temperature change relative to the mean CMB temperature $T_{\rm CMB}$ \citep{Sunyaev1970,Sunyaev1972}. The frequency dependency of this spectral distortion is universal in the non-relativistic limit, while its amplitude, given by the Compton $ y $ parameter (proportional to the integral of the gas pressure along the line of sight), depends on the cluster and its spatial profile \citep{Carlstrom2002,Birkinshaw1999}.

The brightness profile of a cluster as a function of the observation frequency $\nu$ can be written as 
\begin{equation}
\frac{\Delta T}{T_{\rm CMB}} \left( \mathbf{x}, \nu\right) =  y(\mathbf{x}) j(\nu) = y_{0}\tilde{T}_{\theta_{\rm s}}(\mathbf{x}) j(\nu),
\end{equation}
where $j(\nu)$ is the universal dependency on frequency of the SZ signal and $ y(\mathbf{x}) $ is the Compton $y$ parameter at position $ \mathbf{x} $, which can be decomposed into $y_0$, the cluster central $y$-value, multiplied by $ \tilde{T}_{\theta_{\rm s}}(\mathbf{x}) $, a normalized cluster spatial profile (normalized so that its central value is 1). 

Let us imagine that we have carried out an SZ survey covering a certain region of the sky at $N_\nu$ observation frequencies $\nu_i$ ($i=1, ..., N_\nu$), producing $N_\nu$ survey maps, and let us denote the instrument beam at observation frequency $\nu_i$ by $ B_{\nu_{i}}(\mathbf{x}) $. Let us further assume there is a cluster in the observed region, at a position $\mathbf{x}_0$, characterized by a profile $ \tilde{T}_{\theta_{\rm s}}(\mathbf{x}) $ and a central $y$-value $y_0$.  The set of survey maps will contain the SZ signal of the cluster plus noise and can be expressed in matrix form as
\begin{equation}\label{eq:SZ_map}
{\mathbf{M}}(\mathbf{x}) = y_{0} \mathbf{F}_{\theta_{\rm s}}(\mathbf{x}-\mathbf{x}_0) + {\mathbf{N}}(\mathbf{x}),
\end{equation}
where $\mathbf{M}(\mathbf{x})$ is a column vector whose $i$th component is the map at observation frequency $\nu_i$: 
$\mathbf{M}(\mathbf{x})=[M_{1}(\mathbf{x}), ..., M_{N_\nu}(\mathbf{x})]^{\rm T}$, $\mathbf{F}_{\theta_{\rm s}}(\mathbf{x})$ is a column vector whose $i$th component is given by $F_{i}(\mathbf{x}) = j(\nu_i) T_{i}(\mathbf{x})$, where $j(\nu_i)$ is the SZ spectral function at frequency $\nu_i$ and $T_{i}(\mathbf{x})$ is the normalized cluster profile convolved with the instrument beam at frequency $\nu_i$ ($T_{i}(\mathbf{x}) = \tilde{T}_{\theta_{\rm s}}(\mathbf{x}) \ast B_{\nu_i}(\mathbf{x})$), and  $\mathbf{N}(\mathbf{x})$ is a column vector whose $i$th component is the noise map at observation frequency $\nu_i$: $\mathbf{N}(\mathbf{x})=[N_{1}(\mathbf{x}), ..., N_{N_\nu}(\mathbf{x})]^{\rm T}$. In this context, noise means anything that is not the SZ signal, that is, instrumental noise and astrophysical foregrounds, such as extragalactic point sources, diffuse Galactic emission, and the primary CMB anisotropy.

There is a clear analogy between the SZ maps defined in Eq. \ref{eq:SZ_map} and the X-ray map defined in Eq. \ref{eq:Xray_map}. Again, if the cluster profile is known, the problem reduces to estimating the amplitude $y_0$ of a known signal from an observed signal contaminated by noise. Therefore, a linear estimator, $\hat{y}_0$, of $y_0$ can be constructed as a linear combination of the observed data (the $N_\nu$ observed maps in this case):
\begin{equation}\label{eq:y0_estim}
\hat{y}_{0} = \sum_{\mathbf{x}} \mathbf{\Psi}_{\theta_{\rm s}}^{\rm T}(\mathbf{x}-\mathbf{x}_0) {\mathbf{M}}(\mathbf{x}),
\end{equation}
where the $N_\nu \times 1 $ column vector $\mathbf{\Psi}_{\theta_{\rm s}}$ can be interpreted as a filter whose $i$th component will filter the map at observation frequency $\nu_i$. 
As in the X-ray case, when we restrict this linear estimator to be unbiased and to have minimum variance, we obtain the following expression for the filter in Fourier space \citep{Haehnelt1996, Herranz2002, Melin2006, Melin2012}:
\begin{equation}\label{eq:filter_sz}
	\mathbf{\Psi}_{\theta_{\rm s}}(\mathbf{k}) = \sigma_{\theta_{\rm s}}^2 \mathbf{P}^{-1}(\mathbf{k})  \mathbf{F}_{\theta_{\rm s}}(\mathbf{k}),
\end{equation}
where
\begin{equation}\label{eq:sigma_sz}
	\sigma_{\theta_{\rm s}}^2 =  \left[ \sum_{\mathbf{k}}   \mathbf{F}_{\theta_{\rm s}}^{\rm T}(\mathbf{k})  \mathbf{P}^{-1}(\mathbf{k})  \mathbf{F}_{\theta_{\rm s}}(\mathbf{k}) \right] ^{-1}
\end{equation}
is the total noise variance after filtering and $\mathbf{P}(\mathbf{k})$ is the noise power spectrum, a $N_\nu \times N_\nu$ matrix whose $ij$ component is given by $\left\langle N_i(\mathbf{k})N_j^\ast(\mathbf{k}')\right\rangle  = P_{ij}(\mathbf{k}) \delta(\mathbf{k}-\mathbf{k}')$.

As in the X-ray case, this approach relies on the knowledge of the cluster brightness profile $\tilde{T}_{\theta_{\rm s}}(\mathbf{x})$, which is not known in practice. Therefore, we again need to use a theoretical profile that represents as well as possible the average brightness profile of the clusters we wish to detect. \citet{Melin2006} used a projected spherical $\beta$-profile with $\beta=2/3$ to describe $\tilde{T}_{\theta_{\rm s}}(\mathbf{x})$ as a function of the characteristic scale radius $\theta_{\rm s}$. The \citet{Planck2013ResXXIX} assumed that the 3D pressure profile of the cluster followed the GNFW profile of \citet{Arnaud2010}, given by Eq.  \ref{eq:pressure_prof}
with the parameters
\begin{equation}\label{eq:sz_param}
\left[ \alpha, \beta, \gamma, c_{500}\right]   = \left[ 1.0510, 5.4905, 0.3081, 1.177\right]. 
\end{equation}
The cluster profile $\tilde{T}_{\theta_{\rm s}}(\mathbf{x})$ can be then obtained by numerically integrating the cluster 3D pressure profile in Eq.  \ref{eq:pressure_prof} along the line of sight. We will also adopt this model for the SZ cluster profile.

Finally, as in the X-ray case, we need to convolve this cluster profile by the instrument beams $ B_{\nu_{i}}(\mathbf{x}) $. We used the six highest frequency \textit{Planck} maps, from 100 to 857 GHz, and assumed that the PSF of the instrument is Gaussian, with FWHM between 9.66 and 4.22 arcmin, depending on the frequency, as shown in Table 6 of \citet{Planck2015ResVIII}.

\section{Joint extraction of galaxy clusters on X-ray and SZ maps}\label{sec:jointdetection}

In this section, the proposed joint X-ray-SZ extraction algorithm is described and evaluated. 

\subsection{Description of the algorithm}\label{ssec:jointalgorithm}

The main idea of our joint extraction algorithm is to consider the X-ray map as an additional SZ map at a given frequency and to introduce it, together with the other SZ maps, into the classical SZ-MMF described in Sect. \ref{sec:szdetection}. To do so, we need to convert our X-ray map into an equivalent SZ map at a reference frequency $ \nu_{\rm ref} $, leveraging the expected $F_{\rm X}/Y_{500}$ relation. The details of this conversion are described in Appendix \ref{app:HealpixMap}. 

Once the X-ray map is expressed in the same units as the SZ maps (we used $\Delta T/T_{\rm{CMB}}$ units), the MMF described in Sect. \ref{sec:szdetection} can be applied almost directly to the complete set of maps (the original $N_{\nu}$ SZ maps obtained at observation frequencies $\nu_1, ..., \nu_{N_\nu}$ and an additional SZ map at the reference frequency $ \nu_{\rm ref} $, obtained from the X-ray map). If there is a cluster in the observed region at position $\mathbf{x}_0$, characterized by an SZ profile $ \tilde{T}_{\theta_{\rm s}}(\mathbf{x}) $, a central $y$-value $y_0$, and an X-ray profile $ \tilde{T}^{\rm{x}}_{\theta_{\rm s}}(\mathbf{x}) $, taking the conversion from the original X-ray map to its equivalent SZ map into account, this set of maps can be expressed in matrix form using Eq. \ref{eq:SZ_map} again, 
where $\mathbf{M}(\mathbf{x})$, $\mathbf{F}_{\theta_{\rm s}}(\mathbf{x})$ and $\mathbf{N}(\mathbf{x})$ are now column vectors with $N_{\nu}+1$ components, defined as
\begin{equation}
\mathbf{M}(\mathbf{x})=[M_{1}(\mathbf{x}), ..., M_{N_\nu}(\mathbf{x}), M_{\rm ref}(\mathbf{x})]^{\rm T},
\end{equation}
\begin{equation} \label{eq:F_joint}
\mathbf{F}_{\theta_{\rm s}}(\mathbf{x})  = 
[j(\nu_{1}) T_{1}(\mathbf{x}), ..., j(\nu_{N_{\nu}}) T_{N_{\nu}}(\mathbf{x}), C j(\nu_{\rm ref}) \tilde{T}^{\rm{x}}_{\theta_{\rm s}}(\mathbf{x})]^{\rm T},
\end{equation}
\begin{equation} 
\mathbf{N}(\mathbf{x})=[N_{1}(\mathbf{x}), ..., N_{N_\nu}(\mathbf{x}), N_{\rm ref}(\mathbf{x})]^{\rm T},
\end{equation}
where $T_{i}(\mathbf{x}) = \tilde{T}_{\theta_{\rm s}}(\mathbf{x}) \ast B_{\nu_i}(\mathbf{x})$, $ T^{\rm{x}}_{\theta_{\rm s}}(\mathbf{x}) = \tilde{T}^{\rm{x}}_{\theta_{\rm s}}(\mathbf{x}) \ast B_{\rm xray}(\mathbf{x})$ and the constant $C$ is the ratio of the integrated fluxes of the normalized SZ and X-ray 3D profiles up to $R_{500}$, that is,
\begin{equation}
C = \frac{\int\limits_{ x <1} p_{\rm SZ}(x) dx}{\int\limits_{ x < 1} p_{\rm X}(x) dx}.
\end{equation}
The subindex 'ref' denotes the component corresponding to the additional map. Therefore, $N_{\rm ref}(\mathbf{x})$ contains two types of noise: the background noise and the random noise in addition to the cluster signal that is due to Poisson fluctuations, as described in Sect. \ref{sec:xraydetection} ($N_{\rm ref}(\mathbf{x}) = N_{\rm sig}(\mathbf{x}) + N_{\rm bk}(\mathbf{x})$).

Again, if the cluster profile is known, the problem reduces to estimating the amplitude $y_0$ of a known signal from an observed signal contaminated by noise and, as before, its best linear unbiased estimator $\hat{y}_0$ is given by Eqs. \ref{eq:y0_estim}-\ref{eq:sigma_sz}, where the new $(N_\nu+1) \times 1 $ column vector $\mathbf{\Psi}_{\theta_{\rm s}}$ can be interpreted as a filter whose $i$th component will filter the map at observation frequency $\nu_i$, 
$\sigma_{\theta_{\rm s}}^2$ 
is, approximately, the background noise variance after filtering, $\mathbf{F}_{\theta_{\rm s}}$ is defined in Eq. \ref{eq:F_joint} and $\mathbf{P}(\mathbf{k})$ is the noise power spectrum, a $(N_\nu+1) \times (N_\nu+1)$ matrix whose $ij$ component is given by $\left\langle N_i(\mathbf{k})N_j^\ast(\mathbf{k}')\right\rangle  = P_{ij}(\mathbf{k}) \delta(\mathbf{k}-\mathbf{k}')$. Considering that the noise in the X-ray map and the SZ maps is uncorrelated, we can write the noise power spectrum as
\begin{equation}\label{eq:P_joint}
\mathbf{P}(\mathbf{k}) = \left[ \begin{array}{c c}
\mathbf{P}_{\rm SZ}(\mathbf{k}) & \mathbf{0}_{N_\nu \times 1} \\
 \mathbf{0}_{1 \times N_\nu} & P_{\rm X}(\mathbf{k})\\
\end{array}\right],
\end{equation}
where $\mathbf{P}_{\rm SZ}(\mathbf{k})$ is the noise power spectrum of the SZ maps, defined in Sect. \ref{sec:szdetection}, $P_{\rm X}(\mathbf{k})$ is the noise power spectrum of the X-ray map, defined in Sect. \ref{sec:xraydetection}, and $\mathbf{0}_{n \times m}$ denotes a vector with $n$ rows and $m$ columns whose elements are all equal to 0.

As in the X-ray case, the filtered map is composed of three terms: the amplitude of the cluster profile $y_{0}$, plus the filtered background noise and the filtered Poisson fluctuations on the X-ray signal. The variance of the filtered background noise is given (approximately) by Eq. \ref{eq:sigma_sz}  and the variance due to the Poisson fluctuations on the signal, after passing through the filter, can be written as
\begin{align}\label{eq:poissonvariance_joint}
	\sigma_{\rm J Poisson}^2 = \frac{u y_{0} C^3 j^3(\nu_{\rm ref}) \sigma_{\theta_{\rm s}}^4}{n^2}\sum_{\mathbf{k}}\sum_{\mathbf{k}'}\frac{T^{\rm{x}^\ast}_{\theta_{\rm s}}(\mathbf{k})T^{\rm{x}}_{\theta_{\rm s}}(\mathbf{k}')}{P_{\rm X}(\mathbf{k}) P_{\rm X}(\mathbf{k}')} T^{\rm{x}}_{\theta_{\rm s}}\left( \mathbf{k}-\mathbf{k}'\right),
\end{align}
where $n^2$ is the total number of pixels in the map, $u$ is the unit conversion factor from counts to the units of the SZ-equivalent X-ray map ($\Delta T/T_{\rm CMB}$ units in this case), and the double sum can be computed efficiently by making use of the cross-correlation theorem, as explained in Appendix \ref{app:sigma_poisson}. The derivation of this expression is completely analogous to that of the X-ray case included in Appendix \ref{app:sigma_poisson}, assuming uncorrelated X-ray and SZ noise. We note that this variance depends on $y_0$, the real value of the central $y$-value. As this is not known in practice, it is necessary to approximate it by its estimated value $\hat{y}_0$. 

Therefore, we can characterize our central $y$-value estimator $\hat{y}_{0}$ as a random variable with mean equal to the true central $y$-value $y_0$ and variance given by the sum of the variances of the filtered background noise and the filtered Poisson fluctuations on the signal. That is,
\begin{equation}\label{eq:totalvariance_joint}
\sigma_{\hat{y}_0}^2 =  \sigma_{\theta_{\rm s}}^2 + \sigma_{\rm J Poisson}^2,
\end{equation}
where we have assumed that the Poisson fluctuations on the signal are independent of the background noise. 

As in the previous cases, the cluster profile is not known, so we will need to approximate it by the theoretical profile that best represents the clusters we wish to detect. In this case, we again assumed the GNFW profile given by Eq. \ref{eq:pressure_prof}, with parameters given by Eqs. \ref{eq:sz_param} and \ref{eq:xray_param} for the components corresponding to the original SZ maps and the additional X-ray map, respectively. The cluster profile is then obtained by numerically integrating these GNFW profiles along the line of sight. Finally, as in the previous cases, we need to convolve this cluster profile by the instrument beams, for which we used the beams defined in Sects. \ref{sec:xraydetection} and \ref{sec:szdetection}.

\begin{table*}[]
	\caption{K-correction for $ T_{\rm ref} = 7$ keV.}             
	\label{table:K(z)}      
	\centering          
	\begin{tabular}{c c c c c c c c c c c c c}            
		\hline  
		\noalign{\smallskip}                  
		z    & 0.01 & 0.02 & 0.05 & 0.10 & 0.15 & 0.20 & 0.3  & 0.4  & 0.5  & 0.6  & 0.7  & 0.8    \\
		K(z) & 1.006& 1.011& 1.028& 1.056& 1.081& 1.106& 1.153& 1.198& 1.238& 1.276& 1.311& 1.344  \\
		\hline                  
	\end{tabular}
\end{table*}

\begin{figure*}[]
	\centering
	\subfigure{\includegraphics[width=1.98\columnwidth]{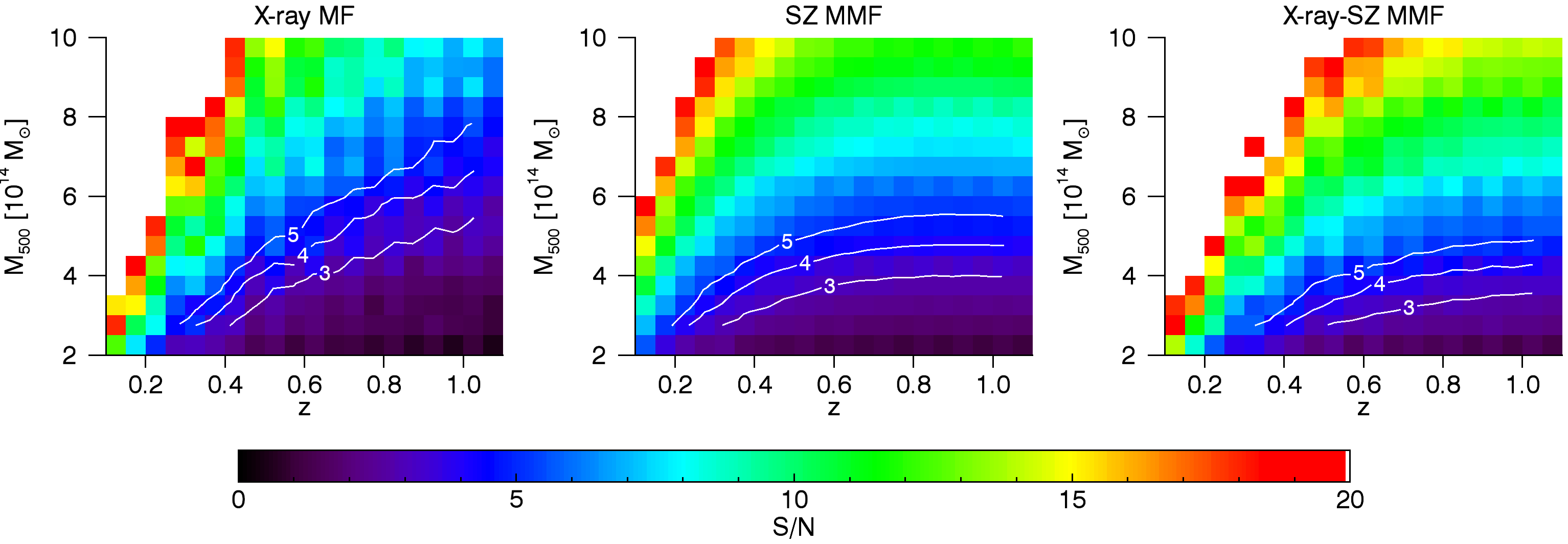}}
	\subfigure[]{\includegraphics[width=1.32\columnwidth]{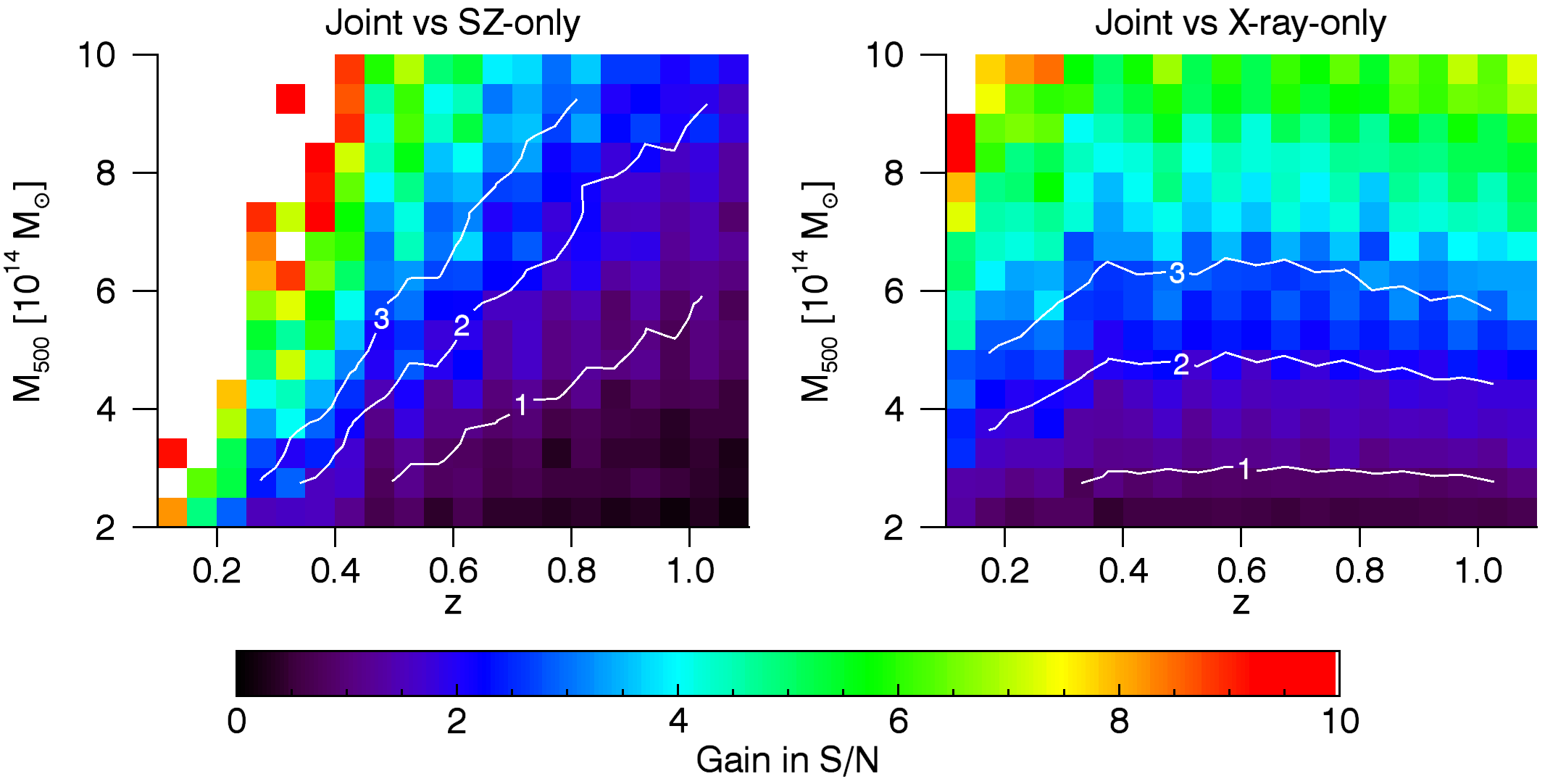}}
	\caption{Top panels: S/N of the clusters extracted with the proposed X-ray matched filter (left panel), the classical SZ MMF (middle panel), and the proposed X-ray-SZ MMF (right panel), averaged in different mass and redshift bins. White lines indicate smoothed isocontours corresponding to a S/N of 3, 4, and 5. White  bins indicate a S/N greater than 20. Bottom panels: Average difference between the S/N obtained with the proposed X-ray-SZ MMF and the S/N obtained with SZ MMF (left panel) and the X-ray MF (right panel).  White lines indicate smoothed isocontours corresponding to a difference in S/N of 1, 2, and 3. White bins indicate a difference in S/N greater than 10.}
	\label{fig:snr_in_mass_z_bins_rpj}
\end{figure*}

\subsection{Performance evaluation: Gain with respect to single-survey extractions}\label{ssec:joint_gain}
The goal of our joint matched filter is to improve the cluster detection rate with respect to the classical SZ MMF approach. Our technique gives a priori a higher S/N than the SZ MMF alone due to the inclusion of the additional map, so it may improve the detectability of galaxy clusters. Clearly, the false-detection rate should also be kept at a low value for a good detector performance. This latter point is beyond the scope of the present work, but we will explore it in a future work. 
It can be proven that the filtered background noise in the proposed algorithm is related to the filtered background noise of the SZ and the X-ray maps by the following relation:
\begin{equation}\label{eq:sigma_background_relation}
\sigma_{\rm JOINT}^{-2} = \sigma_{\rm SZ}^{-2} + \sigma_{\rm Xray}^{-2},
\end{equation}
where $\sigma_{\rm JOINT}$ is given by Eq. \ref{eq:sigma_sz} (with $\mathbf{F}_{\theta_{\rm s}}$ and $\mathbf{P}$ defined in Eqs. \ref{eq:F_joint} and \ref{eq:P_joint}), and $\sigma_{\rm SZ}$ and $\sigma_{\rm Xray}$ are calculated in an analogous way, but just using the $N_{\nu}$ SZ components of $\mathbf{F}_{\theta_{\rm s}}$ and $\mathbf{P}_{\rm SZ}$  in the first case, and the X-ray component of $\mathbf{F}_{\theta_{\rm s}}$ and $\mathbf{P}_{\rm X}$ in the second case. According to the previous relation, if the signal is perfectly estimated, there is always a gain in S/N with respect to using a single-map extraction ($\rm{S/N}_{\rm JOINT}^2 = \rm{S/N}_{\rm SZ}^2 + \rm{S/N}_{\rm Xray}^2$). In practice, however, we may not always see this gain in S/N because of estimation errors. 

In this section we compare the performance of the proposed joint matched filter to that of the classical SZ MMF and the proposed X-ray matched filter in terms of gain in the S/N of the extracted objects, or equivalently, in terms of detection probability (note that when we define a cluster to be detected when its S/N is above a given threshold, the S/N gain will translate into a higher detection rate). This comparison is made both with simulated and real clusters.

\subsubsection{Description of the simulations}
For this comparison, we first carried out a series of experiments in which we injected simulated clusters into real SZ and X-ray maps and extracted them using the three considered methods: the classical SZ MMF described in Sect. \ref{sec:szdetection}, the proposed X-ray matched filter, and the proposed joint matched filter. In the three cases, we assumed that the positions and sizes of the clusters are known, and in the joint MMF case, we further assumed that their redshifts are known. For the injections, we used the six highest frequency \emph{Planck} all-sky maps and the X-ray all-sky HEALPix map that we constructed from RASS data (see Appendix \ref{app:HealpixMap}).

Given the redshift $z$, the mass $M_{500}$, the luminosity $L_{500}$ and the SZ flux $Y_{500}$ of each cluster, the clusters to inject into the X-ray maps were simulated as explained in Sect. \ref{ssec:xray_simu}. The simulation of the corresponding SZ clusters was done similarly: we first created a map containing the SZ cluster profile corresponding to the size $\theta_{500}$ of the cluster (calculated from $z$ and $M_{500}$). This was done by integrating the average profile defined by Eq. \ref{eq:pressure_prof} with the parameters given in Eq. \ref{eq:sz_param}. Then we normalized this map so that its total SZ flux coincided with the SZ flux of the cluster within $5R_{500}$. This total SZ flux was calculated by extrapolating $Y_{500}$ up to $5R_{500}$ using the shape of the cluster profile. 
Finally, we convolved this map with the six different \emph{Planck} beams and applied the SZ spectral function for the corresponding \emph{Planck} frequencies to obtain a set of $N_\nu=6$ images of the simulated cluster. 
These simulated maps were then added to real $10\degr \times 10\degr$ patches centered on the (random) positions of the simulated clusters.

We divided the redshift-mass plane into 32 bins (four mass bins: 2-4, 4-6, 6-8, and 8-10 $\cdot 10^{14} M_{\odot}$ and eight redshift bins: 0.1-0.3, 0.3-0.5, 0.5-0.6, 0.6-0.7, 0.7-0.8, 0.8-0.9, 0.9-1.0, and 1.0-1.1), and for each bin, we injected 1000 clusters at random positions of the sky, with $z$ and $M_{500}$ uniformly distributed in the bin. $L_{500}$ was calculated from the L-M relation in \cite{Arnaud2010,PlanckEarlyXI}, including the scatter $\sigma_{\rm{log} L}=0.183$, and $Y_{500}$ was calculated from the nominal value of $L_{500}$ (from L-M relation without scatter), assuming a given $L_{500}-Y_{500}$ relation. For these simulations we assumed the relation found by the \citet{PlanckIntI2012}:
\begin{equation}\label{eq:FxY500relation}
\frac{F_X \left[ \mbox{erg s$^{-1}$ cm$^{-2}$}\right] }{Y_{500} \left[ \mbox{arcmin}^{2}\right] } = 4.95 \cdot 10^{-9} \cdot E(z)^{5/3} (1+z)^{-4} K(z).
\end{equation}
The K-correction was obtained by interpolating in Table \ref{table:K(z)}, which was calculated using a \textsc{mekal} model for a reference temperature of $ T_{\rm ref} $ = 7 keV (corresponding approximately to the median of the clusters above $z=0.3$). Since we did not include scatter in the Y-M relation (because it is much smaller than the L-M scatter), the $\sigma_{\rm{log} L}=0.183$ scatter translates directly into a $F_{\rm X}/Y_{500}$ scatter. Then we applied the three extraction methods at each cluster position. For the X-ray-SZ matched filter, we assumed the same $F_{\rm X}/Y_{500}$ relation used in the injection (Eq. \ref{eq:FxY500relation}), with the real redshift of the clusters, to convert the X-ray map into an equivalent SZ map.  

\subsubsection{Results of the simulations}

\begin{figure*}[]
	\centering
	\subfigure[]{\includegraphics[width=1.99\columnwidth]{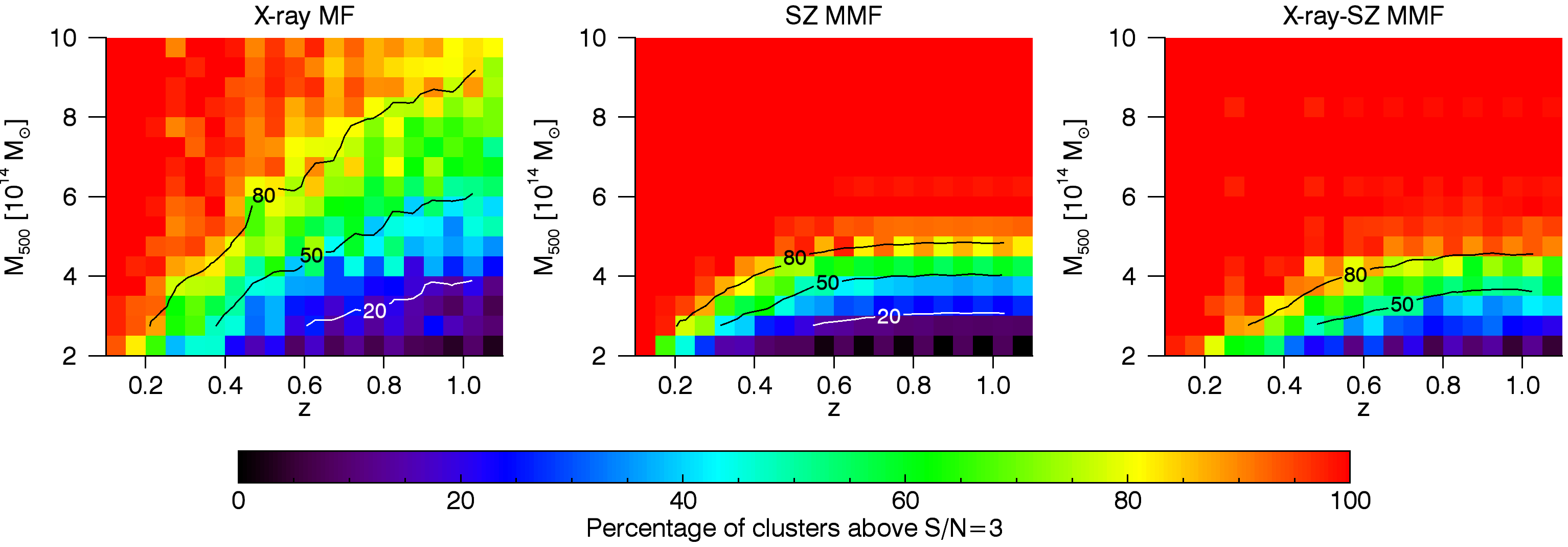}}
	\subfigure[]{\includegraphics[width=1.99\columnwidth]{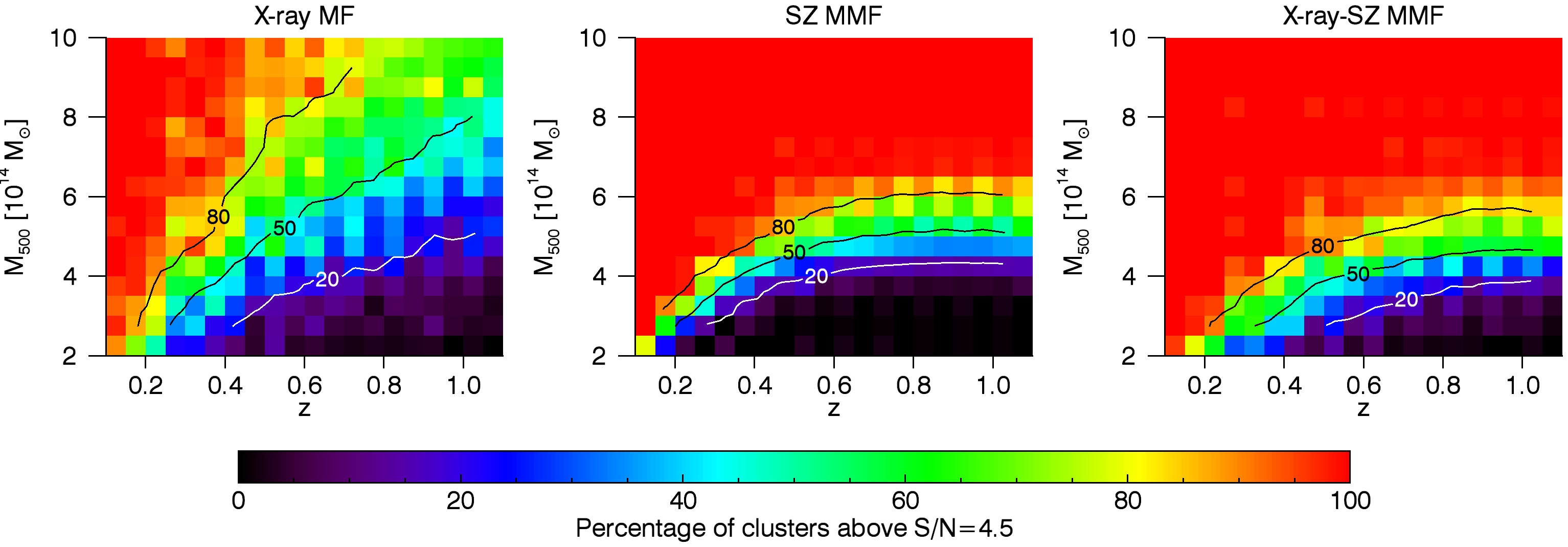}}
	\caption{Estimated completeness of the proposed X-ray matched filter (left panels), the classical SZ MMF (middle panels) and the proposed X-ray-SZ MMF (right panels) for a S/N threshold of 3.0 (top panels) and 4.5 (bottom panels), in different redshift-mass bins.}
	\label{fig:detections_in_mass_z_bins_rpj}
\end{figure*}
\begin{figure*}[]
	\centering
	\subfigure[]{\includegraphics[width=.99\columnwidth]{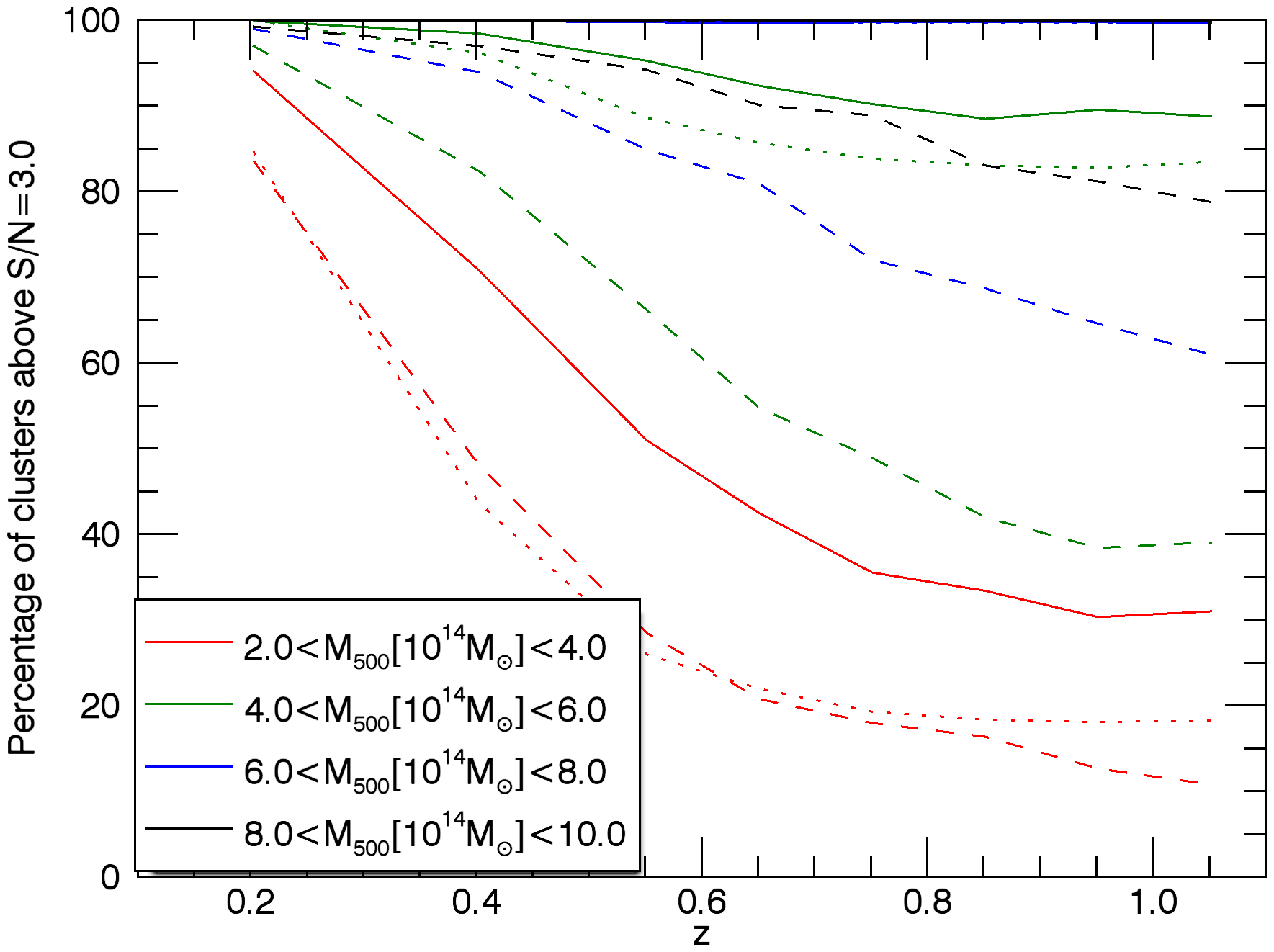}}
	\subfigure[]{\includegraphics[width=.99\columnwidth]{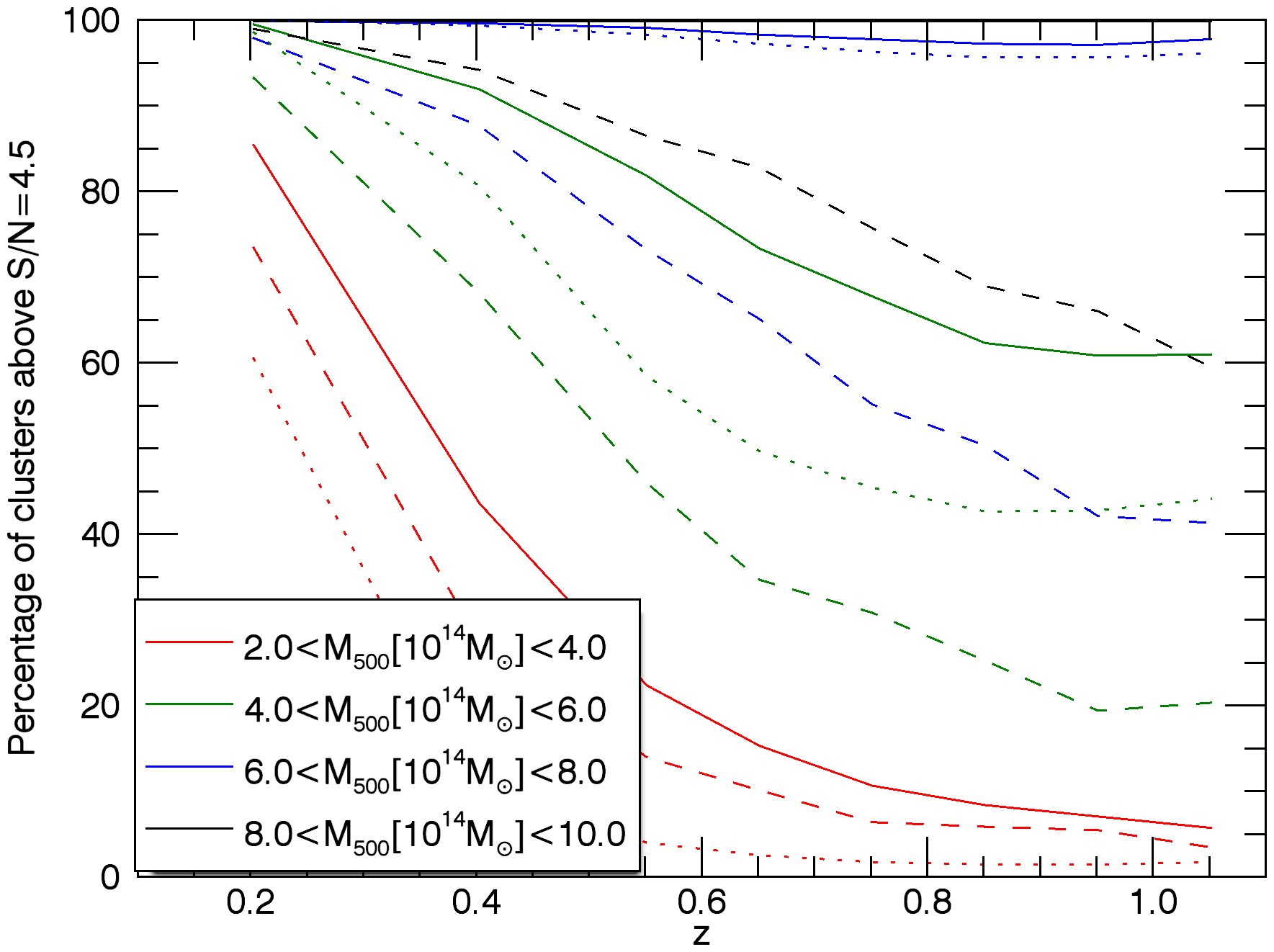}}
	\caption{Estimated completeness of the proposed X-ray matched filter (dashed lines), the classical SZ MMF (dotted lines), and the proposed X-ray-SZ MMF (solid lines) as a function of redshift for a S/N threshold of 3.0 (left panel) and 4.5 (right panel). The different colors correspond to different mass bins, as indicated in the legend.}
	\label{fig:detections_vs_z_simumz_rpj}
\end{figure*}

Figure \ref{fig:snr_in_mass_z_bins_rpj} shows the average S/N of the injected clusters using the three different methods and the gain in S/N of the joint MMF with respect to the single-survey methods. For this computation, we excluded the clusters injected in the Galactic region, where the foreground emission is very high, using the cosmology mask defined by the \cite{Planck2015ResXXVII}. We first note in this figure the different performance that we achieve using the X-ray information from RASS and the SZ information from \emph{Planck}. While the dependence on redshift and mass is quite steep for the X-ray case, for the SZ case it becomes flatter at high redshift (because the SZ signal is insensitive to the $(1+z)^4$ dimming), which makes high-redshift clusters easier to detect using the SZ information. For this particular configuration, for example, the S/N=5 and S/N=3 isocontours of these two cases cross at $z=0.5$ and $z=0.6$, respectively, meaning that below this redshift the X-ray information provided by the RASS survey is more significant than the one provided by the \emph{Planck} survey, while for higher redshift SZ maps become more helpful. This figure also shows that adding the X-ray information to the SZ maps improves the S/N over the whole range of redshifts and masses, although the gain with respect to using only the SZ information decreases with increasing redshift and with decreasing mass, as expected, since the amount of information brought by the X-ray map diminishes in these cases. For example, the typical S/N for a cluster with $M_{500} = 5 \cdot 10^{14} M_{\odot}$ and $z=0.5$ is around 5.4 for the X-ray MF, around 5.2 for the SZ MMF, and around 7.5 for the X-ray-SZ MMF, while the same cluster at $z=1.0$ will typically have a S/N around 2.8 for the X-ray MF, around 4.4 for the SZ MMF, and around 5.2 for the X-ray-SZ MMF.

Figures \ref{fig:detections_in_mass_z_bins_rpj} and \ref{fig:detections_vs_z_simumz_rpj} show the percentage of clusters found above S/N=4.5 and S/N=3.0 in different mass and redshift bins, that is, the estimated completeness, for the three different methods. These figures show that adding the X-ray information improves the detection rate over the whole range of redshifts and masses, although the gain with respect to using only the SZ information decreases with redshift, as expected. From the 20\%, 50\%, and 80\% completeness levels overplotted in Fig. \ref{fig:detections_in_mass_z_bins_rpj}, the advantage of using the joint algorithm is clearly visible because the detection limit is pushed towards higher redshift and lower mass clusters. We checked that the SZ results are compatible with the theoretical expectations using the ERF error function approximation \citep{Planck2013ResXXIX} and with the results presented by the \cite{Planck2015ResXXVII}.  
We reach a 100\% detection rate for the highest mass bin with the SZ and the joint MMF. 
Interestingly, a low, but non-zero, fraction of the low-mass and high-redshift clusters can also be detected. We checked that these are real clusters and not only noise peaks or other objects in the maps because the extraction at the same positions but without the injected clusters provides only one false detection in the last bin (0.1\%) with the SZ MMF and no false detection with the other two methods.

\begin{figure}[]
	\centering
	\includegraphics[width=\columnwidth]{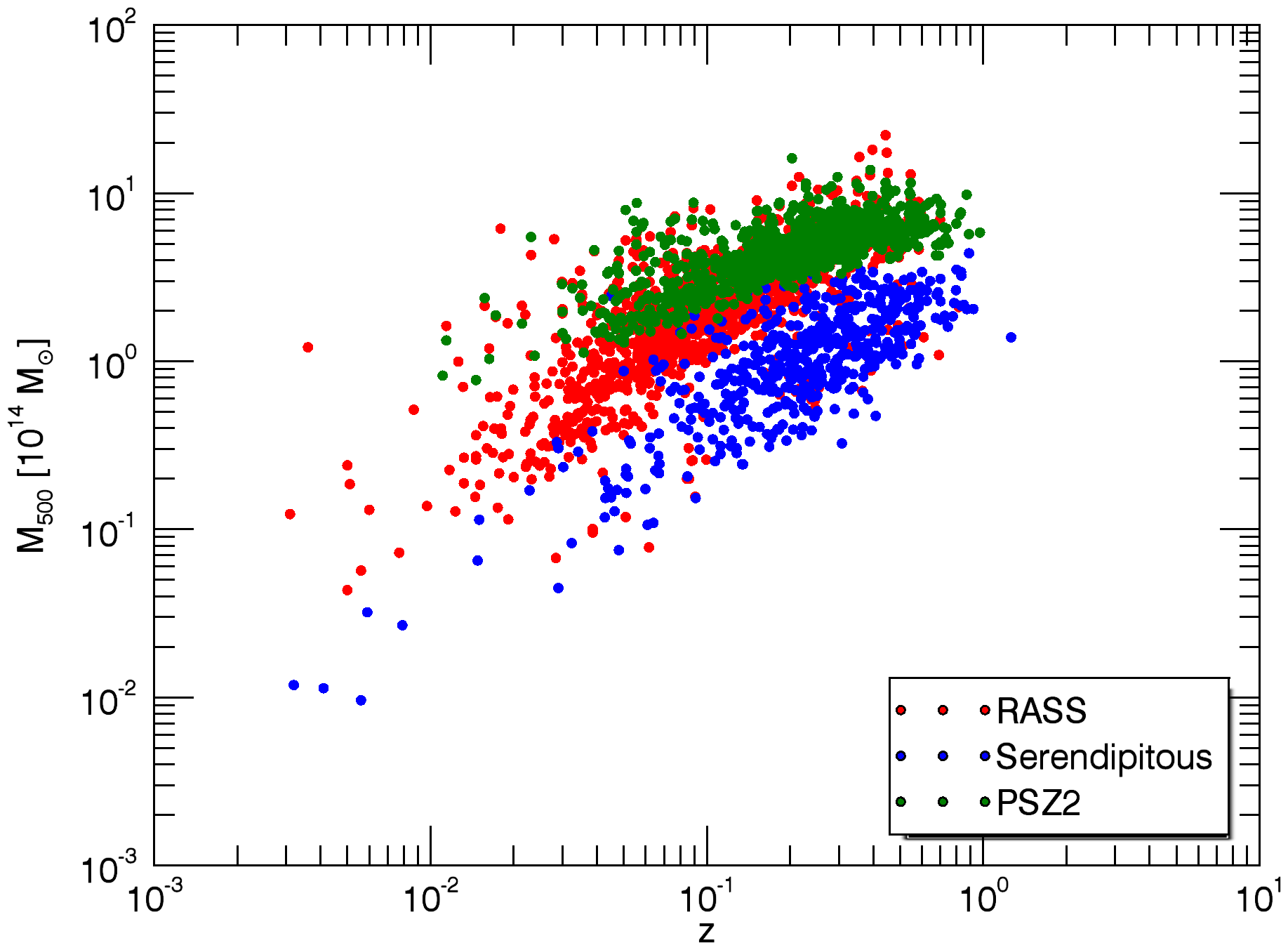}
	\caption{MCXC and PSZ2 cluster samples in the mass-redshift plane. The MCXC sample is divided into RASS and serendipitous subsamples.}
	\label{fig:clusters_in_MZ_plane}
\end{figure}

\subsubsection{Extraction of real clusters}
The analysis presented above illustrates the advantage of using the proposed X-ray-SZ MMF on simulated clusters. However, the simulations are based on some assumptions that may not hold in the real world, for example, an ideal cluster profile or a perfectly known $F_{\rm X}/Y_{500}$ relation. To check whether the expected behavior is maintained in a more realistic scenario, we carried out another set of experiments in which we extracted known real clusters, using the three methods. In particular, we extracted the 1743 clusters of the MCXC sample \citep{Piffaretti2011}, for which we already presented the X-ray extraction results in Sect. \ref{ssec:xray_real}, and the 926 confirmed clusters with redshift estimates of the second \emph{Planck} SZ (PSZ2) catalogue \citep{Planck2015ResXXVII} that were detected with the MMF3 method, which we used as basis of our algorithm (described in Sect. \ref{sec:szdetection}). Figure \ref{fig:clusters_in_MZ_plane} shows these clusters in the mass-redshift plane. These two samples were chosen to analyse the possible differences in the performance of the proposed joint X-ray-SZ on X-ray selected clusters, as the MCXC clusters, and on SZ selected clusters, as the PSZ2 clusters.

\begin{figure}[]
	\centering
	\subfigure[]{\includegraphics[width=\columnwidth]{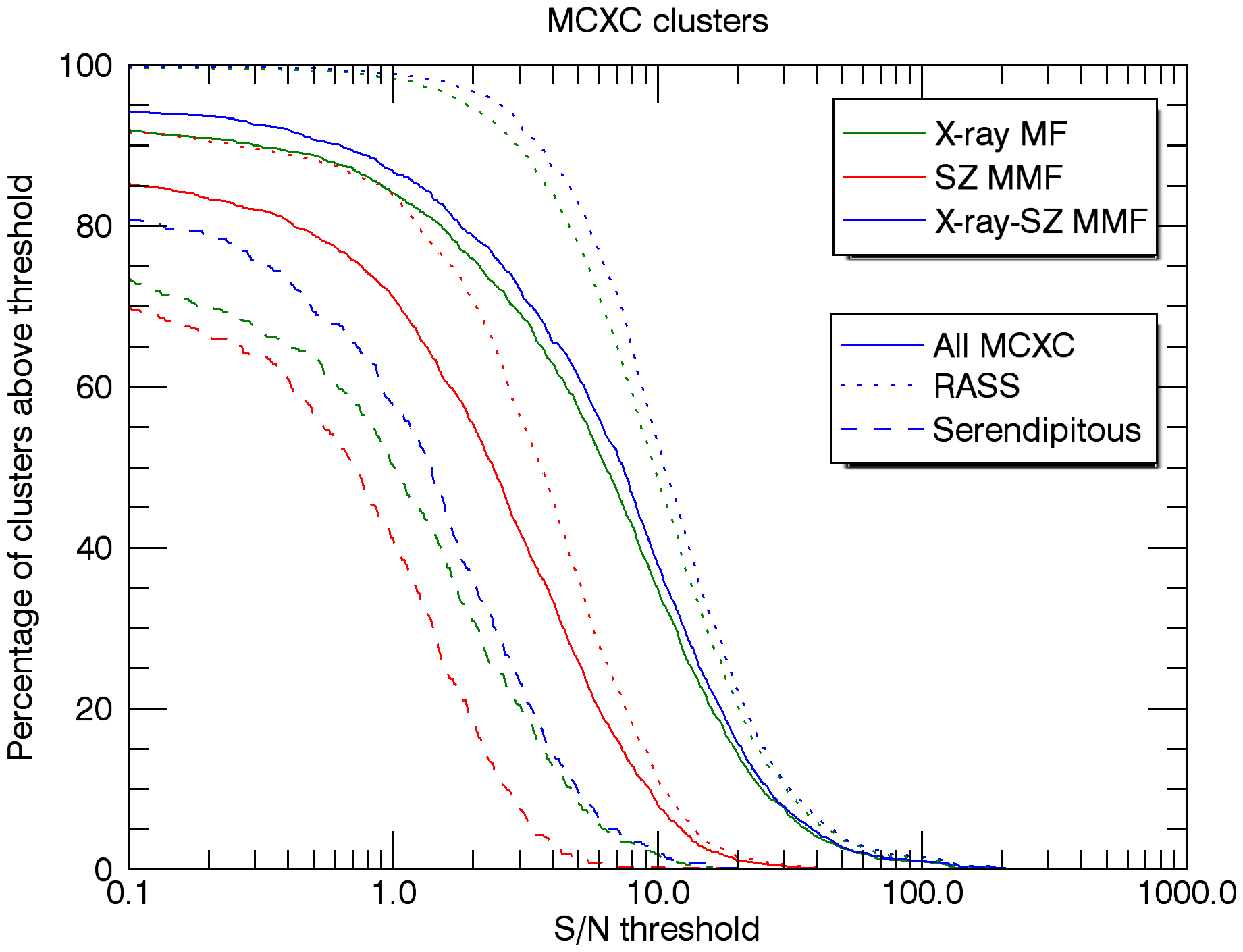}\label{fig:detections_vs_snr_realrass_rpj}}
	\subfigure[]{\includegraphics[width=\columnwidth]{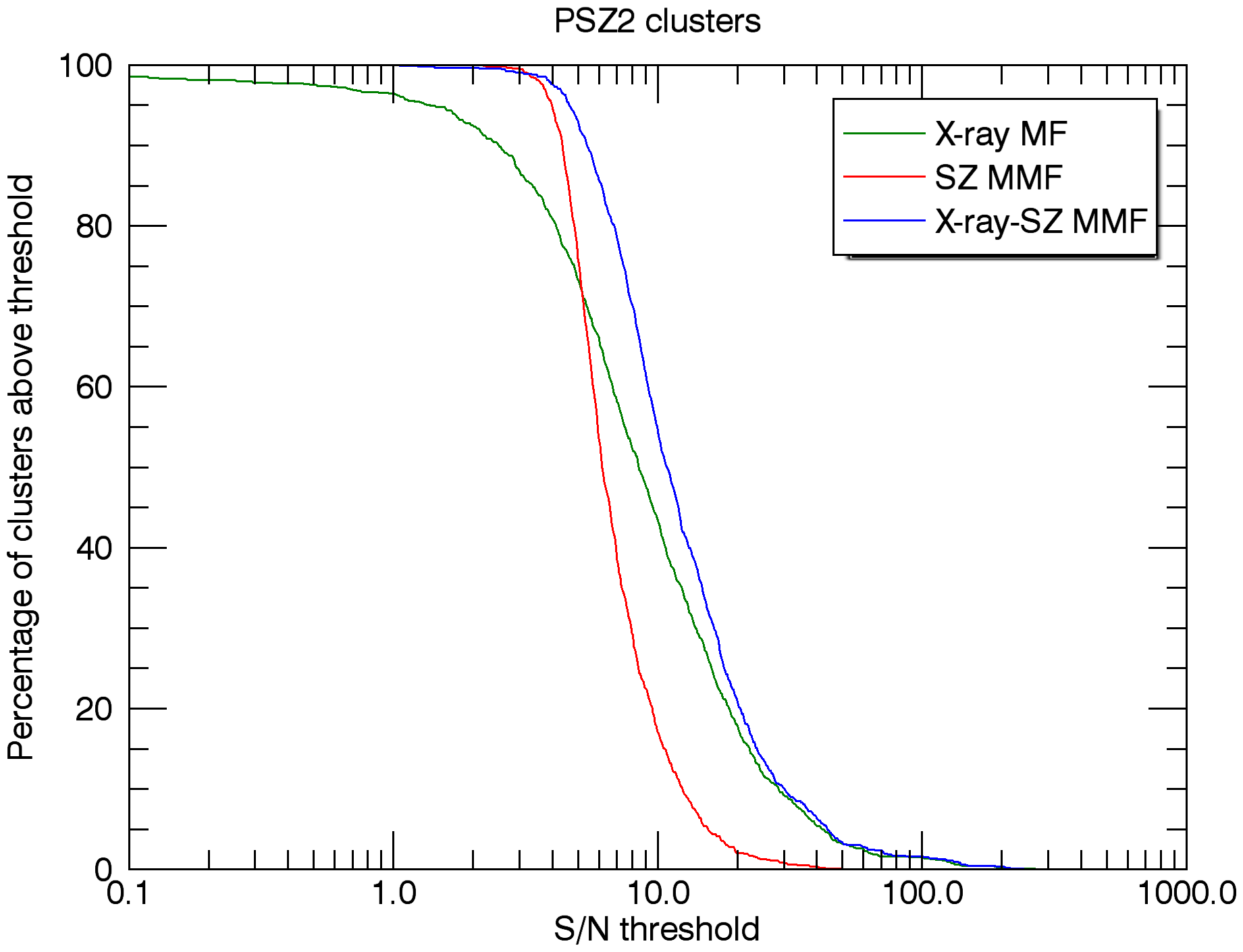}\label{fig:detections_vs_snr_realmmf3_rpj}}
	\caption{Percentage of MCXC (top panel) and PSZ2 (bottom panel) clusters extracted with the proposed X-ray matched filter (green), the classical SZ MMF (red), and the proposed X-ray-SZ MMF (blue) whose S/N is above a certain S/N threshold. In the top panel, the complete MCXC sample (solid lines) is divided into RASS (dotted lines) and serendipitous clusters (dashed lines).}
	\label{fig:detections_vs_snr_realrassmmf3_rpj}
\end{figure}

\begin{figure*}[]
	\centering
	\includegraphics[width=1.6\columnwidth]{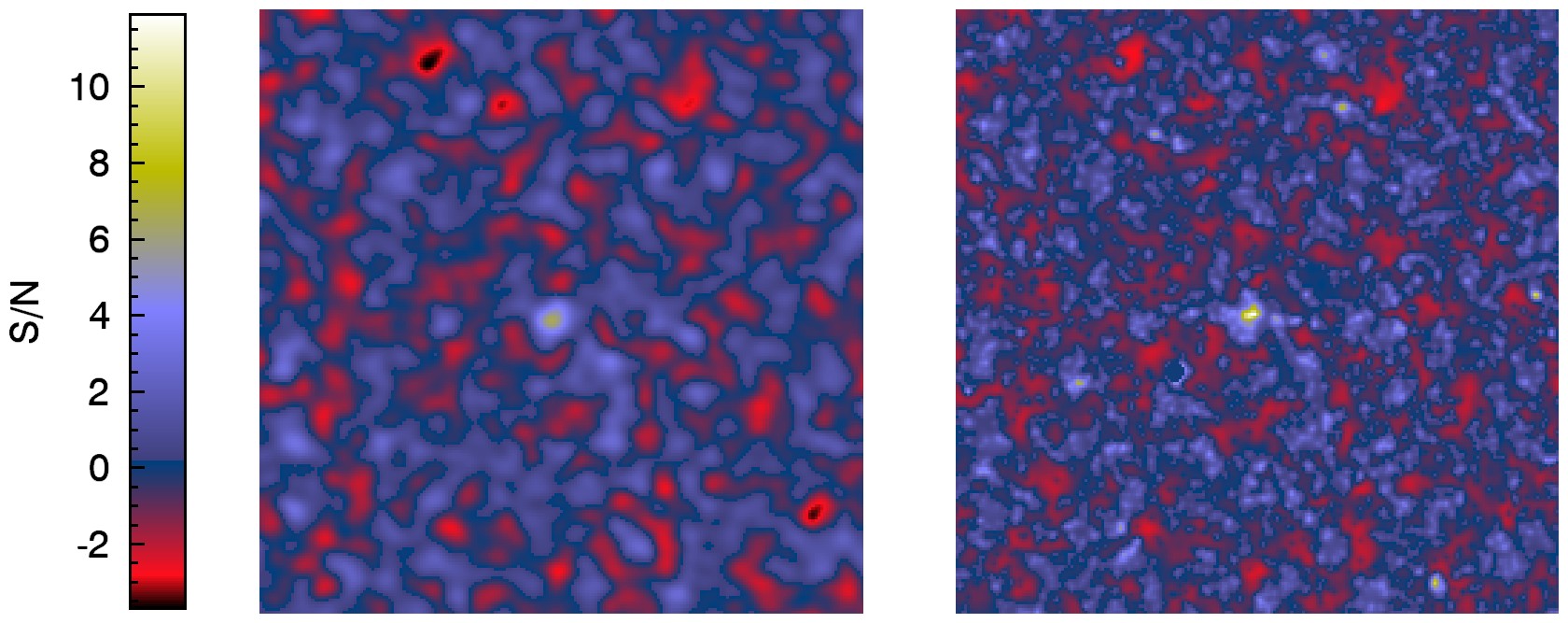}
	\caption{S/N maps obtained in the extraction of cluster PSZ2 G156.26+59.64 using the SZ-MMF (left) and the X-ray-SZ MMF (right). The S/N at the cluster position improves from 5.38 to 11.92. The angular size of the images is 3$\degr$. The cluster redshift is $z=0.59$. We have masked an X-ray source to the bottom-left of the cluster.}
	\label{fig:nice_cluster}
\end{figure*}

The MCXC clusters were extracted on $10\degr \times 10\degr$ patches centered on the cluster position, fixing the cluster size to the true value. In this case, we assumed the following $F_{\rm X}/Y_{500}$ relation to convert the X-ray map into an equivalent SZ map:
\begin{equation}\label{eq:FxY500relation_PXCC}
\frac{F_X \left[ \mbox{erg s$^{-1}$ cm$^{-2}$}\right] }{Y_{500} \left[ \mbox{arcmin}^{2}\right] } = 7.41 \cdot 10^{-9} \cdot E(z)^{5/3} (1+z)^{-4} K(z).
\end{equation}
This expression was obtained from the $D^2_{\rm A}Y_{500}-L_{500}$ relation found by the \cite{PlanckEarlyX} for this cluster sample, with the approximation $\hat{\alpha}_{\rm L}=1$ for a pivot luminosity of 1~erg/s. We note the difference in the normalization with respect to Eq. \ref{eq:FxY500relation}. In Sect. \ref{sec:discussion} we investigate the effects of the assumed $F_{\rm X}/Y_{500}$ relation in detail.

For the PSZ2 sample, we used the same extraction procedure as for the MCXC clusters, but in this case we assumed the general $F_{\rm X}/Y_{500}$ relation defined in Eq. \ref{eq:FxY500relation}, since there is no better relation available for this particular sample. We also allowed for a small freedom (2.5 arcmin) in the position to account for the lower position accuracy provided by the PSZ2 catalogue.

Due to the small number of clusters in these samples, we cannot statistically analyse the results in different mass and redshift bins, as we did for the simulations, but we can show the global behaviour.
Figure \ref{fig:detections_vs_snr_realrass_rpj} shows how the proposed joint algorithm increases the detection probability with respect to the X-ray matched filter and the classical SZ MMF for the MCXC clusters. In comparison with the extraction of the MCXC clusters using the X-ray matched filter (see Sect. \ref{ssec:xray_real}), the joint algorithm is able to detect slightly more clusters: 97\%, 93\%, and 87\% of the RASS clusters (and 36\%, 23\%, and 14\% of the serendipitous clusters) for S/N thresholds of 2, 3, and 4, respectively. As expected, for this X-ray selected sample, including the SZ information slightly improves the detection probability, while adding the X-ray map to the classical SZ MMF has a strong effect. Figure \ref{fig:detections_vs_snr_realmmf3_rpj} shows the percentage PSZ2 clusters detected above a given S/N threshold for the three detection methods. For this SZ selected sample, the proposed joint algorithm again increases the detection probability with respect to the X-ray matched filter and the classical SZ MMF alone. 
In Appendix \ref{app:SNRgain} we present a more detailed comparison of the estimated S/N obtained for the MCXC clusters and the PSZ2 clusters using the three detection methods as a function of redshift and mass, which shows that there is a S/N gain for most of the clusters, even for high redshift. We therefore conclude that in comparison with the single-survey filters, the proposed X-ray-SZ MMF provides a better significance of the extracted clusters, helping to detect clusters up to higher redshift and lower mass. Of course, to use the proposed technique as a blind detection tool, the detection probability not only has to be high, but the false-detection rate must also be kept at a low value. The latter point is beyond the scope of this paper, but will be investigated in future work.

As a visual illustration of the gain we can achieve, Fig. \ref{fig:nice_cluster} shows the S/N maps obtained in the extraction of cluster PSZ2 G156.26+59.64, a massive cluster at $z = 0.59$, using the SZ MMF and the X-ray-SZ MMF. In these images the improvement in S/N is obvious.

\subsection{Performance evaluation: Photometry}\label{ssec:joint_photometry}

Since the joint matched filter provides an estimate of the flux of the detected clusters by combining the X-ray and SZ information and assuming a given underlying relation, it is important to check whether the provided photometry is accurate and how it depends on the assumptions. In this section we assess the performance of the proposed joint matched filter in terms of photometry, through simulations and through extraction of real clusters.

\begin{figure}[]
	\centering
	\includegraphics[width=.99\columnwidth]{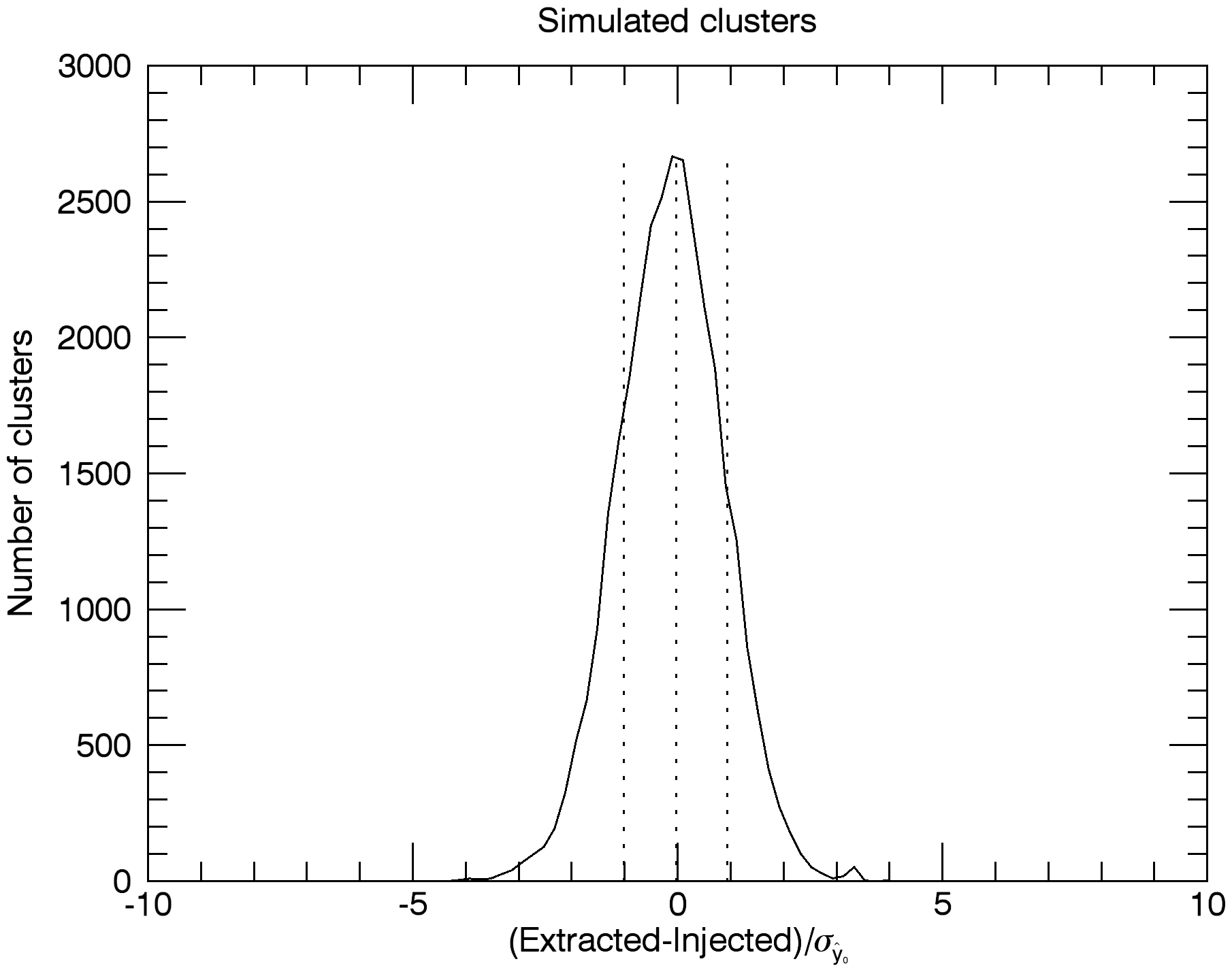}
	\caption{Histogram of the difference between the extracted and the injected $L_{500}$, divided by the estimated $\sigma_{\hat{y}_0}$, for the simulated clusters (as described in Sect. \ref{ssec:joint_photometry}) extracted with the proposed X-ray-SZ MMF. The central vertical line shows the median value, while the other two vertical lines indicate the region inside which 68$\% $ of the clusters lie.}
	\label{fig:hist_joint_simuMZnoscatter}
\end{figure}

First, with the aim of checking that the photometry results are correct in the ideal setting, we carried out an injection experiment, similar to the one presented in the previous section, but without scatter in the L-M relation. We checked that the extracted flux follows the injected flux very well, with a linear fit that is very close to the unity-slope line ($y=1.0067(\pm0.0015)x-9.8(\pm0.9)\cdot10^{-2}$). Figure \ref{fig:hist_joint_simuMZnoscatter} shows the histogram of the difference between the extracted and the injected flux, divided by the estimated standard deviation $\sigma_{\hat{y}_0}$ corresponding to this experiment. This histogram shows that there is no bias and that the estimated error bars describe the dispersion on the results well (as 68$\%$ of the extractions fall in an interval that is close to $\pm1\sigma_{\hat{y}_0}$). We also checked that the extraction behaves correctly for all the values of redshift, size, flux, and S/N and that these parameters do not introduce any systematic error or bias in the results.

From this experiment, we conclude that the combined X-ray-SZ matched filter performs well when applied to ideal simulated clusters. However, the simulations are based on some assumptions that may not hold in the real world (ideal cluster profiles, known $L_{500}/Y_{500}$ relation with no scatter). Therefore, the final step in this paper is to check the performance of the proposed filter on real clusters. In the following, we present the photometry results from the extraction of known clusters on real SZ and X-ray data, assuming that we know their position, size, and redshift. In particular, we focused on the MCXC and the PSZ2 samples presented in Sect. \ref{ssec:joint_gain}.

\begin{figure*}[]
	\centering
	\subfigure[]{\includegraphics[width=.99\columnwidth]{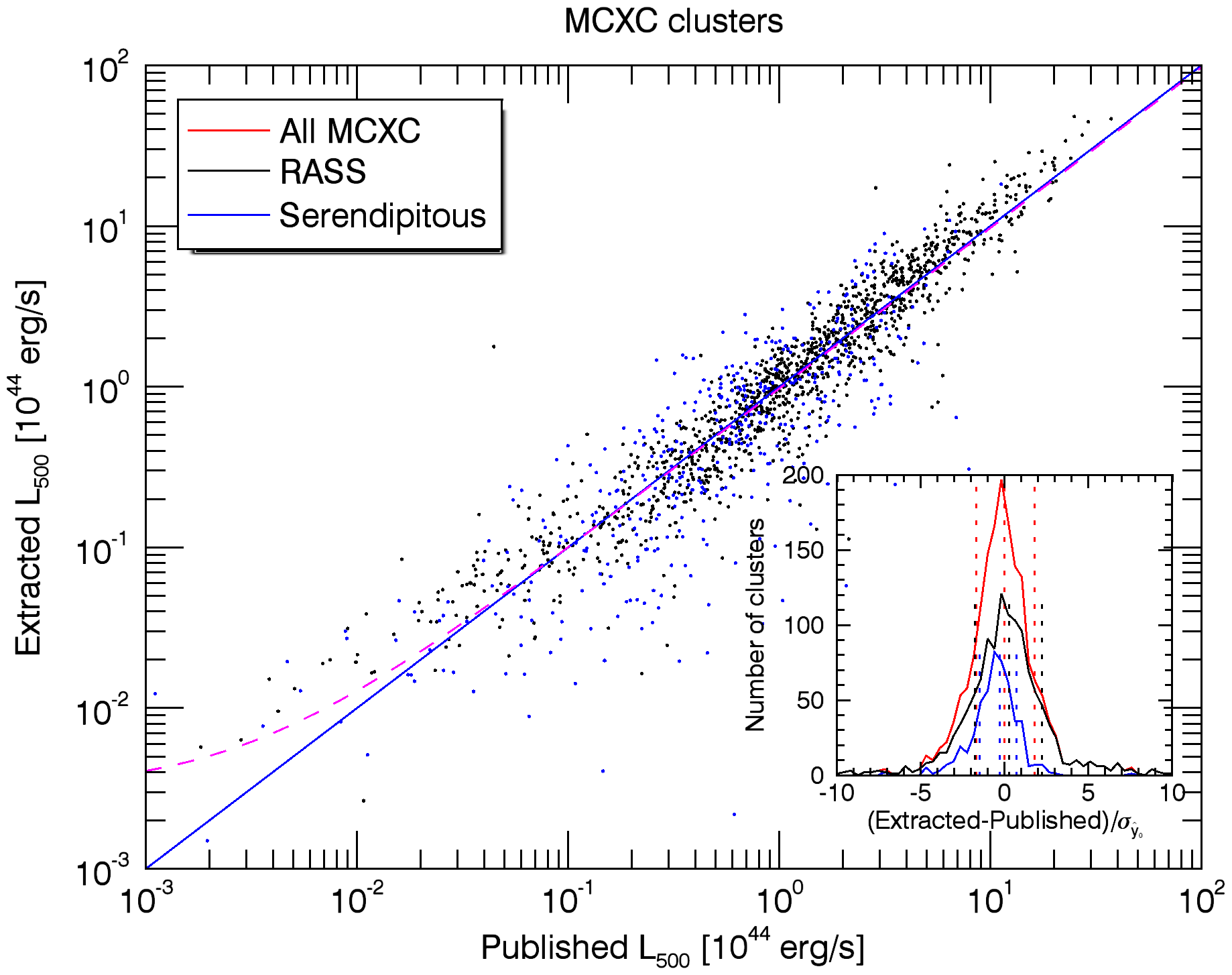}\label{fig:extraction_joint_pxcc}}
	\subfigure[]{\includegraphics[width=.99\columnwidth]{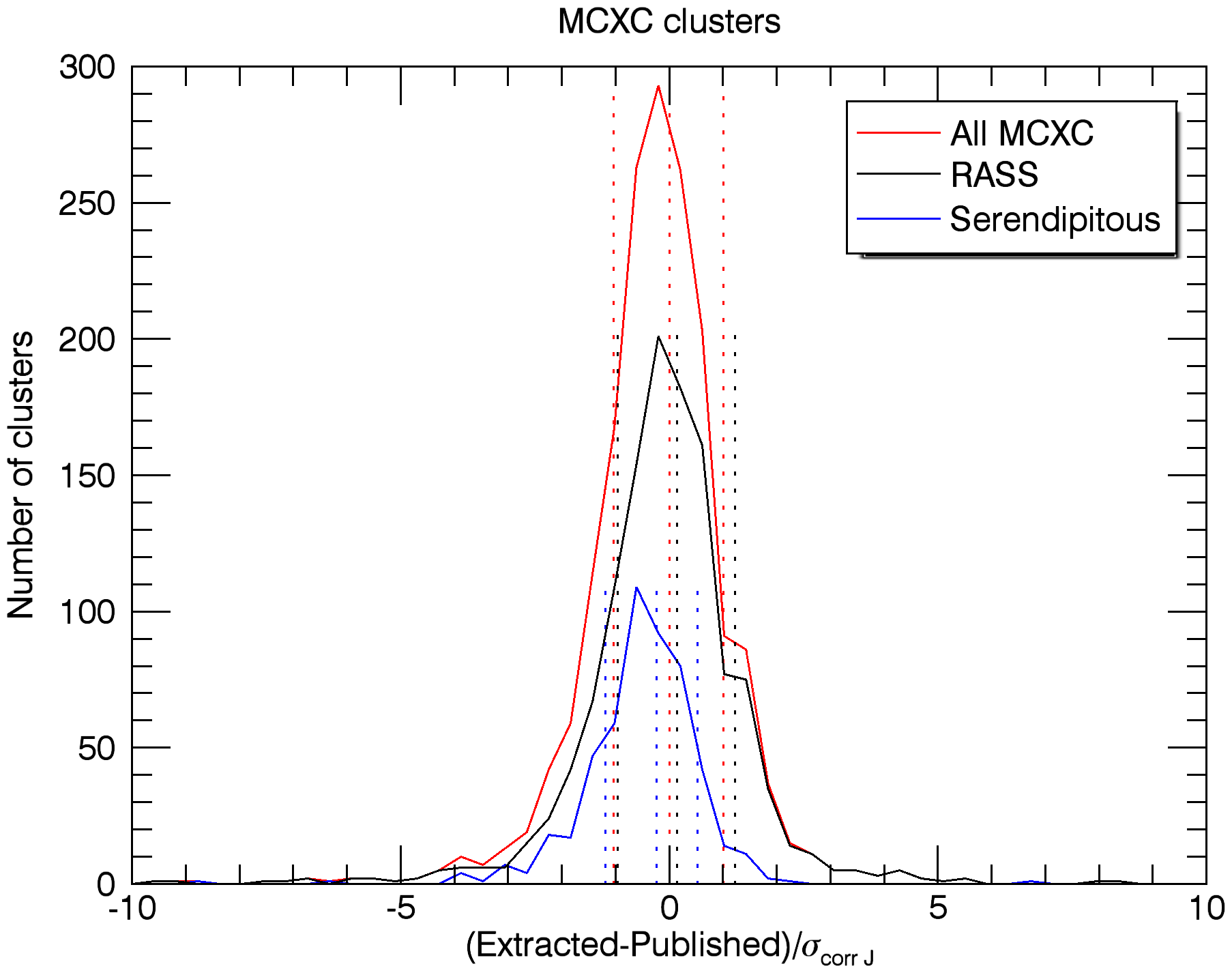}\label{fig:hist_joint_pxcc_corr}}
	\subfigure[]{\includegraphics[width=.99\columnwidth]{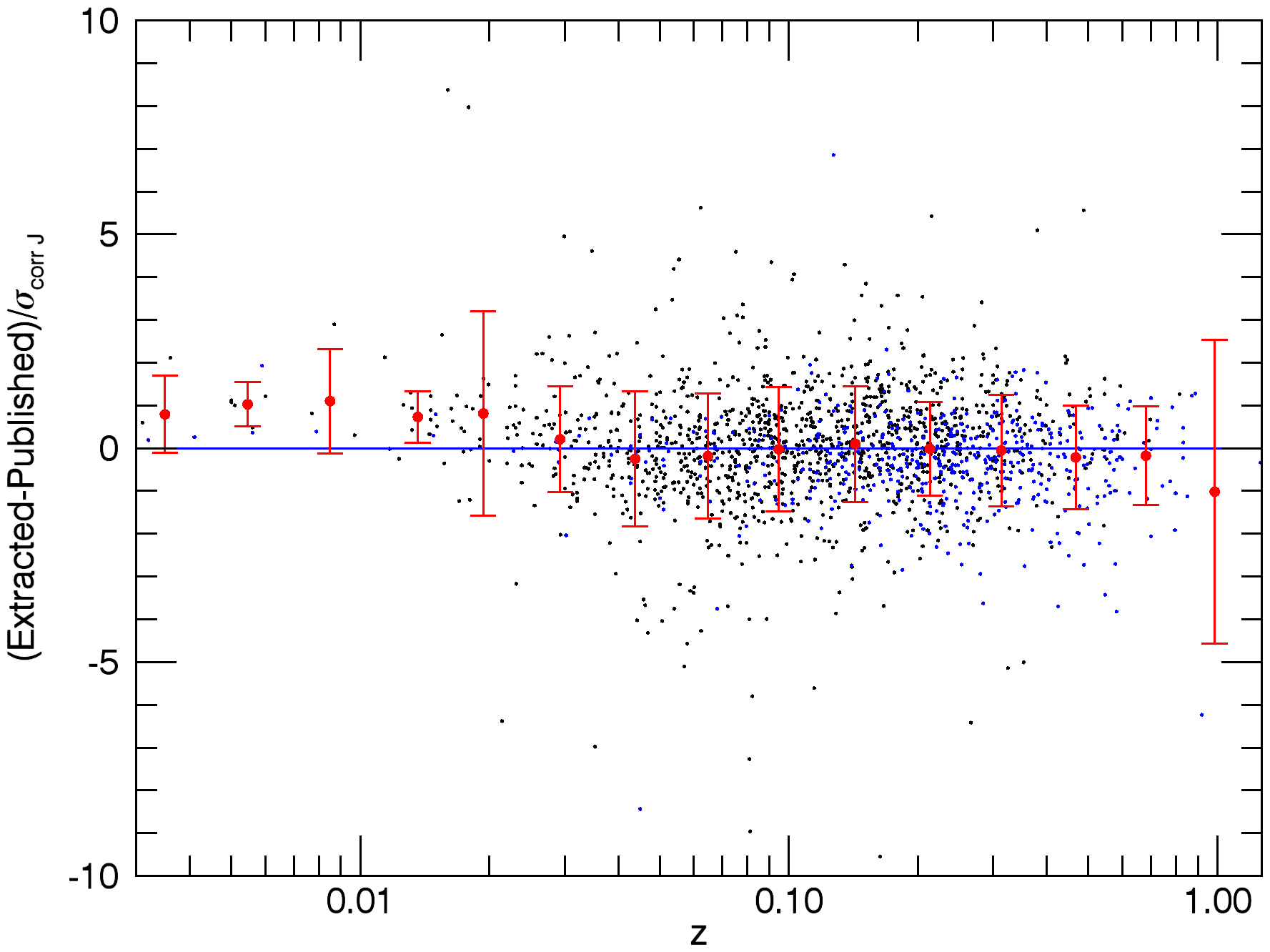}\label{fig:realpxcc_vs_z_joint}}
	\subfigure[]{\includegraphics[width=.99\columnwidth]{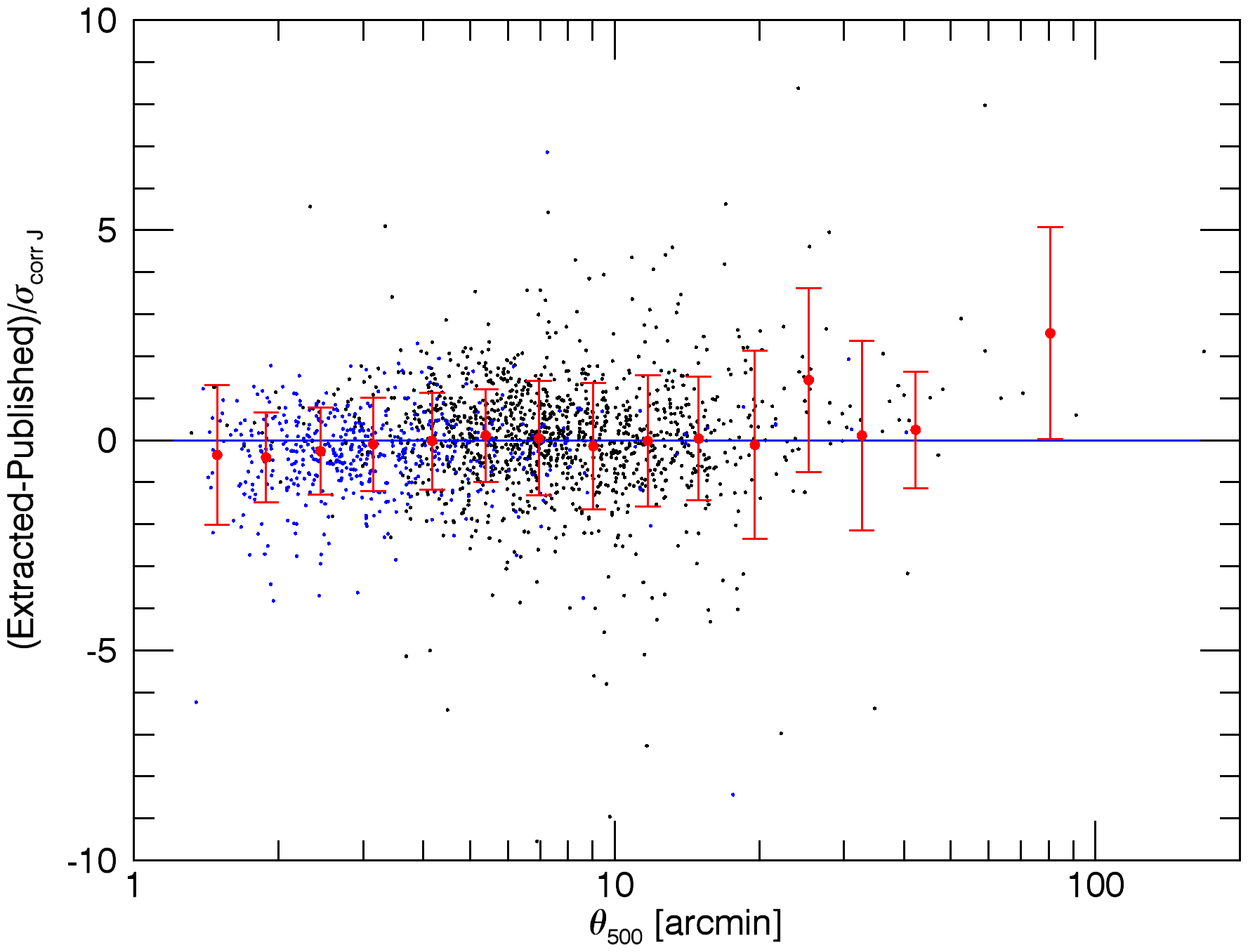}\label{fig:realpxcc_vs_theta_joint}}
	\subfigure[]{\includegraphics[width=.99\columnwidth]{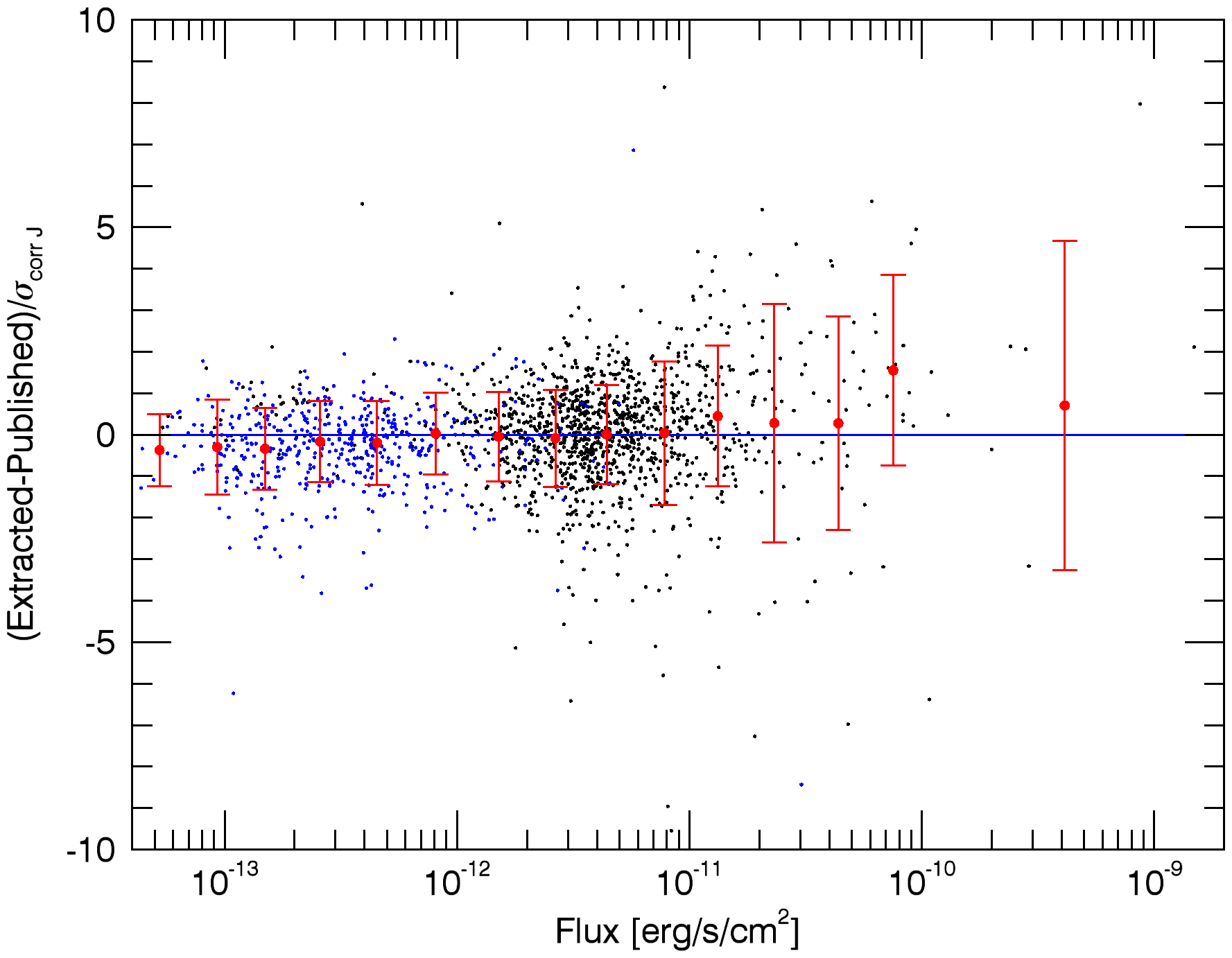}\label{fig:realpxcc_vs_flux_joint}}
	\subfigure[]{\includegraphics[width=.99\columnwidth]{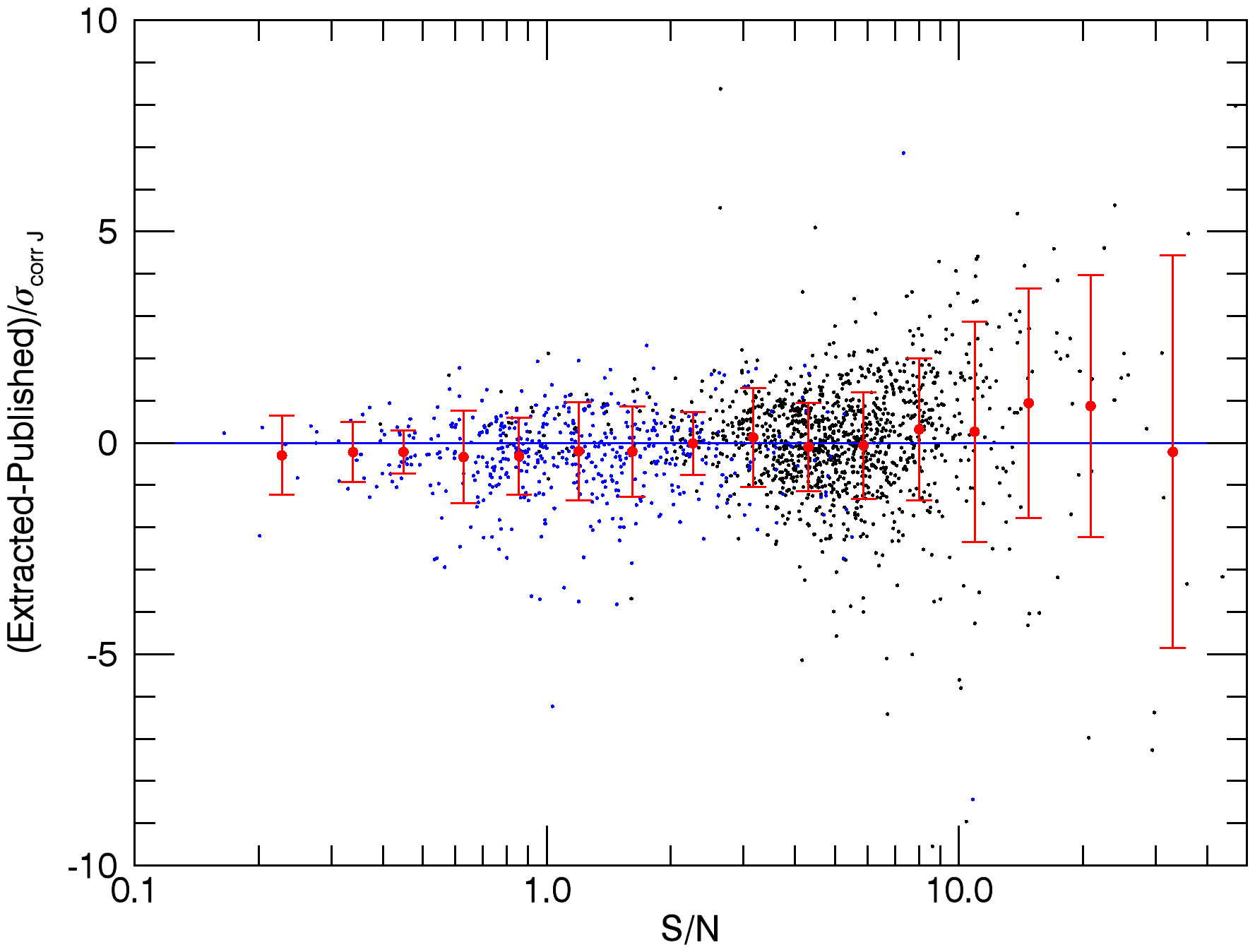}\label{fig:realpxcc_vs_snr_joint}}
	\caption{Photometry results of the extraction of the MCXC clusters using the proposed X-ray-SZ filter with the average cluster profile and assuming the position, size, and redshift of the clusters are known. The top left panel is analogous to the top left panel of Fig. \ref{fig:realpxcc}, and its subpanel shows the histogram of the difference between the extracted and the published $L_{500}$, divided by the estimated $\sigma_{\hat{y}_0}$ (analogously to Fig. \ref{fig:hist_pxcc}). The five other panels are analogous to those in Fig. \ref{fig:realpxcc}, but in this case, the standard deviation used to normalize the difference between the extracted and the published $L_{500}$ is already corrected for the effect of the profile mismatch ($\sigma_{\rm corr J}$). 
	}
	\label{fig:joint_pxcc}
\end{figure*}

\begin{figure*}[]
	\centering
	\subfigure[]{\includegraphics[width=.99\columnwidth]{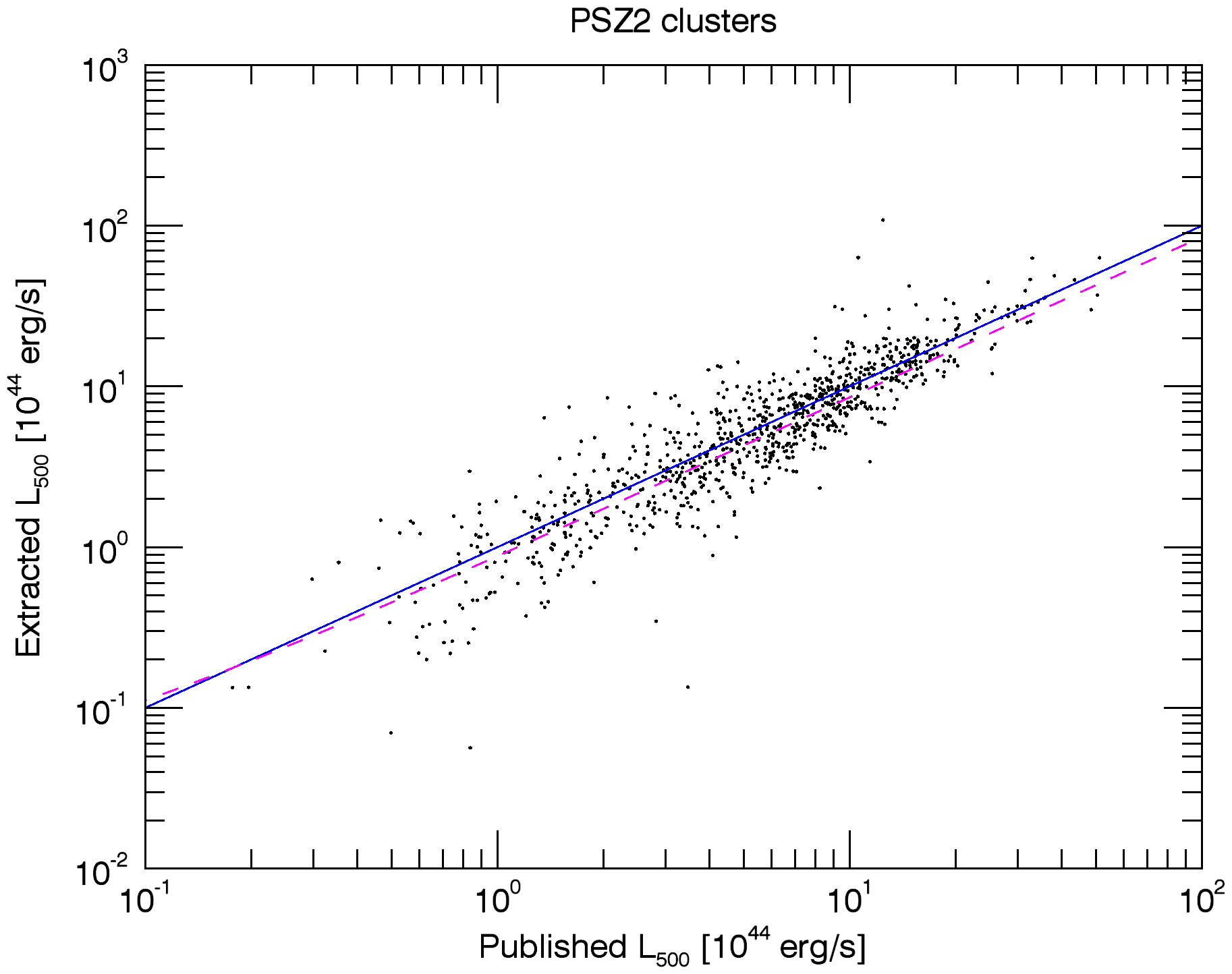}	\label{fig:extraction_joint_mmf3}}
	\subfigure[]{\includegraphics[width=.99\columnwidth]{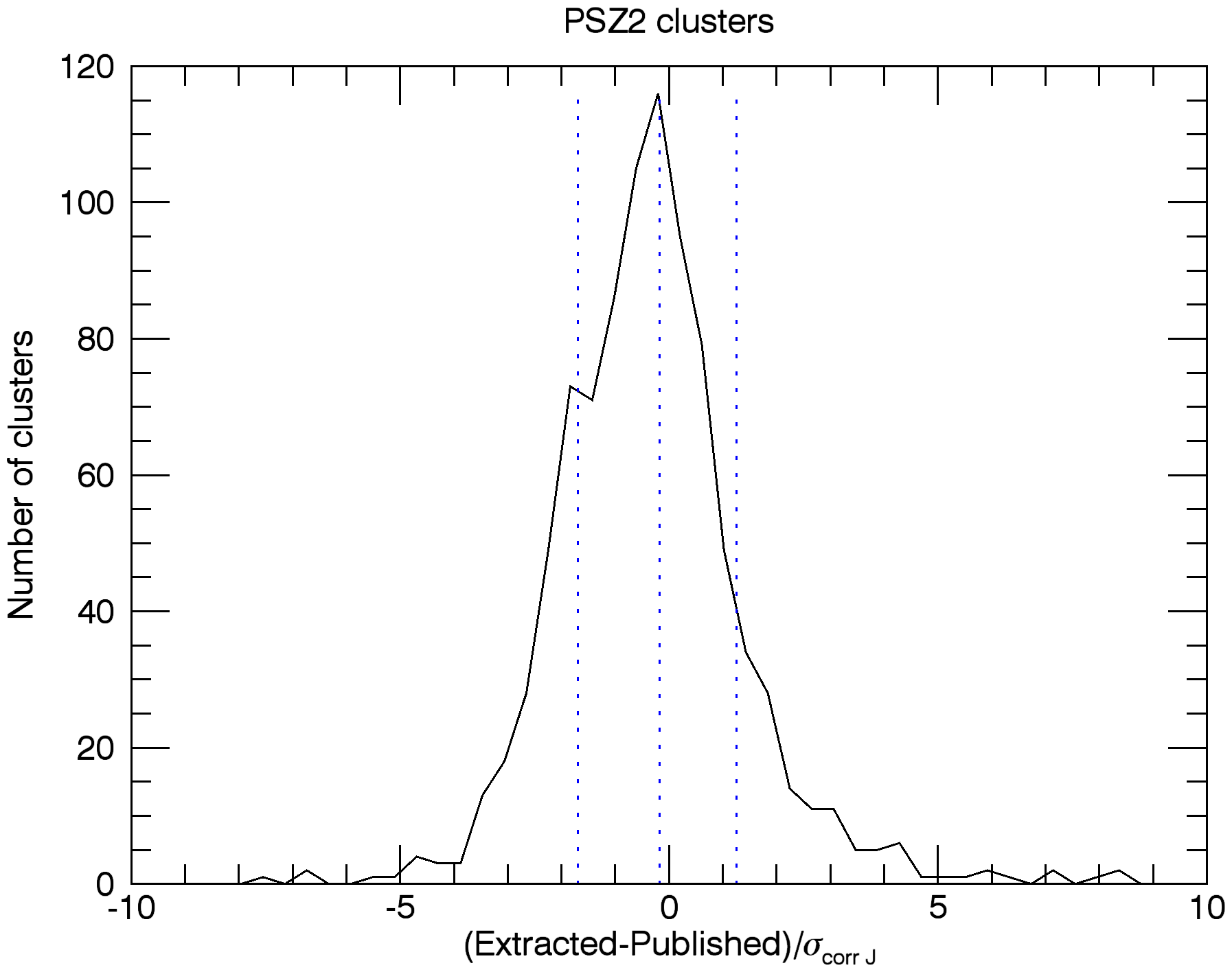}\label{fig:hist_joint_mmf3}}
	\subfigure[]{\includegraphics[width=.99\columnwidth]{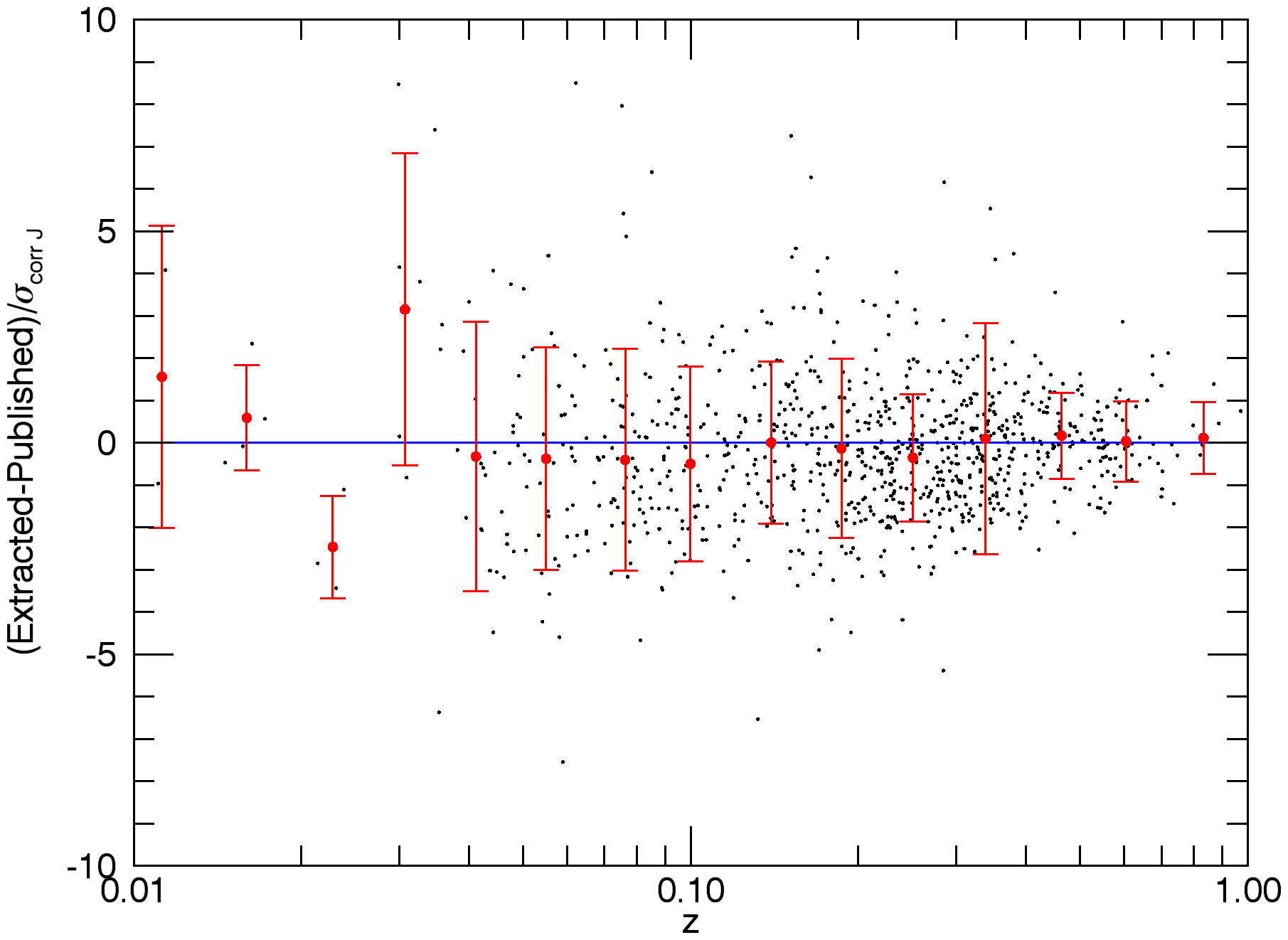}\label{fig:realmmf3_vs_z_joint}}
	\subfigure[]{\includegraphics[width=.99\columnwidth]{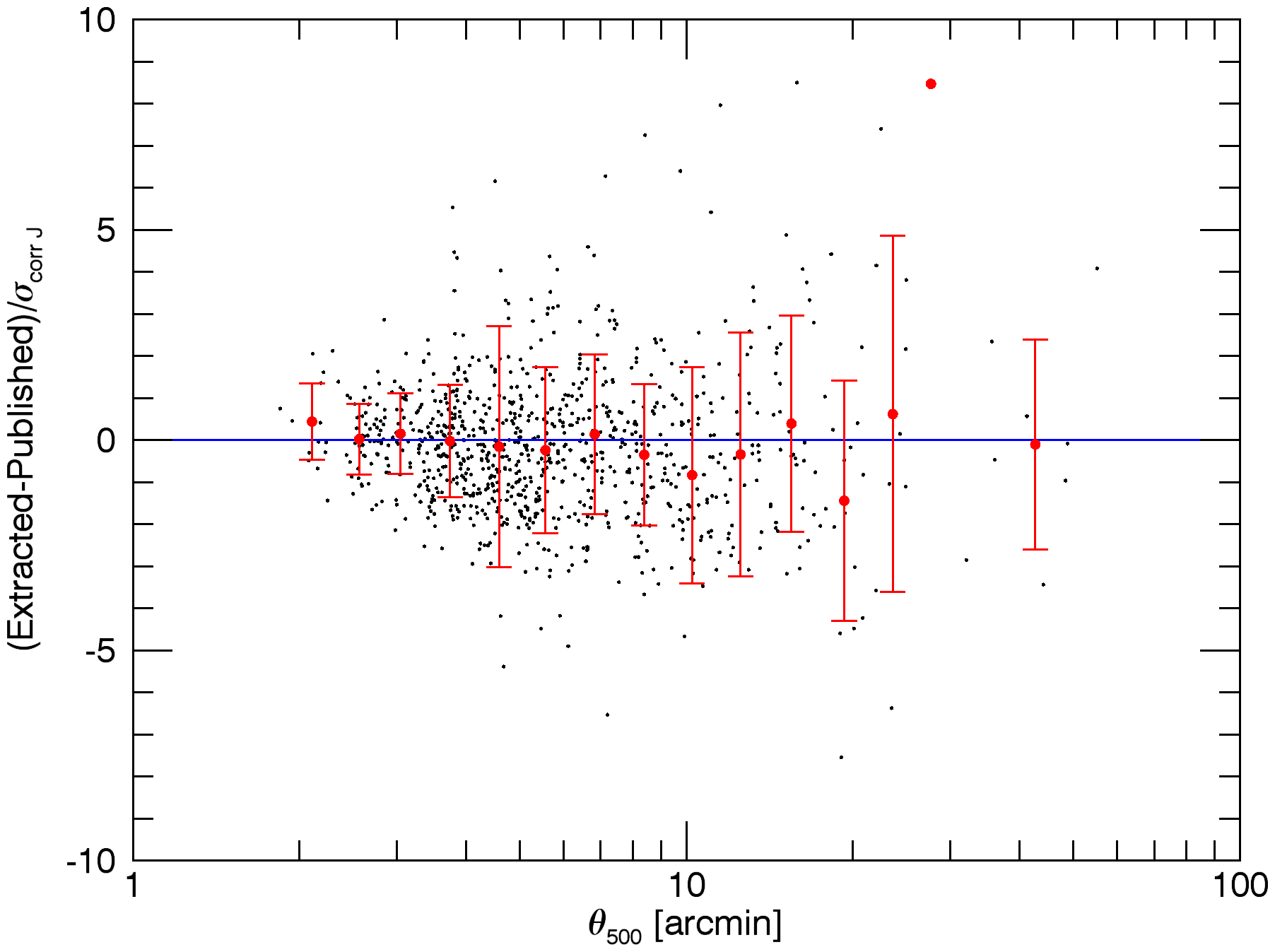}\label{fig:realmmf3_vs_theta_joint}}
	\subfigure[]{\includegraphics[width=.99\columnwidth]{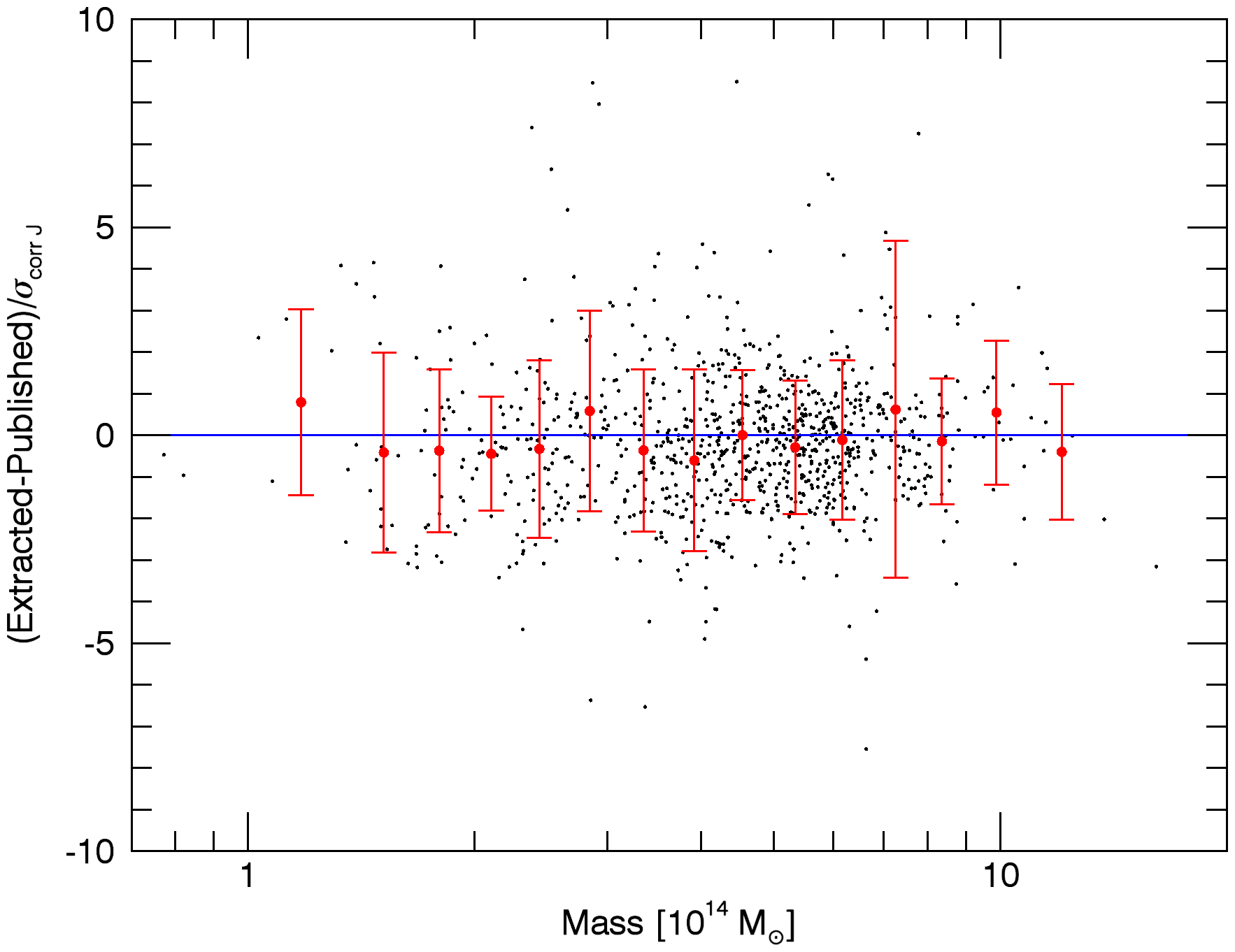}\label{fig:realmmf3_vs_flux_joint}}
	\subfigure[]{\includegraphics[width=.99\columnwidth]{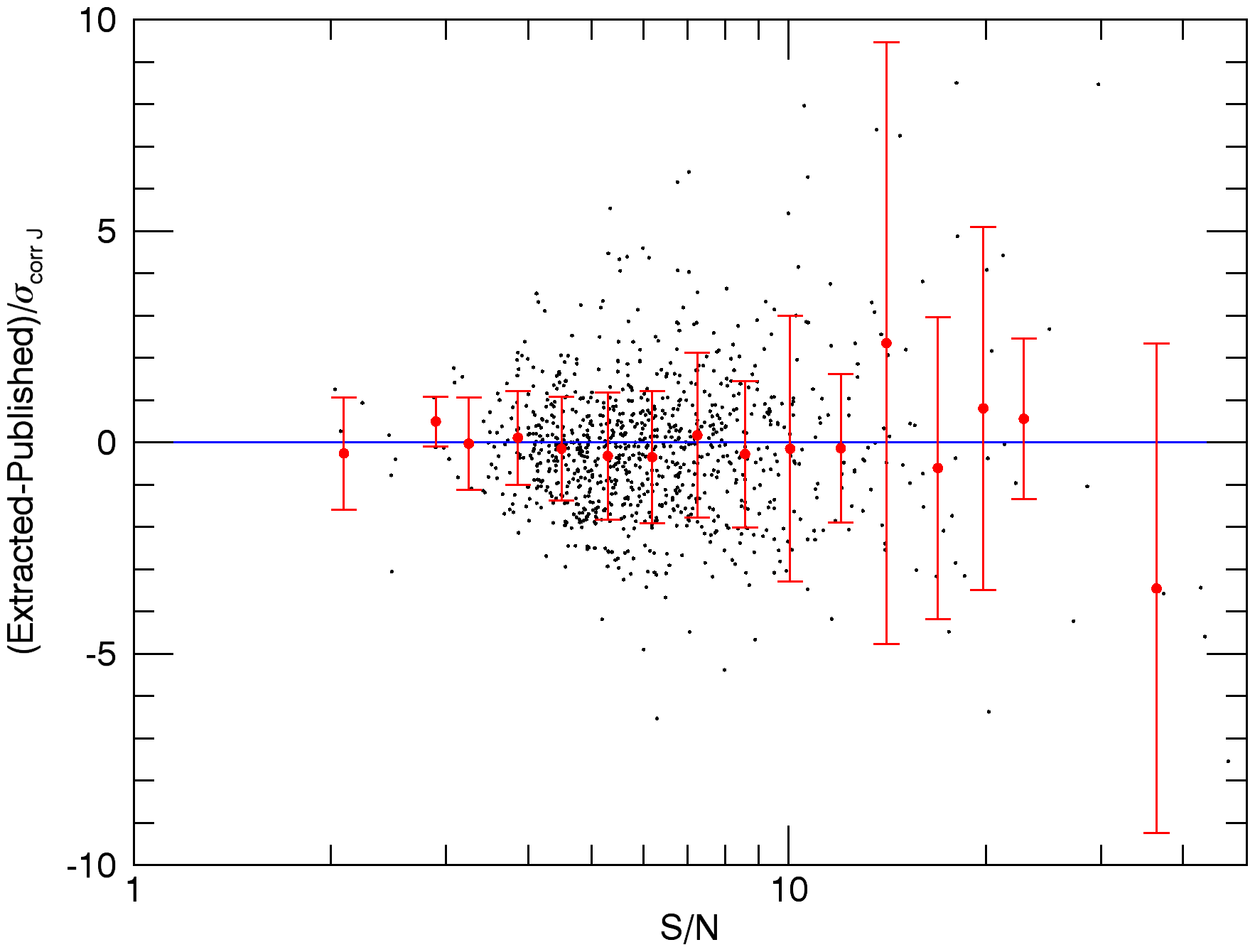}\label{fig:realmmf3_vs_snr_joint}}
	\caption{Photometry results of the extraction of the PSZ2 clusters using the proposed X-ray-SZ filter with the average cluster profile and assuming the position, size, and redshift of the clusters are known. The six panels are analogous to those in Fig. \ref{fig:joint_pxcc}, with panel e showing the difference between the extracted and the published $L_{500}$, divided by the corrected standard deviation $\sigma_{\rm corr J}$ as a function of the mass of each cluster instead of as a function of the flux as in Fig. \ref{fig:realpxcc_vs_flux_joint}.
	}
	\label{fig:realmmf3_joint}
\end{figure*}

\begin{table}
	\caption{Main properties of the histograms in Figs. \ref{fig:extraction_joint_pxcc},  \ref{fig:hist_joint_pxcc_corr}, and \ref{fig:hist_joint_mmf3}, corresponding to the extraction of clusters using the proposed X-ray-SZ MMF. The first and second columns correspond to the histograms of the extraction results for the real MCXC clusters, before (Fig. \ref{fig:extraction_joint_pxcc}) and after (Fig. \ref{fig:hist_joint_pxcc_corr}) correction for the profile mismatch effect. The third column corresponds to the histogram of the extraction results for the real PSZ2 clusters after correction for the profile mismatch effect (Fig. \ref{fig:hist_joint_mmf3}).}
	\label{table:simupxcc_joint}
	\centering 
	\setlength{\tabcolsep}{4pt}
	\begin{tabular}{c c c c}
		\hline
		\noalign{\smallskip}
		& Fig. \ref{fig:extraction_joint_pxcc} & Fig. \ref{fig:hist_joint_pxcc_corr} & Fig. \ref{fig:hist_joint_mmf3}\\
		\noalign{\smallskip}
		\hline
		\noalign{\smallskip}
		Median 				& +0.00066  & +0.00035  & -0.171\\
		Mean        		& +0.093    & -0.020    & -0.132\\
		Skewness    		& +3.119    & -0.514    & +3.333\\
		Kurtosis    		& +63.518   & +7.315    & +52.664\\
		Standard deviation 	& 3.319     & 1.374     & 2.138\\
		68\% lower limit 	& -1.689    & -1.041    & -1.689\\
		68\% upper limit 	& +1.775    & +1.001    & +1.251\\
		\noalign{\smallskip}
		\hline
	\end{tabular}
\end{table}

Figure \ref{fig:joint_pxcc} shows the extraction results for the MCXC sample. Figure \ref{fig:extraction_joint_pxcc} shows the extracted value of $L_{500}$ for each cluster as a function of the published value. As in the simulations, we used the redshift in the catalogue to convert from flux to $L_{500}$.  
The extracted flux follows the published flux quite well, but the dispersion is larger than in the simulations. The best linear fit to these data is given by $y=0.972(\pm0.004)x+3.03(\pm0.28)\cdot10^{-3}$, which is very close to the unity-slope line, as shown in the figure. 
The subplot shows the histogram of the difference between the extracted and the published flux, divided by the estimated standard deviation $\sigma_{\hat{y}_0}$. Some of the properties of this histogram are summarized in Table \ref{table:simupxcc_joint}. This histogram again shows that there is no bias, but that this time the estimated error bars are not large enough to describe the dispersion of the results (as the 68$\%$ of the extractions fall in an interval that is almost $\pm2\sigma_{\hat{y}_0}$). This additional dispersion may come from the difference between the profile used for extraction and the real profile of each particular cluster, as in the X-ray case, but also from the scatter in the real $L_{500}/Y_{500}$ relation (we refer to Appendix \ref{app:simuPXCCjoint} for an illustration based on simulations of how this scatter increases the dispersion in the extracted flux).

Since in practice the exact profile of the clusters we will detect and their $L_{500}/Y_{500}$ are unknown, we will multiply the estimated value of $\sigma_{\hat{y}_0}$, given in Eq. \ref{eq:totalvariance_joint}, by a correction factor to embed the additional dispersion into the estimated standard deviation, as we did in the X-ray case. In particular, we will use the following expression, which was obtained similarly to the correction factor for the X-ray case,
\begin{equation}\label{eq:sigma_joint_corrected}
\sigma_{\rm corr J}^2 =  \sigma_{\hat{y}_0}^2  \cdot \left( 1+0.1 \frac{\theta_{500}}{1 \rm arcmin}\right) ^2.
\end{equation}
As for the X-ray case, this correction factor is not universal
. However, the correction is stronger in the regime we are not interested in (clusters with large apparent sizes), therefore we consider it as a good approximation for our purposes.

By applying this empirical correction, we obtain the histogram in Fig. \ref{fig:hist_joint_pxcc_corr} (see main properties on Table \ref{table:simupxcc_joint}), which again shows that there is no bias and that the corrected error bars now describe the dispersion on the results well (as the 68$\%$ of the extractions fall in the interval $\pm1\sigma$ ). Figures \ref{fig:realpxcc_vs_z_joint} to \ref{fig:realpxcc_vs_snr_joint} show the difference between the extracted and the published flux, divided by the corrected $\sigma_{\hat{y}_0}$ as a function of the redshift, the size, the flux, and the S/N of each cluster. The extraction behaves correctly for almost all the values of these parameters (except for clusters with very large apparent size or that are very close, which are not the objects of our interest), and that no systematic error is introduced in our region of interest (more distant clusters).

Figure \ref{fig:realmmf3_joint} shows the extraction results for the PSZ2 sample. Figure \ref{fig:extraction_joint_mmf3} shows the extracted value of $L_{500}$ for each cluster as a function of the published value. As before, we used the redshift in the catalogue to convert from flux to $L_{500}$. 
Figure \ref{fig:hist_joint_mmf3} shows the histogram of the difference between the extracted and the published flux, divided by $\sigma_{\rm corr J}$, the estimated standard deviation corrected according to Eq. \ref{eq:sigma_joint_corrected}. Some of the properties of this histogram are summarized in Table \ref{table:simupxcc_joint}. This histogram shows a small negative bias and a dispersion that is higher than expected. The bias arises because the assumed $F_{\rm X}/Y_{500}$ relation is not perfectly suited for this sample of clusters. The additional dispersion may again stem from the scatter in the relation and from the profile mismatch, and the correction that we found for the MCXC clusters is not perfect in this case.
Figures \ref{fig:realmmf3_vs_z_joint} to \ref{fig:realmmf3_vs_snr_joint} show the difference between the extracted and the published flux, divided by the corrected $\sigma_{\hat{y}_0}$ as a function of the redshift, the size, the mass, and the S/N of each cluster. In the range we are interested in, these parameters do not introduce additional biases.

\subsection{Considerations on the $F_{\rm X}/Y_{500}$ relation}\label{sec:discussion}

\begin{figure*}[]
	\centering
	\subfigure{\includegraphics[width=\columnwidth]{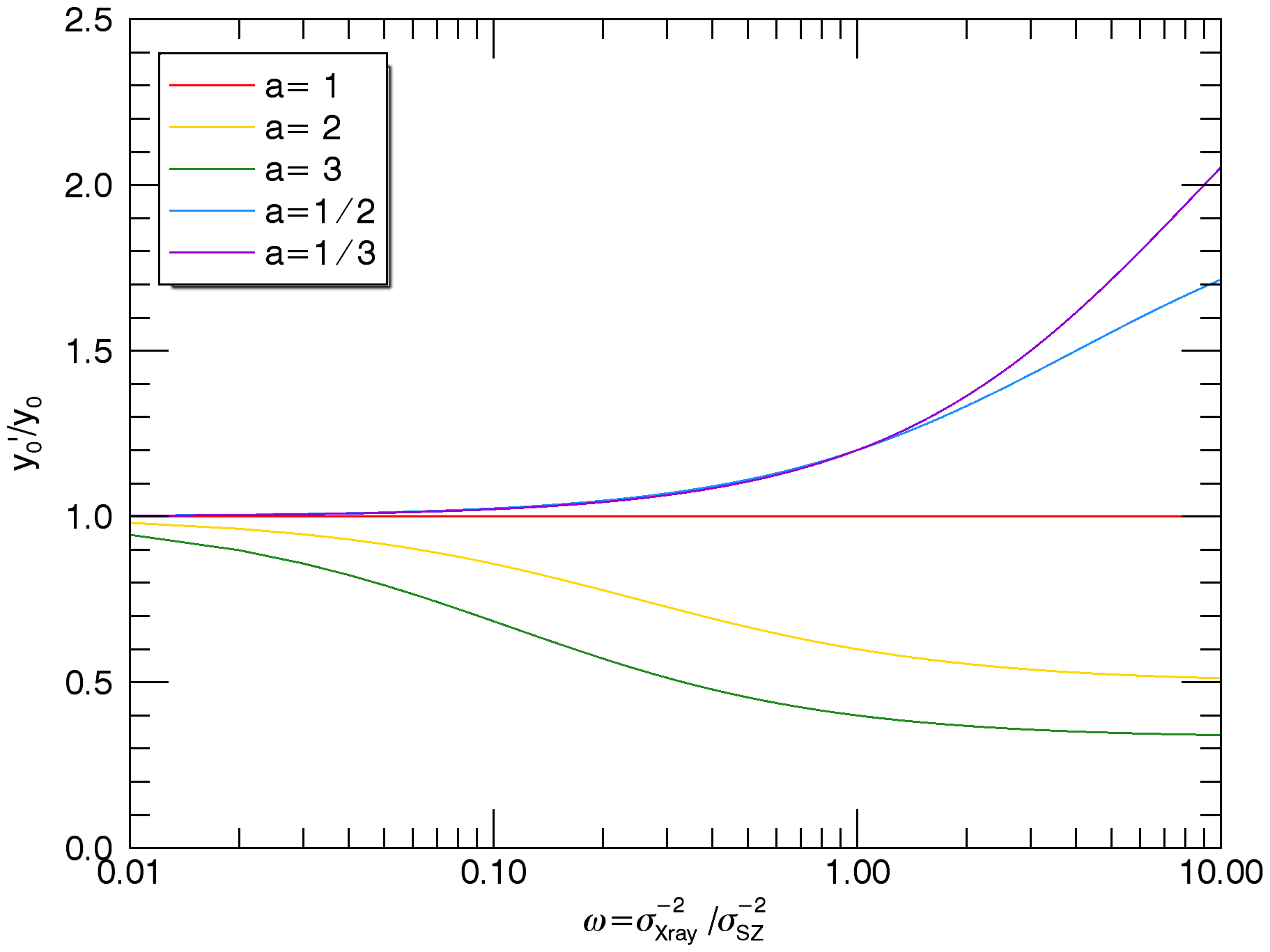}}
	\subfigure{\includegraphics[width=\columnwidth]{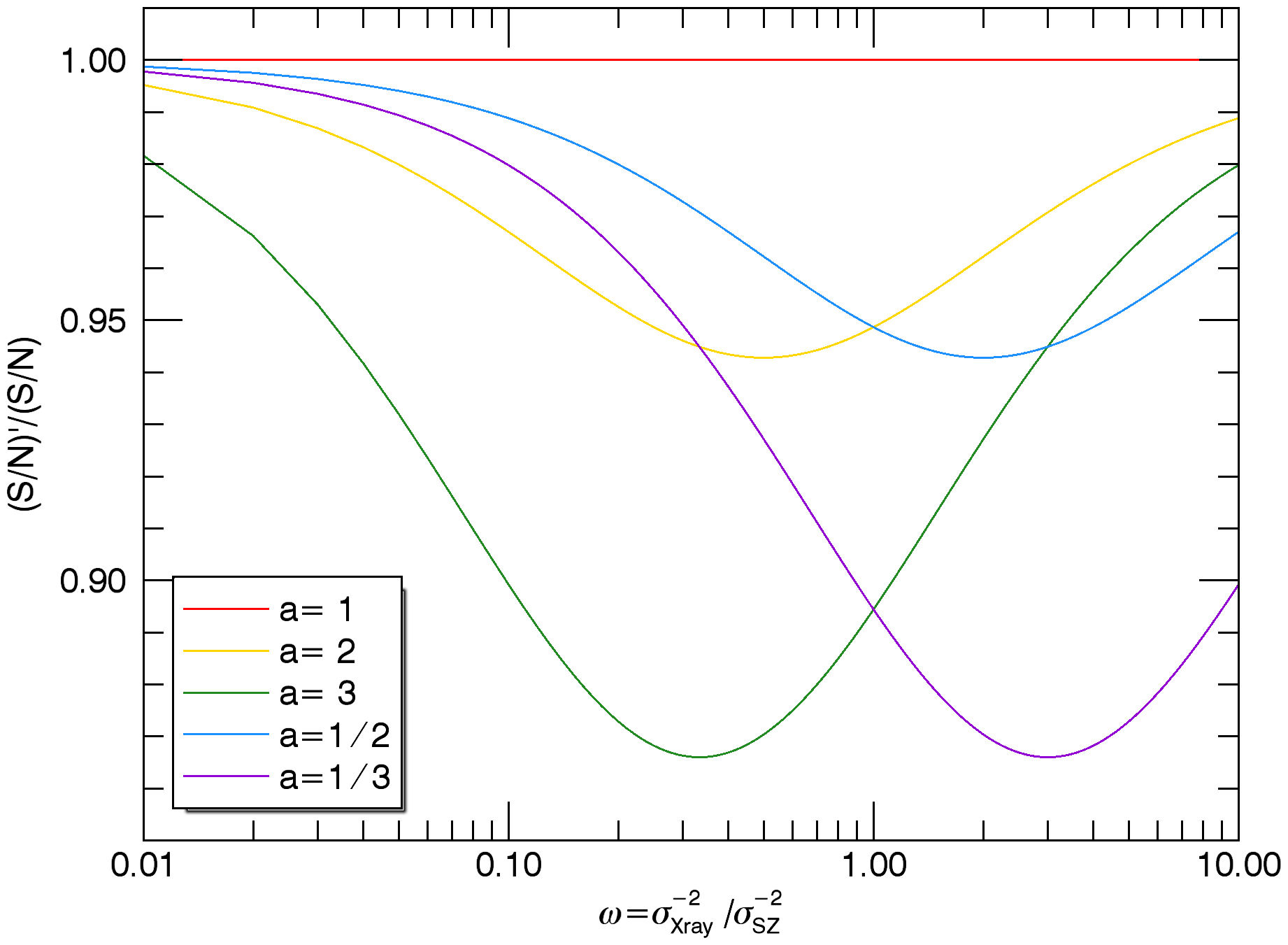}}
	\caption{Expected values of the extracted flux (left panel) and the estimated S/N (right panel) when an incorrect $F_{\rm X}/Y_{500}$ relation is assumed. The figure shows the ratio with respect to the expected value corresponding to the true relation as a function of the relative contribution of the X-ray and SZ background noises $\omega$ (low $\omega$ values represent better SZ maps and high $\omega$ values correspond to a better X-ray map). Different values of $a = (F_{\rm X}/Y_{500})_{\rm assumed} / (F_{\rm X}/Y_{500})_{\rm true}$ are shown in different colors, according to the legend.}
	\label{fig:effects_wrongfx2y500}
\end{figure*}

A key point of the joint algorithm is the expected $F_{\rm X}/Y_{500}$ relation that is used to convert the X-ray map into an additional single-frequency SZ map. If the assumed relation does not correspond to the true relation for a given cluster, the estimated flux for that cluster will be incorrect, and so will its estimated S/N. We have already pointed out the effects on photometry in the previous section, where we showed with the extraction of the MCXC clusters that although we know the average $F_{\rm X}/Y_{500}$ relation for this cluster sample, the intrinsic dispersion in this relation produces an additional scatter in the estimated flux. We also showed with the extraction of the PSZ2 clusters that if we do not precisely know the average $F_{\rm X}/Y_{500}$ relation for the clusters we extract, we will get a bias in the extracted flux. In this section, we quantitatively analyse the effects of using an incorrect $F_{\rm X}/Y_{500}$ relation on the photometry and also on the estimated S/N.

Let us consider a given cluster for which the assumed relation differs from the true relation by a factor $a$ defined as follows:
\begin{equation}\label{eq:factor_a}
a=\frac{(F_{\rm X}/Y_{500})_{\rm assumed}}{(F_{\rm X}/Y_{500})_{\rm true}}.
\end{equation}
It can be proven that the expected values for the extracted joint flux and the S/N of that cluster are given by
\begin{equation}\label{eq:y0_wrongFxY500}
y'_0 =  y_0\frac{1+a\omega}{1+a^2\omega}
\end{equation}
\begin{equation}\label{eq:snr_wrongFxY500}
\rm{S/N}' =  \rm{S/N}\frac{1+a\omega}{\sqrt{(1+a^2\omega)(1+\omega)}},
\end{equation}
where 
\begin{equation}\label{eq:omega}
\omega=\frac{\sigma_{\rm Xray}^{-2}}{\sigma_{\rm SZ}^{-2}},
\end{equation}
and $y_{0}$ and $\rm{S/N}=y_0/\sigma_{\theta_{\rm s}}$ are the expected values when the true $F_{\rm X}/Y_{500}$ relation is known.

Figure \ref{fig:effects_wrongfx2y500} shows the effect of using an incorrect $F_{\rm X}/Y_{500}$ on the extracted flux and the estimated S/N. The joint flux can be severely affected, especially when the noise of the X-ray map is small in comparison to the noise of the SZ maps. For example, when we use a relation that is a factor of two higher than the true one, the estimated flux can be divided by two. The figure also shows that when the assumed relation is smaller than the true relation (a<1), the joint flux is overestimated (so the SZ flux obtained from it will be also overestimated, and the corresponding estimated X-ray flux will be underestimated), while when the assumed relation is higher (a>1), the joint flux is underestimated. For the S/N the effect of using an incorrect $F_{\rm X}/Y_{500}$ is not as significant as for the flux. The S/N is always underestimated with respect to the S/N that we would obtain with the correct $F_{\rm X}/Y_{500}$ relation. However, the ratio between the two is bounded: there is a minimum at $\omega=1/a$, whose value is $2\sqrt{a}/(1+a)$ (note that this value is the same for $a$ and $1/a$). For example, if we are incorrect by a factor of two in the $F_{\rm X}/Y_{500}$ relation, we will obtain down to 94\% of the S/N, while if we are incorrect by a factor of three, we will obtain down to 87\% of the S/N. Given that we do not expect to deviate from the true relation by much more than this, we can conclude that the estimated S/N will be affected by at most a few percent, which means that the effect on the detection probability will be weak as well.

\begin{figure}[]
	\centering
	\includegraphics[width=\columnwidth]{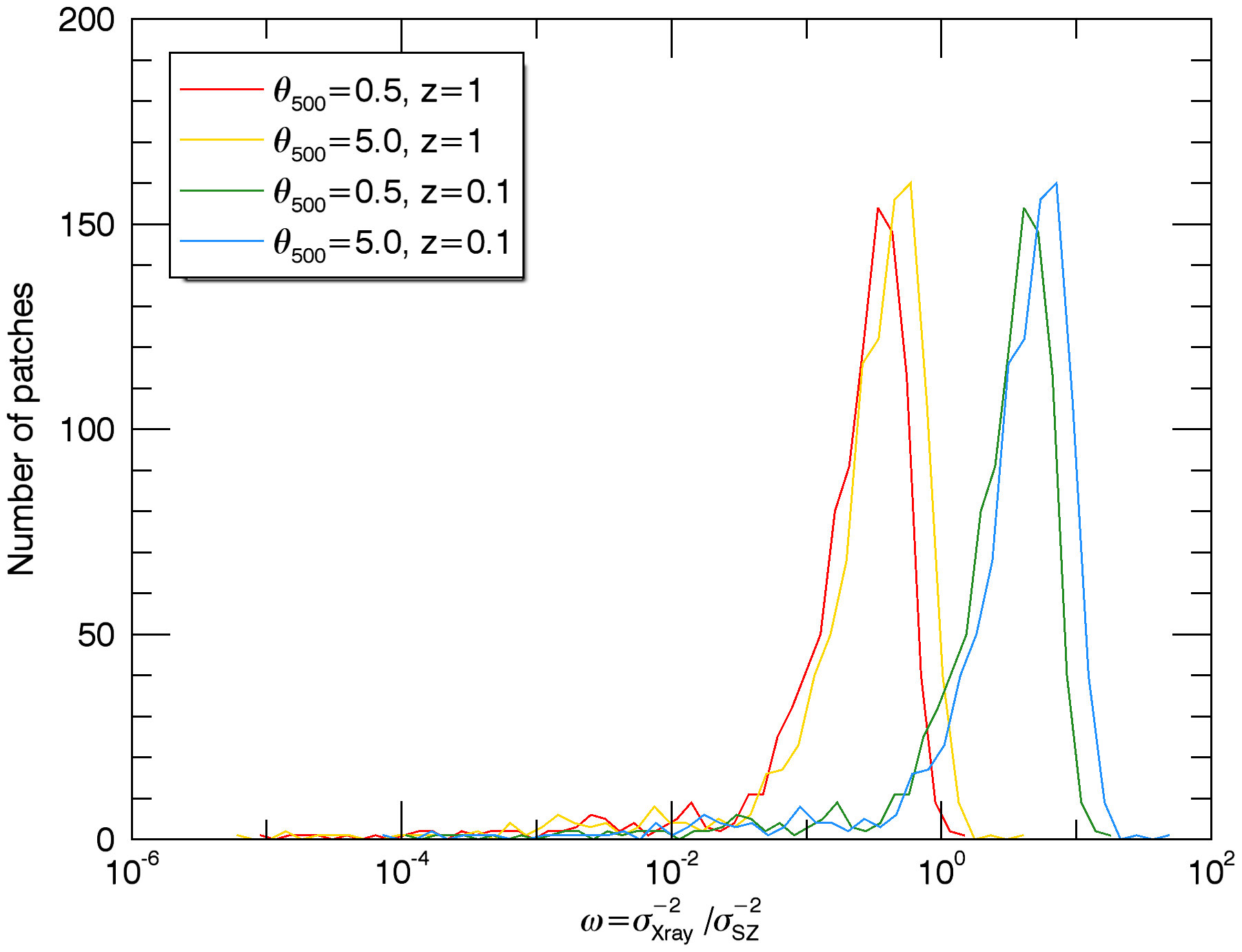}
	\caption{Histogram of the relative contribution of the X-ray and SZ background noises $\omega$ for the RASS and \emph{Planck} maps.}
	\label{fig:hist_sigmaXR_sigmaSZ}
\end{figure}

To estimate the effect of using an incorrect $F_{\rm X}/Y_{500}$ relation with this figure, we still have two questions to answer. First, what is the range of $\omega$ values that we have in our maps? and second, what is the range of values that we expect for $a$?. Figure \ref{fig:hist_sigmaXR_sigmaSZ} answers the first question, showing the regime of $\omega$ values that we have in our maps, that is, the six highest frequency \emph{Planck} all-sky maps and the X-ray all-sky map that we constructed from RASS data. The histograms were calculated by filtering 1000 random patches of the sky with different filter sizes. The value of $\omega$ slightly depends on the size of the filter, while it depends much more on the reference redshift that we use to convert the X-ray map into an equivalent SZ map, since $\sigma_{\rm Xray}^{-2}$ is proportional to $(F_{\rm X}/Y_{500})^2(z)$.

\begin{figure}[]
	\centering
	\includegraphics[width=\columnwidth]{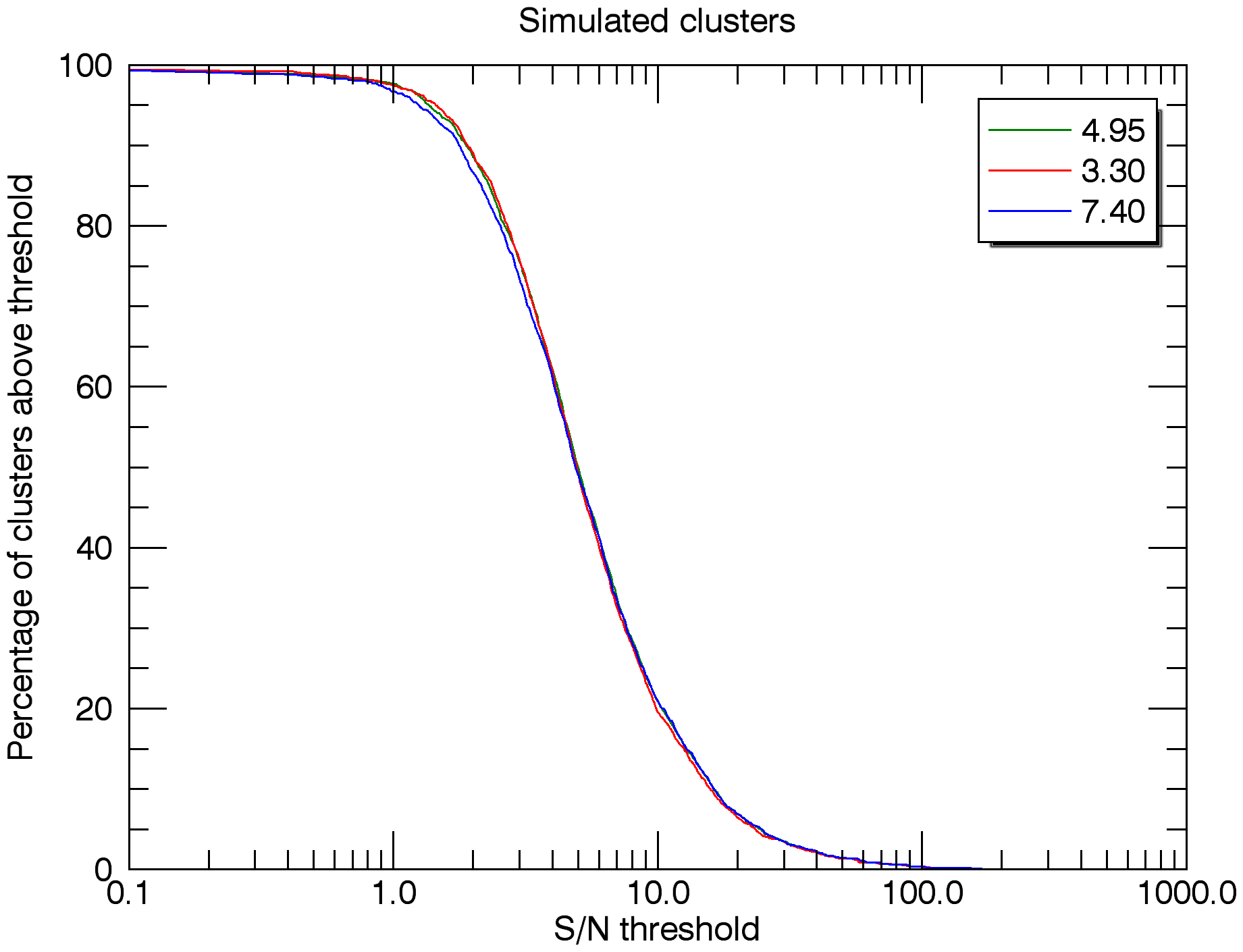}
	\caption{Percentage of simulated clusters (as described in Sect. \ref{sec:discussion}) whose extracted S/N, using the proposed X-ray matched filter and assuming the position, size, and redshift of the clusters are known, is above a given S/N threshold. The different colors correspond to different normalizations of the $F_{\rm X}/Y_{500}$ relation assumed for the extraction, as indicated in the legend.}
	\label{fig:detections_vs_snr_simumassf_fx2y500}
\end{figure}

To illustrate the effect of the $F_{\rm X}/Y_{500}$ relation that is assumed in the extraction, we carried out an experiment in which we simulated clusters according to the Tinker mass function \citep{Tinker2008}, with $M_{500} > 3 \cdot 10^{14} M_{\odot}$ and $0<z<1$. In this simulation, we injected clusters following, on average, the $F_{\rm X}/Y_{500}$ relation in Eq. \ref{eq:FxY500relation}, and we added a scatter of $\sigma_{\rm{log} L}=0.183$, which is the typical scatter found in the L-M relation \citep{Arnaud2010,PlanckEarlyXI}. A total of 1787 clusters were injected in the maps and extracted using the proposed joint matched filter, assuming three different $F_{\rm X}/Y_{500}$ relations for the extraction (similar to Eq. \ref{eq:FxY500relation}, but with different normalizations: 3.30, 4.95 and 7.40). Although assuming a $F_{\rm X}/Y_{500}$ relation different from the real one introduces, as discussed previously, errors in the estimated flux, it has no strong effect on the S/N of the detected clusters and therefore does not affect the detection probability of the method, as illustrated in Fig. \ref{fig:detections_vs_snr_simumassf_fx2y500}. We also checked that the assumed $F_{\rm X}/Y_{500}$ does not have an effect on the estimated S/N of the joint extraction of the real MCXC clusters, which makes the detection robust against possible errors in the assumed relation. 

Regarding the second question, the $F_{\rm X}/Y_{500}$ relation for a given cluster is not a fixed quantity: it depends on the redshift and on the luminosity of the cluster, and even for a given redshift and luminosity, it also has some intrinsic dispersion. 
 In the following, we summarize how these variables are expected to affect the joint extraction results: 
\begin{itemize}
	\item Redshift: When we search for clusters whose redshift we do not know, we assume a reference redshift $z_{\rm ref}$ for all the potential clusters in our maps. For example, if we take an intermediate redshift of 0.5 as reference, and assuming the $F_{\rm X}/Y_{500}$ only depends on the redshift as in Eq. \ref{eq:FxY500relation}, then the values of $a$ would range between 0.5 and 1.7 for clusters at z=0.1 and z=1, respectively. Since the $F_{\rm X}/Y_{500}$ relation decreases with redshift, if the real redshift of a cluster is higher (or lower) than $z_{\rm ref}$, then $a>1$ ($a<1$), so that its SZ flux will be underestimated (or overestimated). Regarding the S/N, the effect of using a reference redshift that is different from the cluster redshift is limited. For example, if we take an intermediate redshift of 0.5 as reference, then the S/N for a cluster at z=1 will only be 96.6\% (at most) of the one we would obtain with the correct redshift.
	\item Luminosity: The $F_{\rm X}/Y_{500}$ relation is calculated by approximating the $Y_{500}-L_{500}$ relation by a unity-slope relation for a pivot luminosity. If this reference luminosity is higher than the true luminosity, the assumed relation will be smaller ($a<1$). For example, the $F_{\rm X}/Y_{500}$ relation in Eq. \ref{eq:FxY500relation} comes from the $Y_{500}-L_{500}$ relation derived by the \cite{PlanckEarlyXI}, for a pivot luminosity of $7\cdot10^{44}$ erg/s. For clusters with luminosities ranging from $10^{42}$ to $10^{46}$ erg/s, the difference between Eq. \ref{eq:FxY500relation} and the $F_{\rm X}/Y_{500}$ value derived from the original $Y_{500}-L_{500}$ relation (without approximations) ranges between $a=0.88$ and $a=1.05$. This clearly is a second-order effect, and it could be further reduced by using an iterative process (assume luminosity, then extract flux for the corresponding relation, calculate new relation corresponding to estimated luminosity, re-extract, and so on), which means that we do not need to worry about it.
	\item Intrinsic scatter: Even if we knew the redshift and luminosity of our cluster, there is an intrinsic dispersion in the $F_{\rm X}/Y_{500}$ relation. This will introduce an additional scatter in the estimated flux (see Appendix \ref{app:scatter_fx_Y500} for an illustration), as we have mentioned before, and slightly reduce the S/N of the clusters.
 
\end{itemize}

\begin{figure}[]
	\centering
	\includegraphics[width=\columnwidth]{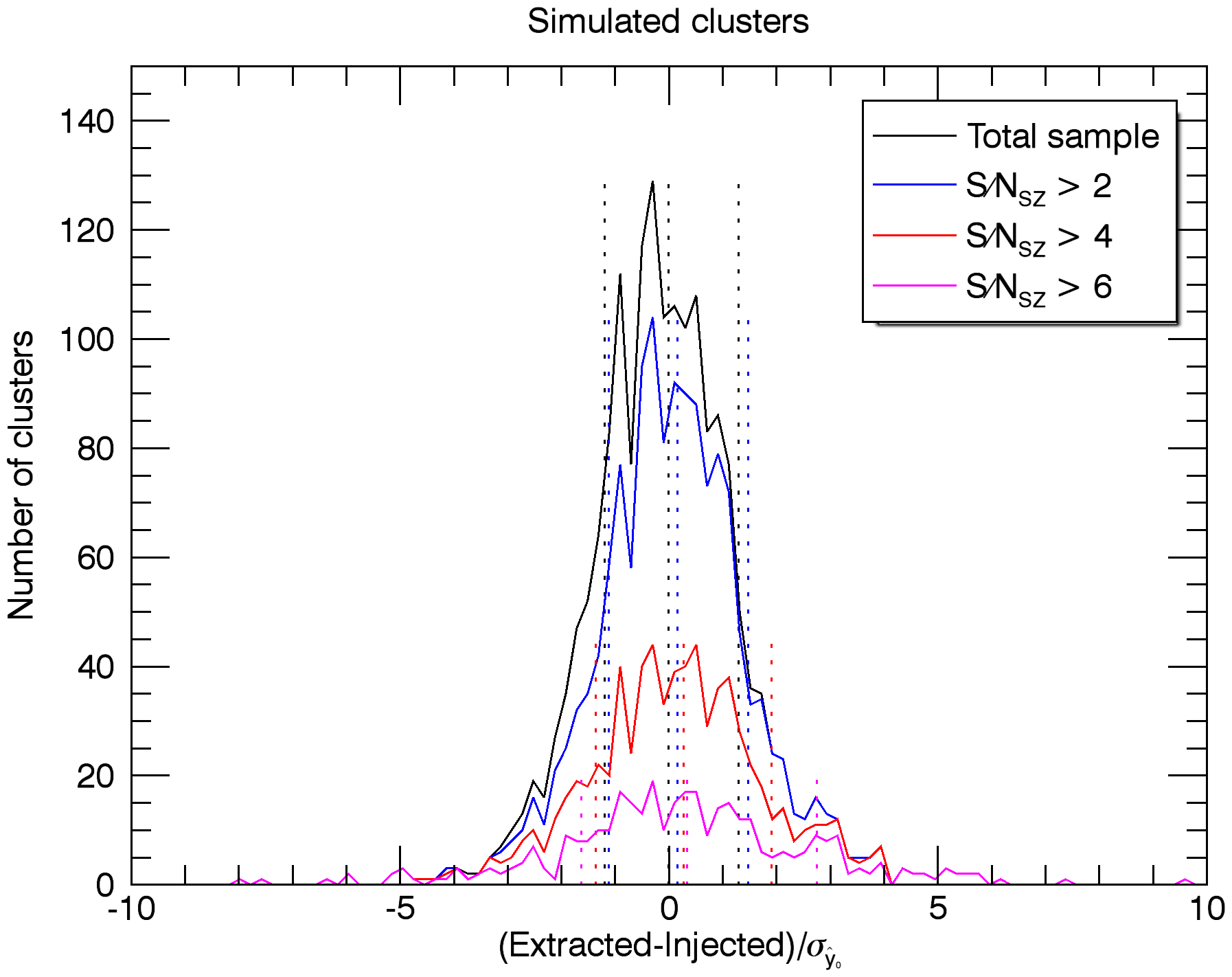}
	\caption{Histogram of the difference between the extracted and the injected $L_{500}$, divided by the estimated $\sigma_{\hat{y}_0}$ (scaled to $L_{500}$ units) for the simulated clusters (as described in Sect. \ref{sec:discussion}) extracted with the proposed X-ray-SZ MMF and applying different selection criteria. The central vertical line shows the median value, while the other two vertical lines indicate the region inside which 68$\%$ of the clusters lie.}
	\label{fig:hist_joint_simumassf}
\end{figure}

Finally, we should note that the average $F_{\rm X}/Y_{500}$ depends on the sample, therefore selection effects play a role in its average value. If we take an average $F_{\rm X}/Y_{500}$ relation that does not correspond to the true average $F_{\rm X}/Y_{500}$ relation of the clusters we detect, the flux of the sample will be biased. 
To illustrate the effect of the 
last point, 
we analysed the photometry results from the last simulation (clusters following the Tinker mass function). 
Figure \ref{fig:hist_joint_simumassf} shows how the histogram of the difference between the extracted and the injected value, divided by the estimated standard deviation $\sigma_{\hat{y}_0}$, changes depending on the selection of the sample. For the complete sample, we can see that there is no bias, but because of the scatter in the $L_{500}-Y_{500}$ relation, the estimated error bars are not enough to fully describe the dispersion on the results. If we select only those clusters for which the S/N in the \emph{Planck} maps is above a given threshold, an increasing positive bias appears. 

From these considerations we can conclude that the proposed X-ray-SZ matched filter will provide a higher detection probability than single-survey filters, even if we have some error in the assumed $F_{\rm X}/Y_{500}$ relation, since the S/N of the detected clusters is only slightly affected by these errors. However, not knowing the value of the $F_{\rm X}/Y_{500}$ relation has a stronger effect on the photometry, which means that the fluxes estimated by the joint algorithm will not be precise. It is therefore better to keep the independent flux estimates from the two single-survey filters instead of using the combined flux that depends on our knowledge of the $F_{\rm X}/Y_{500}$ relation.

We remark that the effects described in this section are averaged effects, that is, they refer to the expected values of the flux and S/N. In addition, there will be a scatter caused by the filtered noise, which will be smaller for clusters with higher flux.

\section{Conclusions}\label{sec:conclusions}

In this paper we have proposed a matched-filter approach to extract the signal from galaxy clusters using X-ray and SZ maps simultaneously. The method is based on the combination of the classical SZ MMF and an analogous single-frequency matched filter developed for X-ray maps. The combination relies on the physical relation between X-ray and SZ emission, namely the expected $F_{\rm X}/Y_{500}$ relation.

We have shown through the injection of simulated clusters on RASS maps and through the extraction of known clusters that the proposed X-ray matched filter provides correct photometry results for the extracted clusters. The estimated standard deviation accounts correctly for the noise in the maps (background and Poisson contributions), but not for other issues that increase the dispersion of the results, mainly the mismatch between the real profile of the cluster and the one assumed for the extraction. We proposed an empirical correction to account for this additional dispersion.

We have also demonstrated through the injection of simulated clusters on RASS and \emph{Planck} maps and through the extraction of known clusters on these maps that the proposed joint MMF also provides a correct photometry as long as we know the $F_{\rm X}/Y_{500}$ relation of our clusters. The estimated standard deviation in this case correctly describes the noise in the maps (background and Poisson contributions), but does not take into account other effects that may come into play, such as the profile mismatch, the mismatch between the real and the assumed $F_{\rm X}/Y_{500}$ relation, or a possible shift between the X-ray and SZ peaks. An empirical correction for the standard deviation is proposed to incorporate part of these effects. 

We have also checked that the proposed joint algorithm provides, in general, a better S/N than the single-map extractions, which results in an increase of the detection rate with respect to the X-ray MF detection or the SZ MMF detection. Interestingly, the assumed $F_{\rm X}/Y_{500}$ relation does not have a strong effect on the estimated S/N, making the detection robust against possible errors in the assumed relation.

In future work we will assess the performance of the proposed filter as a blind detection tool, dealing with the fact that we do not know the position, the size, and the redshift of the clusters. In particular, we are planning to study the detection probability and the false-detection rate (i.e. the purity) and compare it with other cluster detection methods. We will also analyse the effect of the point sources (or other objects) on the detector performance.

The main challenge to be solved when using the proposed X-ray-SZ MMF for blind detection will be to maintain a high purity. Although adding the X-ray information increases the cluster detection probability, it will also increase the number of false detections, produced by non-cluster X-ray sources (mainly AGNs). To deal with them correctly, we will take into account the S/N of the blind detection candidates when extracted from X-ray-only and SZ-only maps. 

The proposed method, applied on \emph{Planck} and RASS maps, will provide the last and deepest all-sky cluster catalogue before the e-ROSITA mission.

\begin{acknowledgements}
        This research is based on observations obtained with Planck (http://www.esa.int/Planck), an ESA science mission with instruments and contributions directly funded by ESA Member States, NASA, and Canada. This research has made use of the ROSAT all-sky survey data which have been processed at MPE. The authors acknowledge the use of the HEALPix package \citep{Gorski2005}. The authors would like to thank Nabila Aghanim, Etienne Pointecouteau, Amandine Le Brun and Remco van der Burg for helpful comments and suggestions. The authors thank an anonymous referee for constructive comments. The research leading to these results has received funding from the European Research Council under the European Union’s Seventh Framework Programme (FP7/2007-2013) / ERC grant agreement n$^{\circ}$ 340519.  
\end{acknowledgements}

\bibliographystyle{aa} 
\bibliography{Tarrio-biblio}

\begin{thebibliography}{41}
\expandafter\ifx\csname natexlab\endcsname\relax\def\natexlab#1{#1}\fi

\bibitem[{{Abell}(1958)}]{Abell1958}
{Abell}, G.~O. 1958, \apjs, 3, 211

\bibitem[{{Arnaud}(2005)}]{Arnaud2005}
{Arnaud}, M. 2005, in Background Microwave Radiation and Intracluster
  Cosmology, ed. F.~{Melchiorri} \& Y.~{Rephaeli}, 77

\bibitem[{{Arnaud} {et~al.}(2010){Arnaud}, {Pratt}, {Piffaretti},
  {B{\"o}hringer}, {Croston}, \& {Pointecouteau}}]{Arnaud2010}
{Arnaud}, M., {Pratt}, G.~W., {Piffaretti}, R., {et~al.} 2010, \aap, 517, A92

\bibitem[{{Bertin} \& {Arnouts}(1996)}]{Bertin1996}
{Bertin}, E. \& {Arnouts}, S. 1996, \aaps, 117, 393

\bibitem[{{Birkinshaw}(1999)}]{Birkinshaw1999}
{Birkinshaw}, M. 1999, \physrep, 310, 97

\bibitem[{{Bleem} {et~al.}(2015){Bleem}, {Stalder}, {de Haan}, {Aird}, {Allen},
  {Applegate}, {Ashby}, {Bautz}, {Bayliss}, {Benson}, {Bocquet}, {Brodwin},
  {Carlstrom}, {Chang}, {Chiu}, {Cho}, {Clocchiatti}, {Crawford}, {Crites},
  {Desai}, {Dietrich}, {Dobbs}, {Foley}, {Forman}, {George}, {Gladders},
  {Gonzalez}, {Halverson}, {Hennig}, {Hoekstra}, {Holder}, {Holzapfel},
  {Hrubes}, {Jones}, {Keisler}, {Knox}, {Lee}, {Leitch}, {Liu}, {Lueker},
  {Luong-Van}, {Mantz}, {Marrone}, {McDonald}, {McMahon}, {Meyer}, {Mocanu},
  {Mohr}, {Murray}, {Padin}, {Pryke}, {Reichardt}, {Rest}, {Ruel}, {Ruhl},
  {Saliwanchik}, {Saro}, {Sayre}, {Schaffer}, {Schrabback}, {Shirokoff},
  {Song}, {Spieler}, {Stanford}, {Staniszewski}, {Stark}, {Story}, {Stubbs},
  {Vanderlinde}, {Vieira}, {Vikhlinin}, {Williamson}, {Zahn}, \&
  {Zenteno}}]{Bleem2015}
{Bleem}, L.~E., {Stalder}, B., {de Haan}, T., {et~al.} 2015, \apjs, 216, 27

\bibitem[{{Boese}(2000)}]{Boese2000}
{Boese}, F.~G. 2000, \aaps, 141, 507

\bibitem[{{B{\"o}hringer} {et~al.}(2013){B{\"o}hringer}, {Chon}, {Collins},
  {Guzzo}, {Nowak}, \& {Bobrovskyi}}]{Bohringer2013}
{B{\"o}hringer}, H., {Chon}, G., {Collins}, C.~A., {et~al.} 2013, \aap, 555,
  A30

\bibitem[{{B{\"o}hringer} {et~al.}(2000){B{\"o}hringer}, {Voges}, {Huchra},
  {McLean}, {Giacconi}, {Rosati}, {Burg}, {Mader}, {Schuecker}, {Simi{\c c}},
  {Komossa}, {Reiprich}, {Retzlaff}, \& {Tr{\"u}mper}}]{Bohringer2000}
{B{\"o}hringer}, H., {Voges}, W., {Huchra}, J.~P., {et~al.} 2000, \apjs, 129,
  435

\bibitem[{{Carlstrom} {et~al.}(2002){Carlstrom}, {Holder}, \&
  {Reese}}]{Carlstrom2002}
{Carlstrom}, J.~E., {Holder}, G.~P., \& {Reese}, E.~D. 2002, \araa, 40, 643

\bibitem[{{Ebeling} \& {Wiedenmann}(1993)}]{Ebeling1993}
{Ebeling}, H. \& {Wiedenmann}, G. 1993, \pre, 47, 704

\bibitem[{{G{\'o}rski} {et~al.}(2005){G{\'o}rski}, {Hivon}, {Banday},
  {Wandelt}, {Hansen}, {Reinecke}, \& {Bartelmann}}]{Gorski2005}
{G{\'o}rski}, K.~M., {Hivon}, E., {Banday}, A.~J., {et~al.} 2005, \apj, 622,
  759

\bibitem[{{Haehnelt} \& {Tegmark}(1996)}]{Haehnelt1996}
{Haehnelt}, M.~G. \& {Tegmark}, M. 1996, \mnras, 279, 545

\bibitem[{{Hasselfield} {et~al.}(2013){Hasselfield}, {Hilton}, {Marriage},
  {Addison}, {Barrientos}, {Battaglia}, {Battistelli}, {Bond}, {Crichton},
  {Das}, {Devlin}, {Dicker}, {Dunkley}, {D{\"u}nner}, {Fowler}, {Gralla},
  {Hajian}, {Halpern}, {Hincks}, {Hlozek}, {Hughes}, {Infante}, {Irwin},
  {Kosowsky}, {Marsden}, {Menanteau}, {Moodley}, {Niemack}, {Nolta}, {Page},
  {Partridge}, {Reese}, {Schmitt}, {Sehgal}, {Sherwin}, {Sievers}, {Sif{\'o}n},
  {Spergel}, {Staggs}, {Swetz}, {Switzer}, {Thornton}, {Trac}, \&
  {Wollack}}]{Hasselfield2013}
{Hasselfield}, M., {Hilton}, M., {Marriage}, T.~A., {et~al.} 2013, \jcap, 7, 8

\bibitem[{{Herranz} {et~al.}(2002){Herranz}, {Sanz}, {Hobson}, {Barreiro},
  {Diego}, {Mart{\'{\i}}nez-Gonz{\'a}lez}, \& {Lasenby}}]{Herranz2002}
{Herranz}, D., {Sanz}, J.~L., {Hobson}, M.~P., {et~al.} 2002, \mnras, 336, 1057

\bibitem[{{Kalberla} {et~al.}(2005){Kalberla}, {Burton}, {Hartmann}, {Arnal},
  {Bajaja}, {Morras}, \& {P{\"o}ppel}}]{Kalberla2005}
{Kalberla}, P.~M.~W., {Burton}, W.~B., {Hartmann}, D., {et~al.} 2005, \aap,
  440, 775

\bibitem[{{Maturi}(2007)}]{Maturi2007}
{Maturi}, M. 2007, Astronomische Nachrichten, 328, 690

\bibitem[{{Melin} {et~al.}(2012){Melin}, {Aghanim}, {Bartelmann}, {Bartlett},
  {Betoule}, {Bobin}, {Carvalho}, {Chon}, {Delabrouille}, {Diego}, {Harrison},
  {Herranz}, {Hobson}, {Kneissl}, {Lasenby}, {Le Jeune}, {Lopez-Caniego},
  {Mazzotta}, {Rocha}, {Schaefer}, {Starck}, {Waizmann}, \& {Yvon}}]{Melin2012}
{Melin}, J.-B., {Aghanim}, N., {Bartelmann}, M., {et~al.} 2012, \aap, 548, A51

\bibitem[{{Melin} {et~al.}(2006){Melin}, {Bartlett}, \&
  {Delabrouille}}]{Melin2006}
{Melin}, J.-B., {Bartlett}, J.~G., \& {Delabrouille}, J. 2006, \aap, 459, 341

\bibitem[{{Merloni} {et~al.}(2012){Merloni}, {Predehl}, {Becker},
  {B{\"o}hringer}, {Boller}, {Brunner}, {Brusa}, {Dennerl}, {Freyberg},
  {Friedrich}, {Georgakakis}, {Haberl}, {Hasinger}, {Meidinger}, {Mohr},
  {Nandra}, {Rau}, {Reiprich}, {Robrade}, {Salvato}, {Santangelo}, {Sasaki},
  {Schwope}, {Wilms}, \& {German eROSITA Consortium}}]{Merloni2012}
{Merloni}, A., {Predehl}, P., {Becker}, W., {et~al.} 2012, ArXiv e-prints
  [\eprint[arXiv]{1209.3114}]

\bibitem[{{Nagai} {et~al.}(2007){Nagai}, {Kravtsov}, \&
  {Vikhlinin}}]{Nagai2007}
{Nagai}, D., {Kravtsov}, A.~V., \& {Vikhlinin}, A. 2007, \apj, 668, 1

\bibitem[{{Pacaud} {et~al.}(2006){Pacaud}, {Pierre}, {Refregier}, {Gueguen},
  {Starck}, {Valtchanov}, {Read}, {Altieri}, {Chiappetti}, {Gandhi}, {Garcet},
  {Gosset}, {Ponman}, \& {Surdej}}]{Pacaud2006}
{Pacaud}, F., {Pierre}, M., {Refregier}, A., {et~al.} 2006, \mnras, 372, 578

\bibitem[{{Pace} {et~al.}(2008){Pace}, {Maturi}, {Bartelmann}, {Cappelluti},
  {Dolag}, {Meneghetti}, \& {Moscardini}}]{Pace2008}
{Pace}, F., {Maturi}, M., {Bartelmann}, M., {et~al.} 2008, \aap, 483, 389

\bibitem[{{Piffaretti} {et~al.}(2011){Piffaretti}, {Arnaud}, {Pratt},
  {Pointecouteau}, \& {Melin}}]{Piffaretti2011}
{Piffaretti}, R., {Arnaud}, M., {Pratt}, G.~W., {Pointecouteau}, E., \&
  {Melin}, J.-B. 2011, \aap, 534, A109

\bibitem[{{Planck Collaboration}(2011{\natexlab{a}})}]{PlanckEarlyVIII}
{Planck Collaboration}. 2011{\natexlab{a}}, \aap, 536, A8

\bibitem[{{Planck Collaboration}(2011{\natexlab{b}})}]{PlanckEarlyX}
{Planck Collaboration}. 2011{\natexlab{b}}, \aap, 536, A10

\bibitem[{{Planck Collaboration}(2011{\natexlab{c}})}]{PlanckEarlyXI}
{Planck Collaboration}. 2011{\natexlab{c}}, \aap, 536, A11

\bibitem[{{Planck Collaboration}(2012)}]{PlanckIntI2012}
{Planck Collaboration}. 2012, \aap, 543, A102

\bibitem[{{Planck Collaboration}(2013)}]{PlanckIntV2013}
{Planck Collaboration}. 2013, \aap, 550, A131

\bibitem[{{Planck Collaboration}(2014)}]{Planck2013ResXXIX}
{Planck Collaboration}. 2014, \aap, 571, A29

\bibitem[{{Planck Collaboration}(2015{\natexlab{a}})}]{Planck2015ResVIII}
{Planck Collaboration}. 2015{\natexlab{a}}, ArXiv e-prints
  [\eprint[arXiv]{1502.01587}]

\bibitem[{{Planck Collaboration}(2015{\natexlab{b}})}]{Planck2015ResXXVII}
{Planck Collaboration}. 2015{\natexlab{b}}, ArXiv e-prints
  [\eprint[arXiv]{1502.01598}]

\bibitem[{{Scharf} {et~al.}(1997){Scharf}, {Ebeling}, {Perlman}, {Malkan}, \&
  {Wegner}}]{Scharf1997}
{Scharf}, C.~A., {Ebeling}, H., {Perlman}, E., {Malkan}, M., \& {Wegner}, G.
  1997, \apj, 477, 79

\bibitem[{{Schuecker} {et~al.}(2004){Schuecker}, {B{\"o}hringer}, \&
  {Voges}}]{Schuecker2004}
{Schuecker}, P., {B{\"o}hringer}, H., \& {Voges}, W. 2004, \aap, 420, 61

\bibitem[{{Sunyaev} \& {Zeldovich}(1970)}]{Sunyaev1970}
{Sunyaev}, R.~A. \& {Zeldovich}, Y.~B. 1970, Comments on Astrophysics and Space
  Physics, 2, 66

\bibitem[{{Sunyaev} \& {Zeldovich}(1972)}]{Sunyaev1972}
{Sunyaev}, R.~A. \& {Zeldovich}, Y.~B. 1972, Comments on Astrophysics and Space
  Physics, 4, 173

\bibitem[{{Tinker} {et~al.}(2008){Tinker}, {Kravtsov}, {Klypin}, {Abazajian},
  {Warren}, {Yepes}, {Gottl{\"o}ber}, \& {Holz}}]{Tinker2008}
{Tinker}, J., {Kravtsov}, A.~V., {Klypin}, A., {et~al.} 2008, \apj, 688, 709

\bibitem[{{Truemper}(1993)}]{Truemper1993}
{Truemper}, J. 1993, Science, 260, 1769

\bibitem[{{Vikhlinin} {et~al.}(1998){Vikhlinin}, {McNamara}, {Forman}, {Jones},
  {Quintana}, \& {Hornstrup}}]{Vikhlinin1998}
{Vikhlinin}, A., {McNamara}, B.~R., {Forman}, W., {et~al.} 1998, \apj, 502, 558

\bibitem[{{Voges} {et~al.}(1999){Voges}, {Aschenbach}, {Boller},
  {Br{\"a}uninger}, {Briel}, {Burkert}, {Dennerl}, {Englhauser}, {Gruber},
  {Haberl}, {Hartner}, {Hasinger}, {K{\"u}rster}, {Pfeffermann}, {Pietsch},
  {Predehl}, {Rosso}, {Schmitt}, {Tr{\"u}mper}, \& {Zimmermann}}]{Voges1999}
{Voges}, W., {Aschenbach}, B., {Boller}, T., {et~al.} 1999, \aap, 349, 389

\bibitem[{{York} {et~al.}(2000){York}, {Adelman}, {Anderson}, {Anderson},
  {Annis}, {Bahcall}, {Bakken}, {Barkhouser}, {Bastian}, {Berman}, {Boroski},
  {Bracker}, {Briegel}, {Briggs}, {Brinkmann}, {Brunner}, {Burles}, {Carey},
  {Carr}, {Castander}, {Chen}, {Colestock}, {Connolly}, {Crocker}, {Csabai},
  {Czarapata}, {Davis}, {Doi}, {Dombeck}, {Eisenstein}, {Ellman}, {Elms},
  {Evans}, {Fan}, {Federwitz}, {Fiscelli}, {Friedman}, {Frieman}, {Fukugita},
  {Gillespie}, {Gunn}, {Gurbani}, {de Haas}, {Haldeman}, {Harris}, {Hayes},
  {Heckman}, {Hennessy}, {Hindsley}, {Holm}, {Holmgren}, {Huang}, {Hull},
  {Husby}, {Ichikawa}, {Ichikawa}, {Ivezi{\'c}}, {Kent}, {Kim}, {Kinney},
  {Klaene}, {Kleinman}, {Kleinman}, {Knapp}, {Korienek}, {Kron}, {Kunszt},
  {Lamb}, {Lee}, {Leger}, {Limmongkol}, {Lindenmeyer}, {Long}, {Loomis},
  {Loveday}, {Lucinio}, {Lupton}, {MacKinnon}, {Mannery}, {Mantsch}, {Margon},
  {McGehee}, {McKay}, {Meiksin}, {Merelli}, {Monet}, {Munn}, {Narayanan},
  {Nash}, {Neilsen}, {Neswold}, {Newberg}, {Nichol}, {Nicinski}, {Nonino},
  {Okada}, {Okamura}, {Ostriker}, {Owen}, {Pauls}, {Peoples}, {Peterson},
  {Petravick}, {Pier}, {Pope}, {Pordes}, {Prosapio}, {Rechenmacher}, {Quinn},
  {Richards}, {Richmond}, {Rivetta}, {Rockosi}, {Ruthmansdorfer}, {Sandford},
  {Schlegel}, {Schneider}, {Sekiguchi}, {Sergey}, {Shimasaku}, {Siegmund},
  {Smee}, {Smith}, {Snedden}, {Stone}, {Stoughton}, {Strauss}, {Stubbs},
  {SubbaRao}, {Szalay}, {Szapudi}, {Szokoly}, {Thakar}, {Tremonti}, {Tucker},
  {Uomoto}, {Vanden Berk}, {Vogeley}, {Waddell}, {Wang}, {Watanabe},
  {Weinberg}, {Yanny}, {Yasuda}, \& {SDSS Collaboration}}]{York2000}
{York}, D.~G., {Adelman}, J., {Anderson}, Jr., J.~E., {et~al.} 2000, \aj, 120,
  1579

\end{thebibliography}

\begin{appendix}

\section{Derivation of the noise variance after the X-ray matched filter}\label{app:sigma_poisson}
Let us assume that we have an X-ray map of dimension $n \times n$ pixels, which we can express using Eq. \ref{eq:Xray_map_2}. Taking its Fourier transform, we have
\begin{equation}\label{eq:ap:XraymapFT}
{M}(\mathbf{k}) =  s_{0} j_{\rm x} T^{\rm x}_{\theta_{\rm s}}(\mathbf{k}) +N_{\rm sig}(\mathbf{k}) + N_{\rm bk}(\mathbf{k})
.\end{equation}

The matched filter in this case is given by Eqs. \ref{eq:Xray_MF} and \ref{eq:Xray_sigmaMF}. The filtered map in Fourier space is then
\begin{align}\label{eq:ap:filteredmapFT}
	\sum_{\mathbf{k}}^{}\Psi_{\theta_{\rm s}}^\ast(\mathbf{k}) {M}(\mathbf{k}) = & s_{0} j_{\rm x}^2 \sigma_{\theta_{\rm s}}^2 \sum_{\mathbf{k}}^{} \frac{\left| T^{\rm x}_{\theta_{\rm s}}(\mathbf{k}) \right|^2 }{P(\mathbf{k})} + j_{\rm x} \sigma_{\theta_{\rm s}}^2 \sum_{\mathbf{k}}^{} \frac{ T^{\rm x \ast}_{\theta_{\rm s}}(\mathbf{k})  }{P(\mathbf{k})} {N_{\rm sig}}(\mathbf{k}) \nonumber\\
	+ & j_{\rm x} \sigma_{\theta_{\rm s}}^2 \sum_{\mathbf{k}}^{} \frac{ T^{\rm x \ast}_{\theta_{\rm s}}(\mathbf{k})  }{P(\mathbf{k})} N_{\rm bk}(\mathbf{k}),
\end{align}
where the first term on the right-hand side of the equation is equal to $s_{0}$ (the amplitude of the cluster profile), the second term is the filtered Poisson fluctuations on the signal, and the third term is the filtered background noise.

The variance of the filtered background noise can be calculated as follows:
\begin{align}\label{eq:ap:backgr_variance}
	\sigma_{\rm bk}^2 = &\left\langle \left( j_{\rm x}\sigma_{\theta_{\rm s}}^2 \sum_{\mathbf{k}} \frac{ T^{\rm x \ast}_{\theta_{\rm s}}(\mathbf{k})  }{P(\mathbf{k})} N_{\rm bk}(\mathbf{k})\right)   \left( j_{\rm x}\sigma_{\theta_{\rm s}}^2 \sum_{\mathbf{k}'} \frac{ T^{\rm x}_{\theta_{\rm s}}(\mathbf{k}')  }{P(\mathbf{k}')} N_{\rm bk}^\ast(\mathbf{k'})\right)         \right\rangle \nonumber \\
	= & j_{\rm x}^2 \sigma_{\theta_{\rm s}}^4    \sum_{\mathbf{k}} \sum_{\mathbf{k}'} \frac{ T^{\rm x \ast}_{\theta_{\rm s}}(\mathbf{k}) T^{\rm x}_{\theta_{\rm s}}(\mathbf{k}')   }{P(\mathbf{k}) P(\mathbf{k}')}       \left\langle N_{\rm bk}(\mathbf{k}) N_{\rm bk}^\ast(\mathbf{k'})         \right\rangle \nonumber \\
	= & j_{\rm x}^2 \sigma_{\theta_{\rm s}}^4    \sum_{\mathbf{k}} \frac{ \left| T^{\rm x}_{\theta_{\rm s}}(\mathbf{k}) \right|^2 }{P(\mathbf{k}) ^2}        P_{\rm bk}(\mathbf{k}) 
	\approx \sigma_{\theta_{\rm s}}^2,
\end{align}
where we used that $\left\langle N_{\rm bk}(\mathbf{k})N_{\rm bk}^\ast(\mathbf{k}')\right\rangle  = \delta(\mathbf{k}-\mathbf{k}')P_{\rm bk}(\mathbf{k})$ and that $P_{\rm bk}(\mathbf{k}) \approx P(\mathbf{k})$. The validity of this approximation was confirmed empirically by checking that the power spectra of the maps were dominated by the background.

The variance due to the Poisson fluctuations on the signal, after passing through the filter, can be written as
\begin{align}\label{eq:ap:poissonvariance}
	\sigma_{\rm Poisson}^2 = &\left\langle \left( j_{\rm x} \sigma_{\theta_{\rm s}}^2 \sum_{\mathbf{k}} \frac{ T^{\rm x \ast}_{\theta_{\rm s}}(\mathbf{k})  }{P(\mathbf{k})} N_{\rm sig}(\mathbf{k})\right)   \left( j_{\rm x} \sigma_{\theta_{\rm s}}^2 \sum_{\mathbf{k}'} \frac{ T^{\rm x}_{\theta_{\rm s}}(\mathbf{k}')  }{P(\mathbf{k}')} N_{\rm sig}^\ast(\mathbf{k'})\right)         \right\rangle \nonumber \\
	= & j_{\rm x}^2 \sigma_{\theta_{\rm s}}^4    \sum_{\mathbf{k}} \sum_{\mathbf{k}'} \frac{T^{\rm x \ast}_{\theta_{\rm s}}(\mathbf{k})T^{\rm x}_{\theta_{\rm s}}(\mathbf{k}')}{P(\mathbf{k}) P(\mathbf{k}')} \left\langle N_{\rm sig}(\mathbf{k}) N_{\rm sig}^\ast(\mathbf{k'})         \right\rangle \nonumber \\
	= & j_{\rm x}^2 \frac{u s_{0} j_{\rm x} \sigma_{\theta_{\rm s}}^4}{n^2}\sum_{\mathbf{k}}\sum_{\mathbf{k}'}\frac{T^{\rm x \ast}_{\theta_{\rm s}}(\mathbf{k})T^{\rm x}_{\theta_{\rm s}}(\mathbf{k}')}{P(\mathbf{k}) P(\mathbf{k}')} T^{\rm x}_{\theta_{\rm s}}\left( \mathbf{k}-\mathbf{k}'\right) 
\end{align}
where the last equality comes from
\begin{align*}
	& \left\langle N_{\rm sig}(\mathbf{k}) N_{\rm sig}^\ast(\mathbf{k}')\right\rangle =\\
	& =  \frac{1}{n^2}\frac{1}{n^2}\sum_{\mathbf{x}} \sum_{\mathbf{x}'} \left\langle N_{\rm sig}(\mathbf{x}) N_{\rm sig}(\mathbf{x}')\right\rangle e^{-j 2\pi \frac{1}{n} \left( \mathbf{k}^{\rm T} \mathbf{x} - \mathbf{k}'^{\rm T}\mathbf{x}'\right) }\\
	& = \frac{1}{n^2}\frac{1}{n^2}\sum_{\mathbf{x}} \left\langle N_{\rm sig}(\mathbf{x})^2\right\rangle e^{-j 2\pi \frac{1}{n} (\mathbf{k}-\mathbf{k}')^{\rm T} \mathbf{x} }\\
	& = \frac{1}{n^2}\frac{1}{n^2}\sum_{\mathbf{x}} u^2 a T^{\rm x}_{\theta_{\rm s}}(\mathbf{x}) e^{-j 2\pi \frac{1}{n}   (\mathbf{k}-\mathbf{k}')^{\rm T} \mathbf{x} }\\
	& = \frac{1}{n^2} u^2 a T^{\rm x}_{\theta_{\rm s}}(\mathbf{k}-\mathbf{k}') 
	= \frac{1}{n^2} u s_0 j_{\rm x} T^{\rm x}_{\theta_{\rm s}}(\mathbf{k}-\mathbf{k}') ,
\end{align*}
where we used that $\left\langle N_{\rm sig}(\mathbf{x})N_{\rm sig}(\mathbf{x}')\right\rangle  = \delta(\mathbf{x}-\mathbf{x}')\left\langle N_{\rm sig}(\mathbf{x})\right\rangle ^2$, Eq. \ref{eq:Nsig_variance}, and the definition of the 2D Fourier transform,
\begin{equation}
N(\mathbf{k}) = \frac{1}{n^2}\sum_{\mathbf{x}}N(\mathbf{x})e^{-j2\pi \frac{1}{n}   \mathbf{k}^{\rm T} \mathbf{x}},
\end{equation}
where the sum is over all the pixels in the map.

The double sum in Eq. \ref{eq:ap:poissonvariance} can be reduced to a single sum and calculated very efficiently by expressing it as a cross-correlation of two signals and making use of the cross-correlation theorem, as we explain next. 

Let us call $Q(\mathbf{k}) = { T^{\rm x \ast}_{\theta_{\rm s}}(\mathbf{k})  /P(\mathbf{k})}$,  $R(\mathbf{k}') = { T^{\rm x}_{\theta_{\rm s}}(\mathbf{k}')  /P(\mathbf{k}')}$ and $\mathbf{m} = \mathbf{k} - \mathbf{k}'$. Then we can express the double sum in Eq. \ref{eq:ap:poissonvariance} as
\begin{equation} 
	\sum_{\mathbf{k}}\sum_{\mathbf{k}'}\frac{T^{\rm x \ast}_{\theta_{\rm s}}(\mathbf{k})T^{\rm x}_{\theta_{\rm s}}(\mathbf{k}')}{P(\mathbf{k}) P(\mathbf{k}')} T^{\rm x}_{\theta_{\rm s}}( \mathbf{k}-\mathbf{k}') = \sum_{\mathbf{m}} T^{\rm x}_{\theta_{\rm s}}( \mathbf{m}) \sum_{\mathbf{k}}Q(\mathbf{k})R(\mathbf{k-m}).
\end{equation}
The second sum in the right-hand side of this equation is the cross-correlation of $Q^\ast$ and $R$, evaluated at $-\mathbf{m}$, that is,
\begin{equation} 
\sum_{\mathbf{k}}Q(\mathbf{k})R(\mathbf{k-m}) = (R \star Q^\ast)(\mathbf{-m}) = (R \star Q^\ast)^{\rm T}(\mathbf{m}),
\end{equation}
which can be computed very efficiently by making use of the cross-correlation theorem:
\begin{equation} 
(R \star Q^\ast) = \mathcal{F}^{-1}\left( (\mathcal{F}(R))^\ast \mathcal{F}(Q^\ast)\right),
\end{equation}
where $\mathcal{F}$ denotes the Fourier transform. Thus, the double sum in Eq. \ref{eq:ap:poissonvariance} can be calculated very quickly using the following expression: 
\begin{align*} 
& \sum_{\mathbf{k}}\sum_{\mathbf{k}'}\frac{T^{\rm x \ast}_{\theta_{\rm s}}(\mathbf{k})T^{\rm x}_{\theta_{\rm s}}(\mathbf{k}')}{P(\mathbf{k}) P(\mathbf{k}')} T^{\rm x}_{\theta_{\rm s}}( \mathbf{k}-\mathbf{k}') \\ 
& = n^2 \sum_{\mathbf{m}} T^{\rm x}_{\theta_{\rm s}}( \mathbf{m})  \left[ \mathcal{F}^{-1}\left( (\mathcal{F}(R))^\ast \mathcal{F}(Q^\ast)\right)\right]^{\rm T} 
\end{align*}

\section{Generation of an X-ray all-sky HEALPix map}\label{app:HealpixMap}

In this section we describe how we constructed our all-sky HEALPix map using the 1378 individual RASS fields. Each RASS field covers an area of 6.4 $\deg$ x 6.4 $\deg$ (512 x 512 pixels), has a resolution of 0.75 arcmin/pixel and gives information on the number of counts in the total (0.1-2.4 keV), hard (0.5-2.0 keV) and soft (0.1-0.4 keV) bands, and the corresponding exposure time. We used the hard band information and the exposure time to build our X-ray all-sky map.

The procedure to create the all-sky map consisted of assigning each RASS pixel to the nearest HEALPix pixel (to be more precise, the sky coordinates of the center of each RASS pixel were calculated and the HEALPix pixels that contained those sky positions were assigned to them). In doing so, we came across two main difficulties: the different resolution of HEALPix maps and RASS fields, and the overlapping of contiguous RASS fields.

The Hierarchical Equal Area isoLatitude Pixelization (HEALPix) scheme \citep{Gorski2005} is based on a division of the sphere into 12 large pixels that are further subdivided dyadically at the desired resolution, yielding maps of $ N_{\rm pix} = 12 \times 2^{2r} $ pixels at resolution $r$, with a pixel size of $ d_{\rm pix} = \sqrt{4\pi /N_{\rm pix}} = 2^{-r}\sqrt{\pi/3}$. For example, at \textit{Planck} resolution ($r=11$), we have 1.72 arcmin/pixel, while if we increase the resolution ($r=12$), the pixel size is divided by two: 0.86 arcmin/pixel. The resolution of the RASS fields is 0.75 arcmin/pixel, which means that it does not coincide with any of the possible HEALPix resolutions. To construct our all-sky map, we decided to work with the HEALPix resolution closest to the RASS resolution, that is, 0.86 arcmin/pixel. Therefore, we have 1.3 RASS pixels on average per each HEALPix pixel. To conserve the total flux of the map, we constructed the all-sky map by summing for each HEALPix pixel the count rate (counts/second) of the corresponding RASS pixels.

A second difficulty is the fact that RASS fields overlap at least 0.23 degrees with contiguous fields. The overlapping region of two fields contains the same information, but with a different projection (so that overlapping RASS pixels do not exactly match).  
We decided to deal with these regions by considering only one of the fields and discarding the information from the other fields. 
It is worth mentioning that this choice implies that we may lose a small fraction of the flux in the borders of the overlapping regions, since the RASS pixels assigned to a HEALPix pixel in one of these borders may not cover all that HEALPix pixel area. 

At this point we have a HEALPix X-ray count rate map ready to be used in the X-ray matched filter. To use this map together with the \textit{Planck} multifrequency maps to jointly extract clusters using the joint X-ray-SZ MMF, we need to express all the maps in the same units. Therefore, the last step in the map construction consists of converting the X-ray map from counts/second to the equivalent $\Delta T/T_{\rm CMB}$ units. To do so, we need to choose a reference redshift ($ z_{\rm ref} $), a reference frequency ($ \nu_{\rm ref} $) and a reference temperature ($ T_{\rm ref} $), as explained below, and apply the following steps:

\begin{enumerate}

	\item Convert countrate into X-ray flux: We convert counts/s (in the 0.5-2 keV band) into  erg/s/cm$ ^{2} $ (in the 0.1-2.4 keV band) using an all-sky $ N_{\rm H} $ map\footnote{We built a HEALPix $ N_{\rm H} $ map from the $ N_{\rm H} $ map of the Leiden/Argentine/Bonn (LAB) survey \citep{Kalberla2005}.  
		The HEALPix $ N_{\rm H} $ map was created at a low resolution, enough for the $ N_{\rm H} $ map (r=8, $ n_{\rm side} $=256), and then up-sampled to the necessary resolution. The $ N_{\rm H} $ value at each HEALPix pixel was calculated by taking the average of the original $ N_{\rm H} $ data within a radius of 1 degree of the considered HEALPix pixel.
	}
	and interpolating in a conversion table, calculated for a temperature of $ T_{\rm ref}=7$ keV and a redshift of $z=0.3$ using a \textsc{mekal} model and the PSPC-C response file.  
	
	\item Convert X-ray flux into equivalent $ Y_{500} $ integrated flux: We convert the X-ray flux in the 0.1-2.4 keV band (in erg/s/cm$ ^{2} $) into $ Y_{500} $ flux (in arcmin$ ^{2} $), using the expected $ F_{\rm X}/Y_{500} $ relation.  
	It is important to note that this expression depends on the redshift: we express our X-ray flux in the equivalent $ Y_{500} $ flux that we expect for clusters situated at a given redshift. Clearly, the redshifts of the clusters we are going to detect are not known in advance, so that to apply this expression we need to assume a reference redshift $ z_{\rm ref} $. The choice of the reference redshift will have an effect, as discussed in Sect. \ref{sec:discussion}. 
	
	\item Convert from integrated $ Y_{500} $ flux (in arcmin$ ^{2} $) into $y$ flux (adimensional), dividing the flux of each pixel by the HEALPix pixel area.
	
	\item Convert from $y$ units into $ \Delta T/T_{\rm CMB} $ units, using the following expression:
	\begin{equation}\label{eq:yDeltaTTconversion}
	\frac{data \left[  \Delta T/T \mbox{units} \right]}{data \left[ y \mbox{ units} \right]} = x \frac{e^{x}+1}{e^{x}-1}-4,
	\end{equation}      
	where  $x=(h \nu )/(kT_{\rm CMB} )$. This step creates a mono-frequency SZ map corresponding to the calculated flux. To apply this conversion we need to assume a reference frequency $\nu_{\rm ref}$ for the map, which is just a fiducial value, with no effect on the extraction algorithm. In our case, we took $\nu_{\rm ref}$=1000 GHz. 
	
\end{enumerate}

\section{Additional results on the gain of the X-ray-SZ MMF}\label{app:SNRgain}

\begin{figure}[]
	\centering
	\subfigure[]{\includegraphics[width=.99\columnwidth]{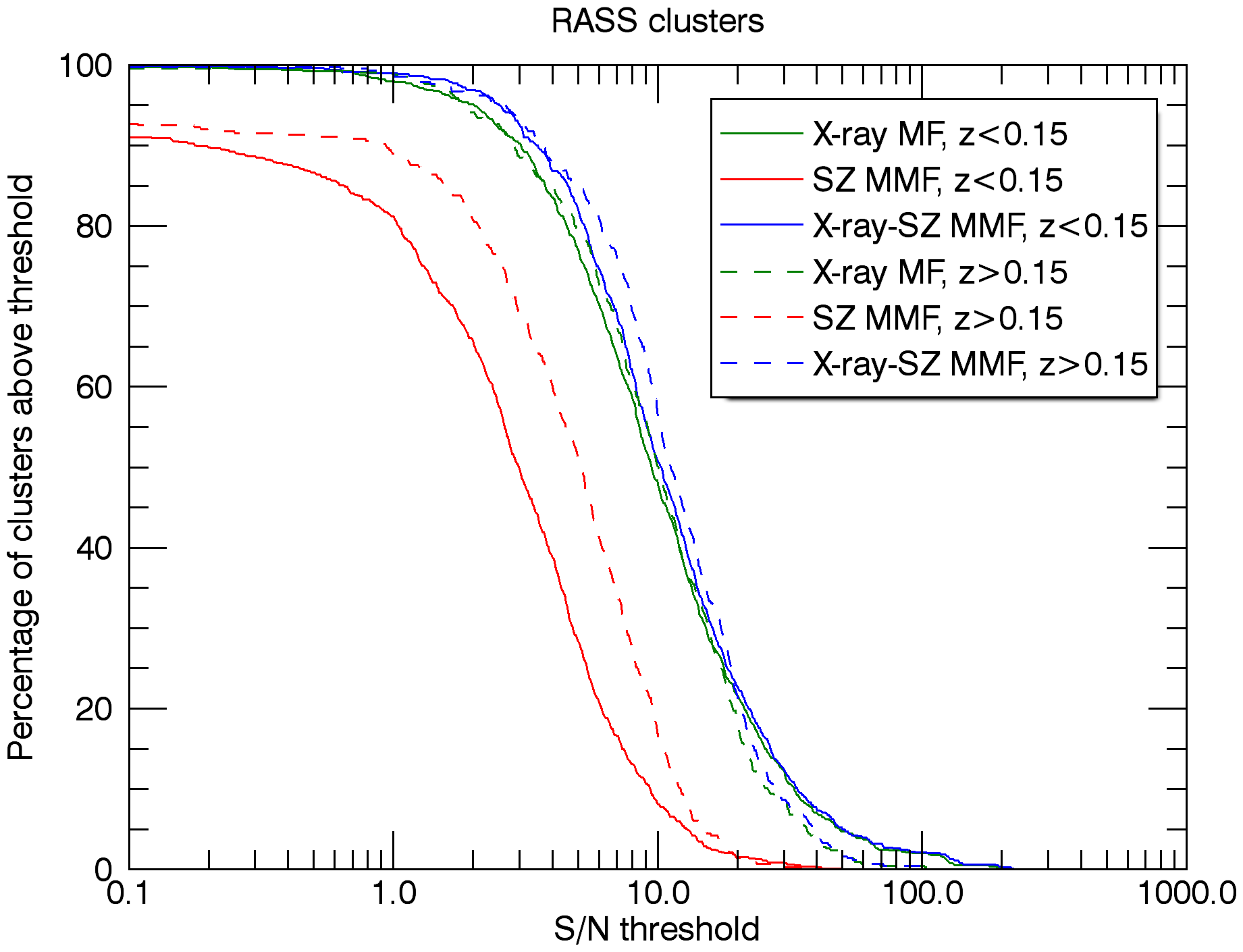}\label{fig:detections_vs_snr_realrass_rpj_2z}}
	\subfigure[]{\includegraphics[width=.99\columnwidth]{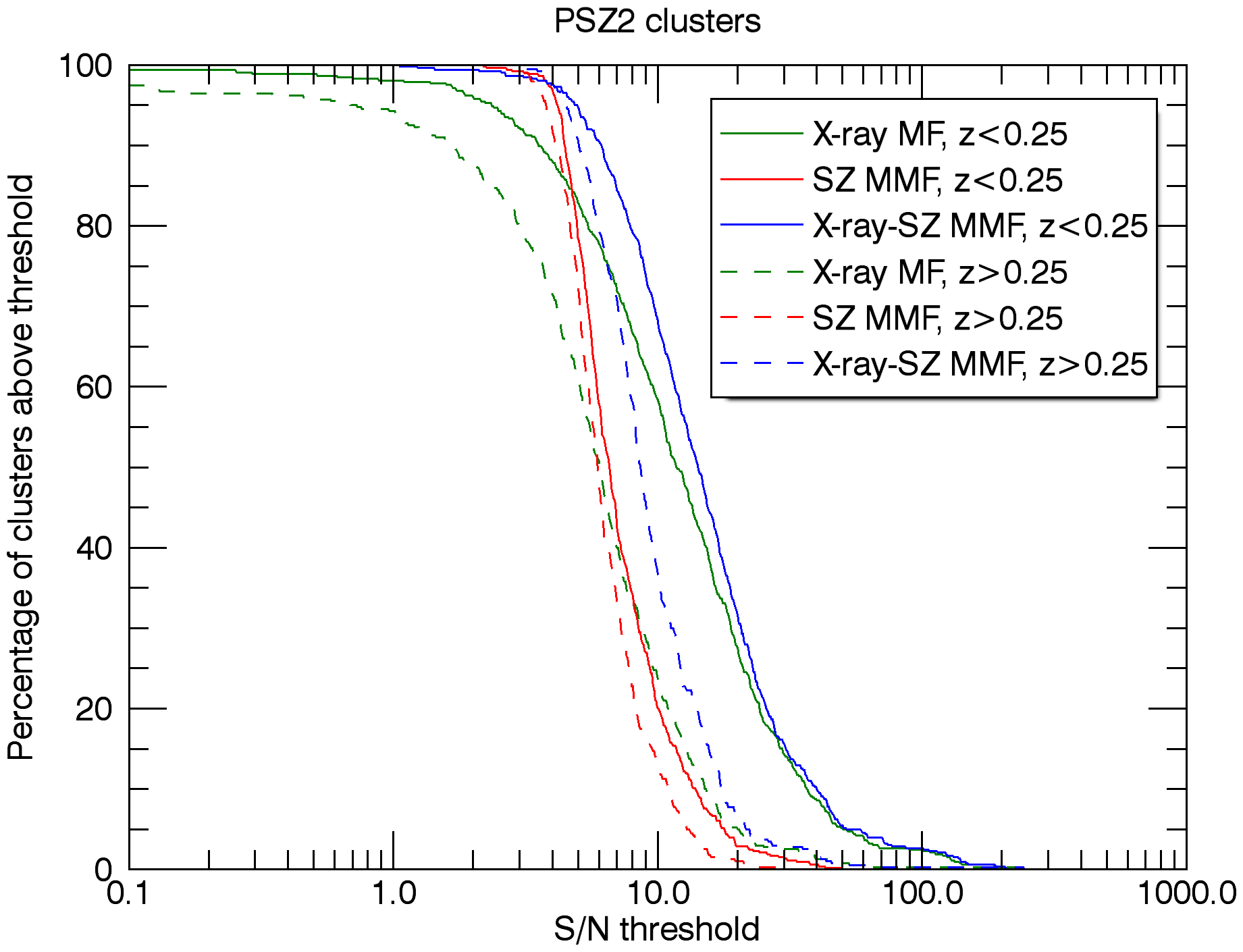}\label{fig:detections_vs_snr_realmmf3_rpj_2z}}
	\caption{Percentage of RASS (top panel) and PSZ2 (bottom panel) clusters extracted with the proposed X-ray matched filter (green), the classical SZ MMF (red) and the proposed X-ray-SZ MMF (blue) whose S/N is above a certain S/N threshold, in two different redshift bins.}
	\label{fig:detections_vs_snr_realrassmmf3_rpj_2z}
\end{figure}

\begin{figure*}[]
	\centering
	\subfigure[]{\includegraphics[width=.74\columnwidth]{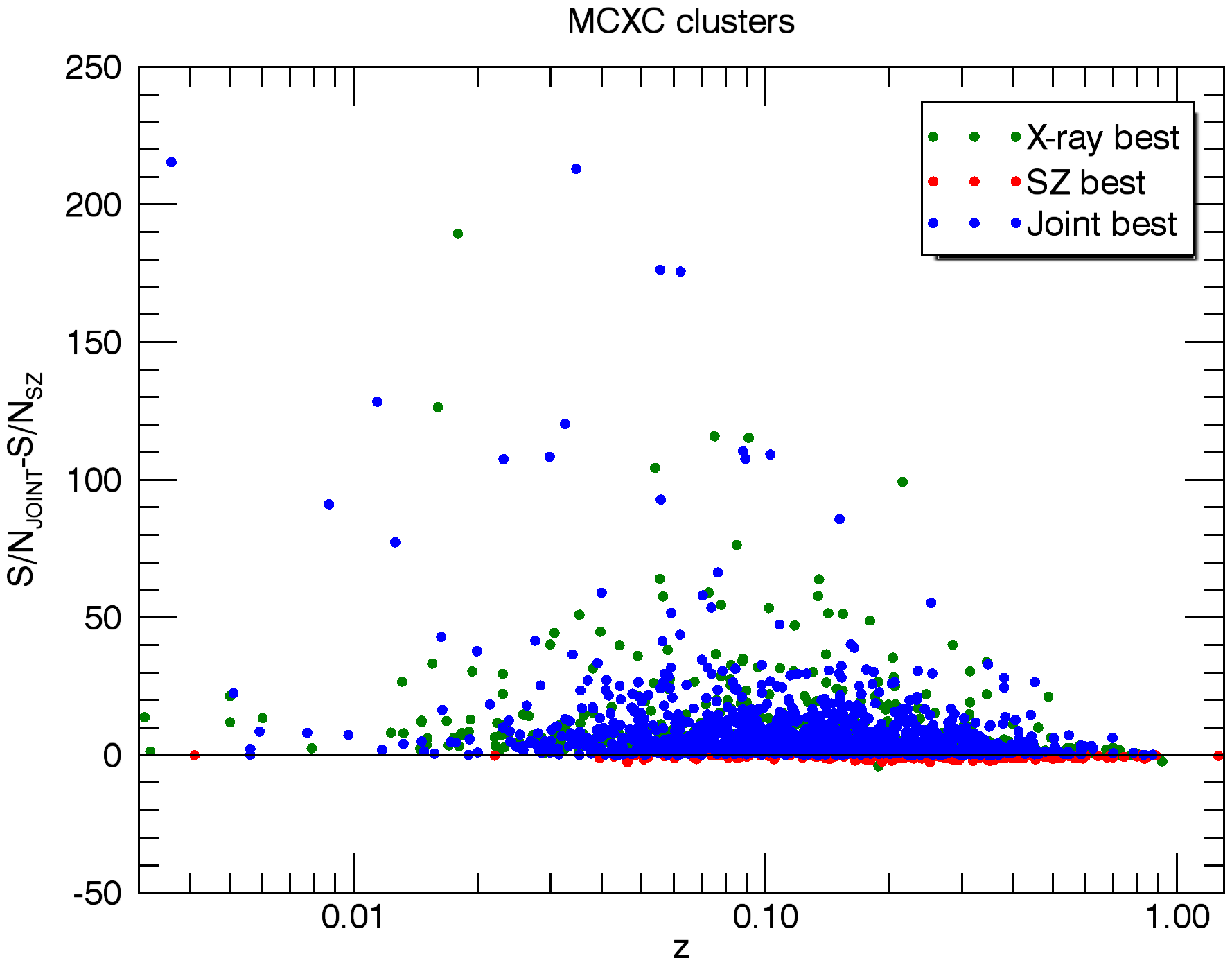}\label{fig:realpxcc_snrgainsz_vs_z}}
	\subfigure[]{\includegraphics[width=.74\columnwidth]{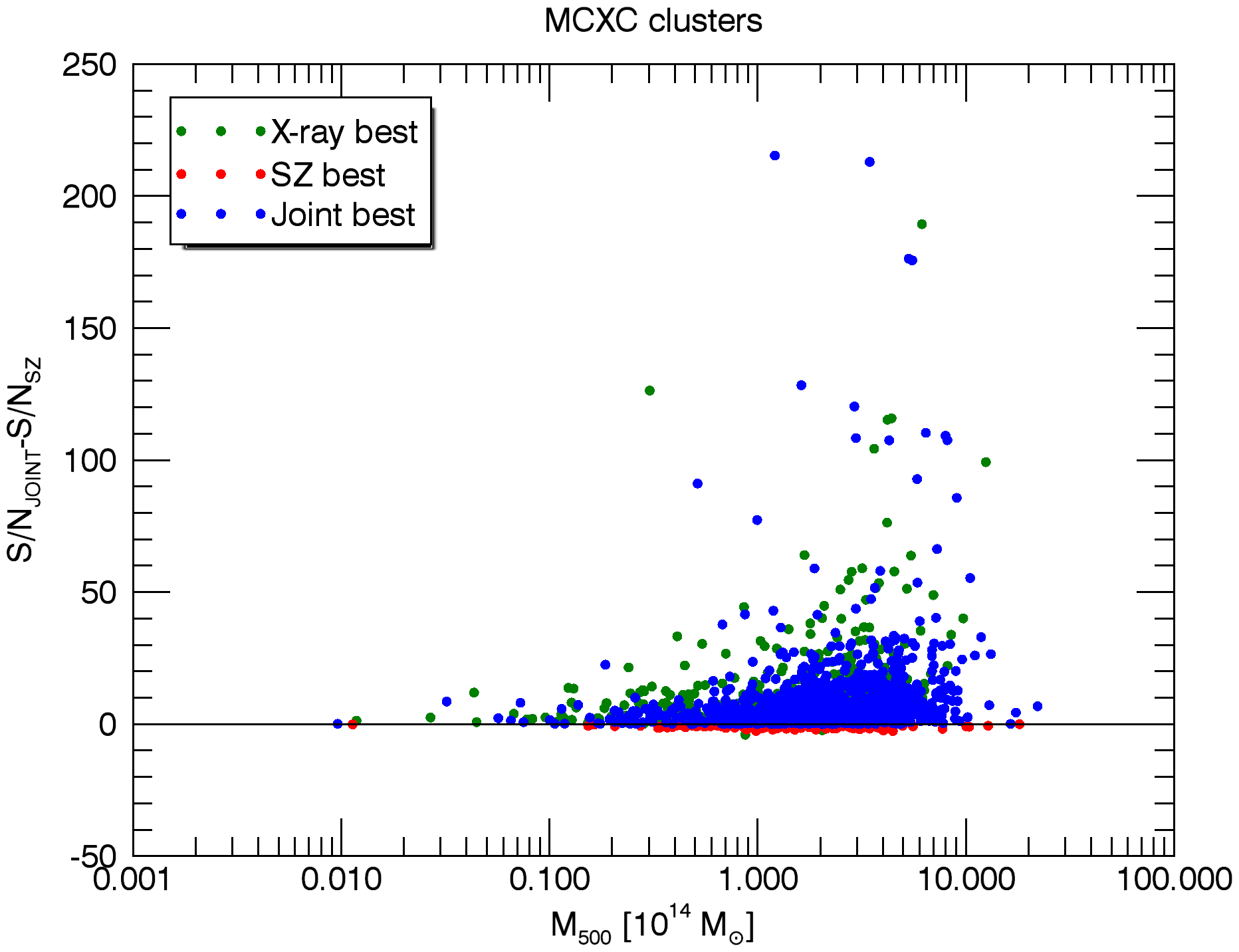}\label{fig:realpxcc_snrgainsz_vs_l500}}
	\subfigure[]{\includegraphics[width=.74\columnwidth]{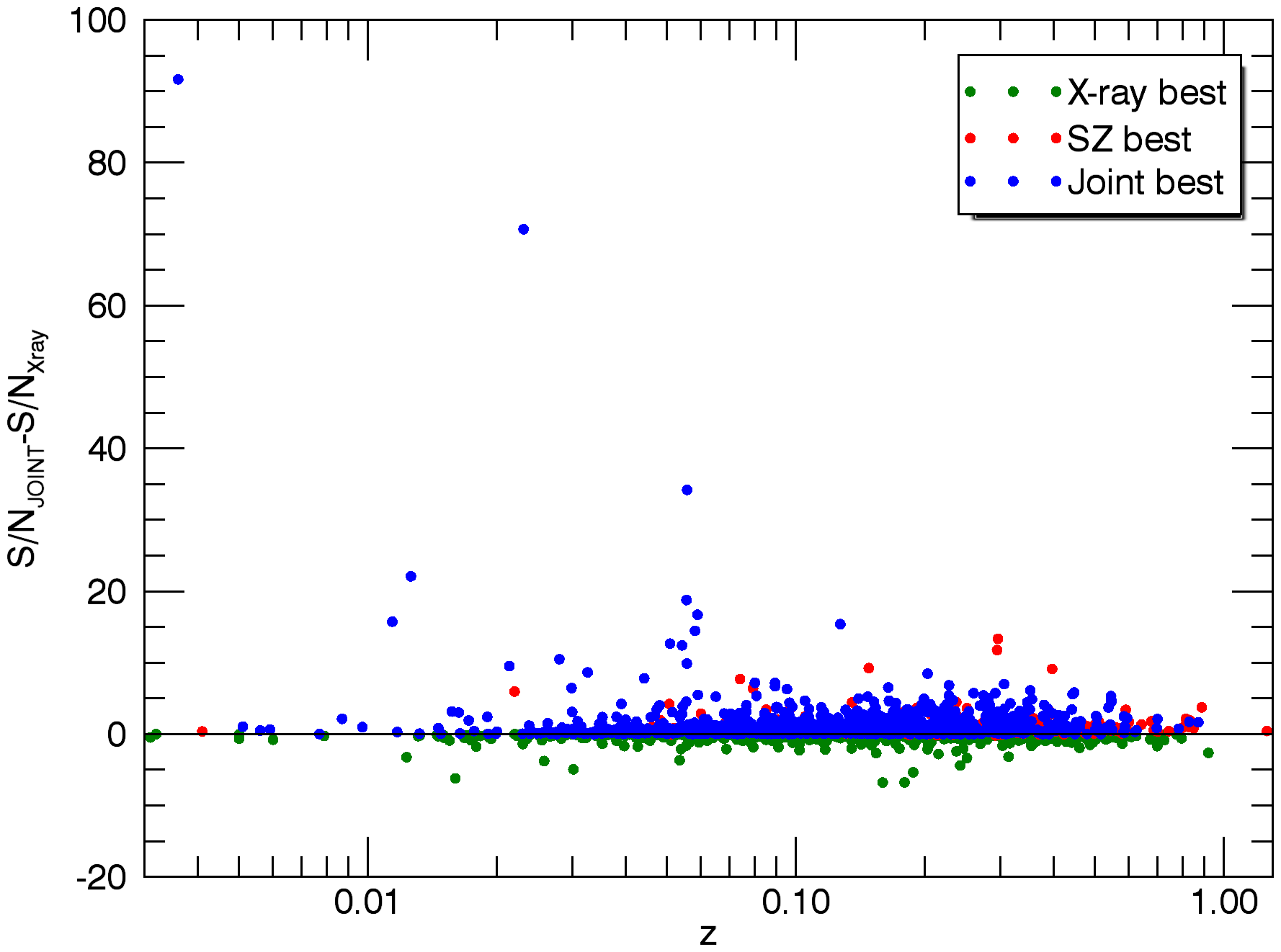}\label{fig:realpxcc_snrgainxray_vs_z}}
	\subfigure[]{\includegraphics[width=.74\columnwidth]{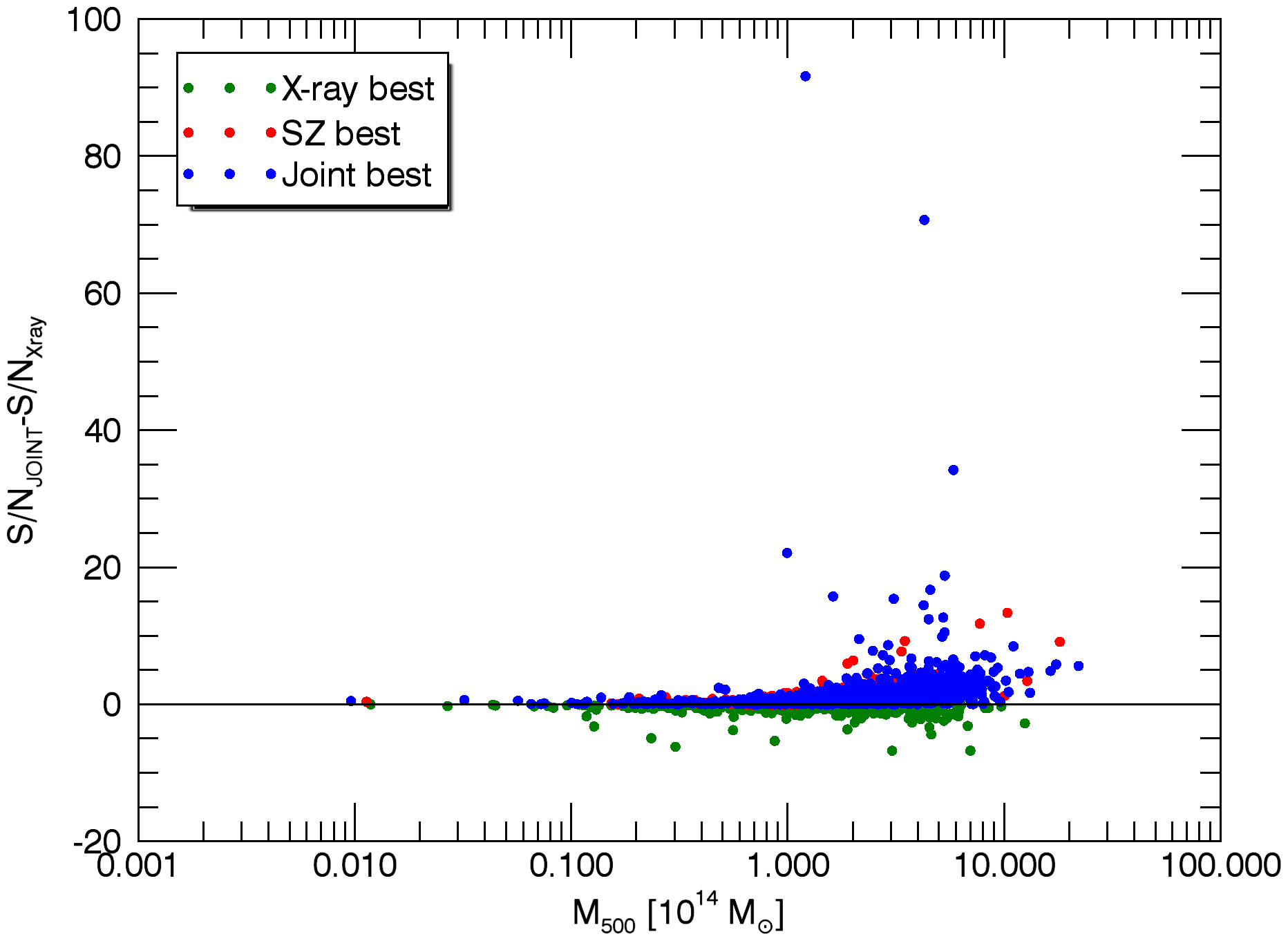}\label{fig:realpxcc_snrgainxray_vs_l500}}
		\subfigure[]{\includegraphics[width=.74\columnwidth]{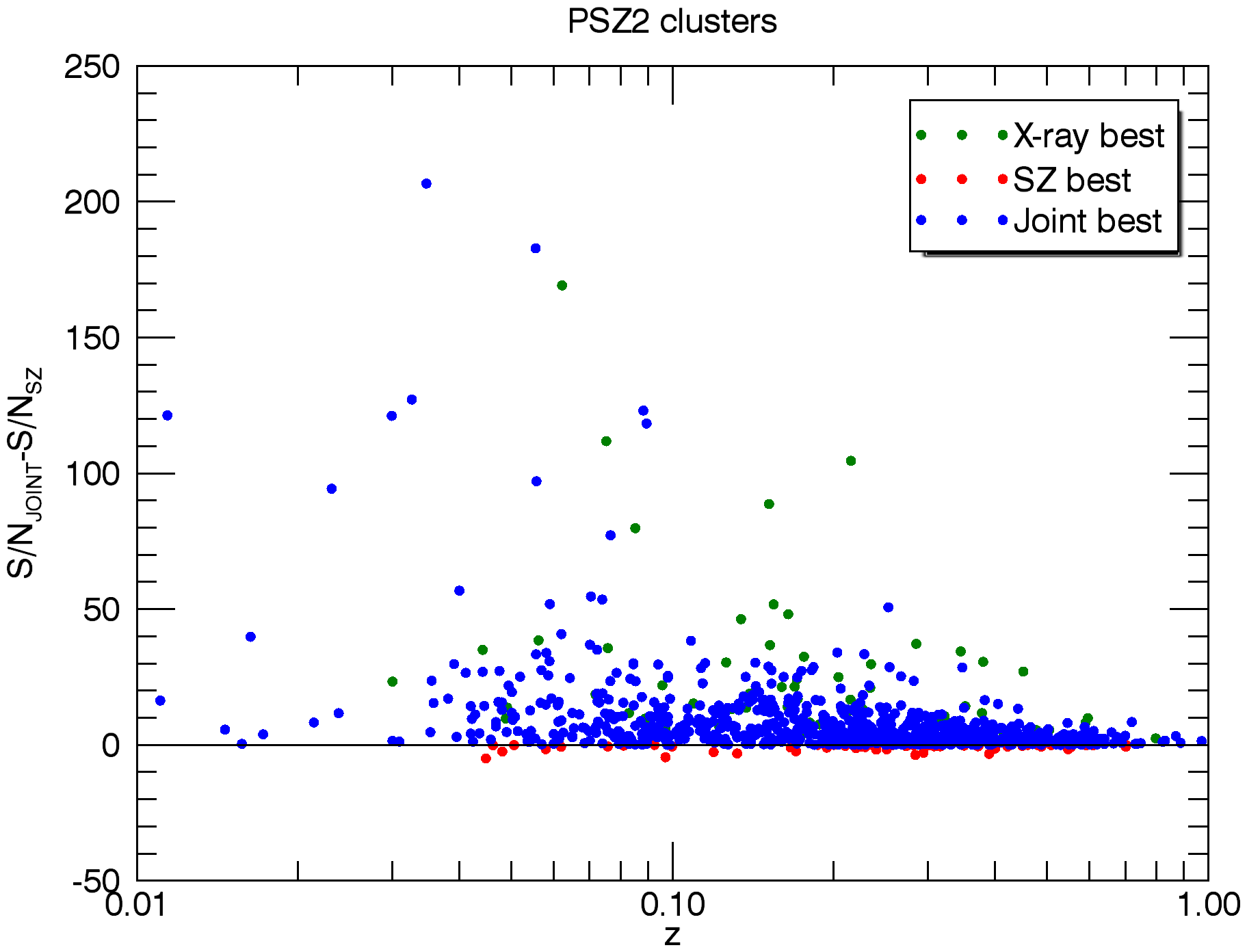}\label{fig:realmmf3c_snrgainsz_vs_z}}
		\subfigure[]{\includegraphics[width=.74\columnwidth]{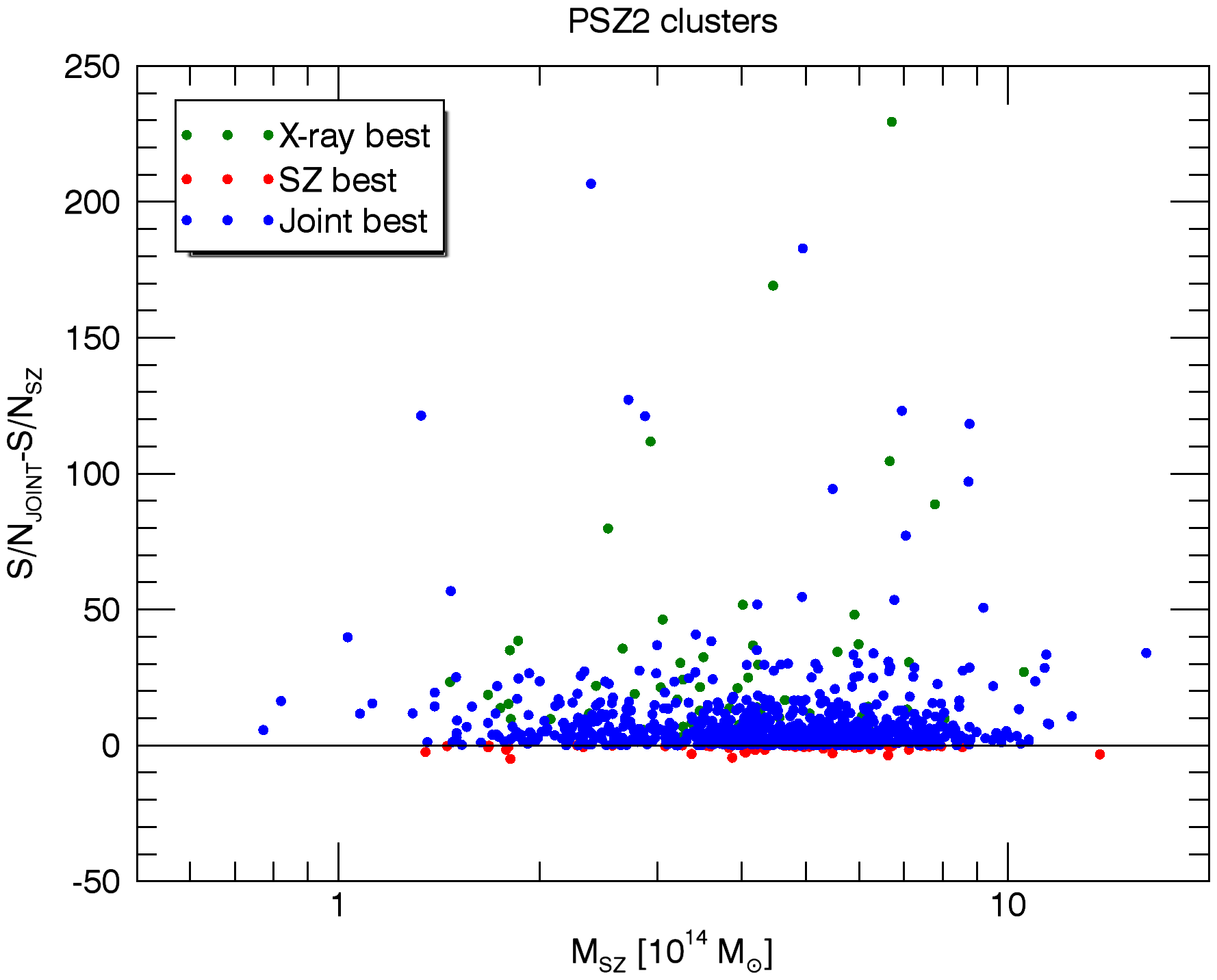}\label{fig:realmmf3_snrgainsz_vs_l500}}
		\subfigure[]{\includegraphics[width=.74\columnwidth]{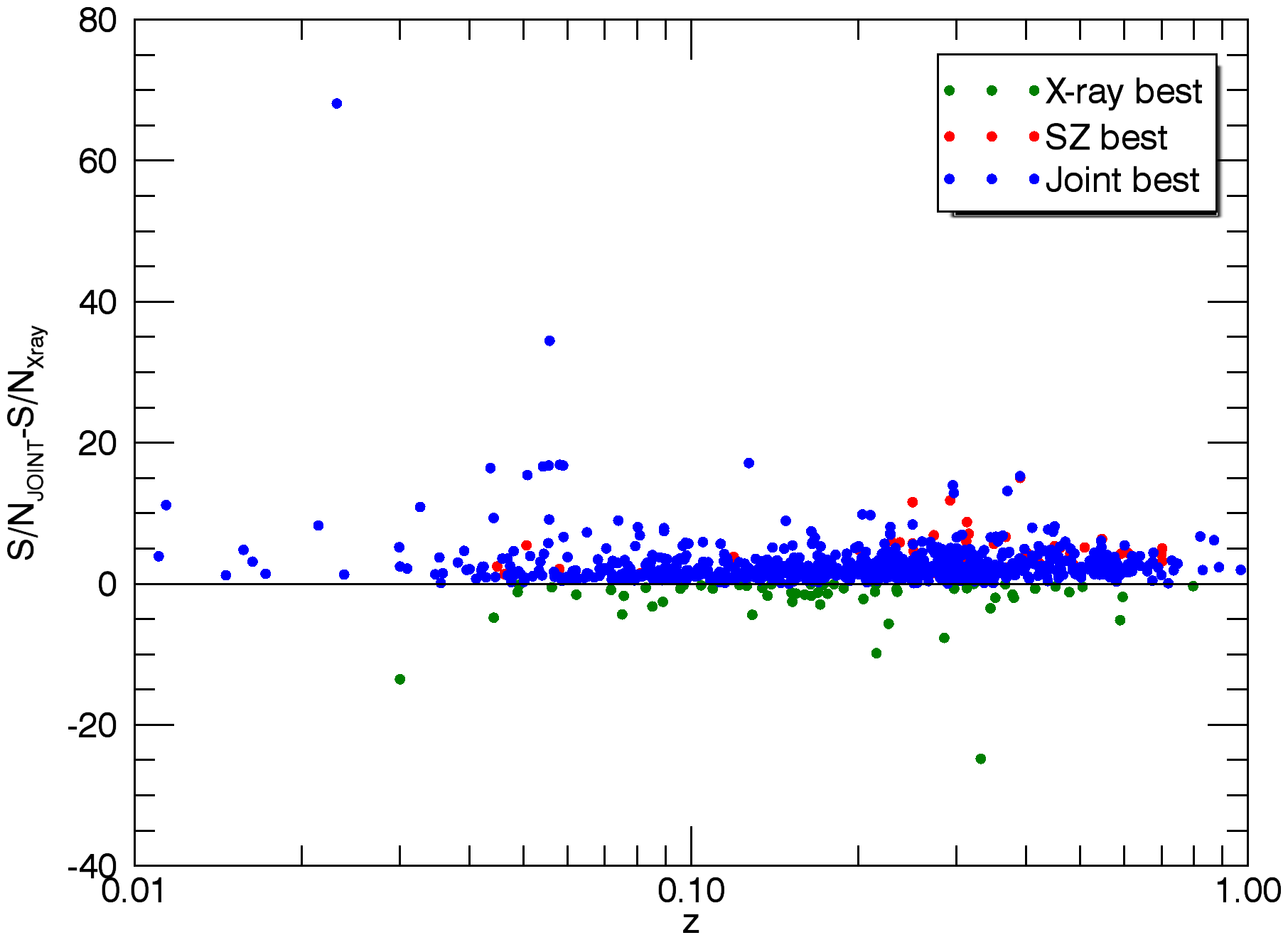}\label{fig:realmmf3_snrgainxray_vs_z}}
		\subfigure[]{\includegraphics[width=.74\columnwidth]{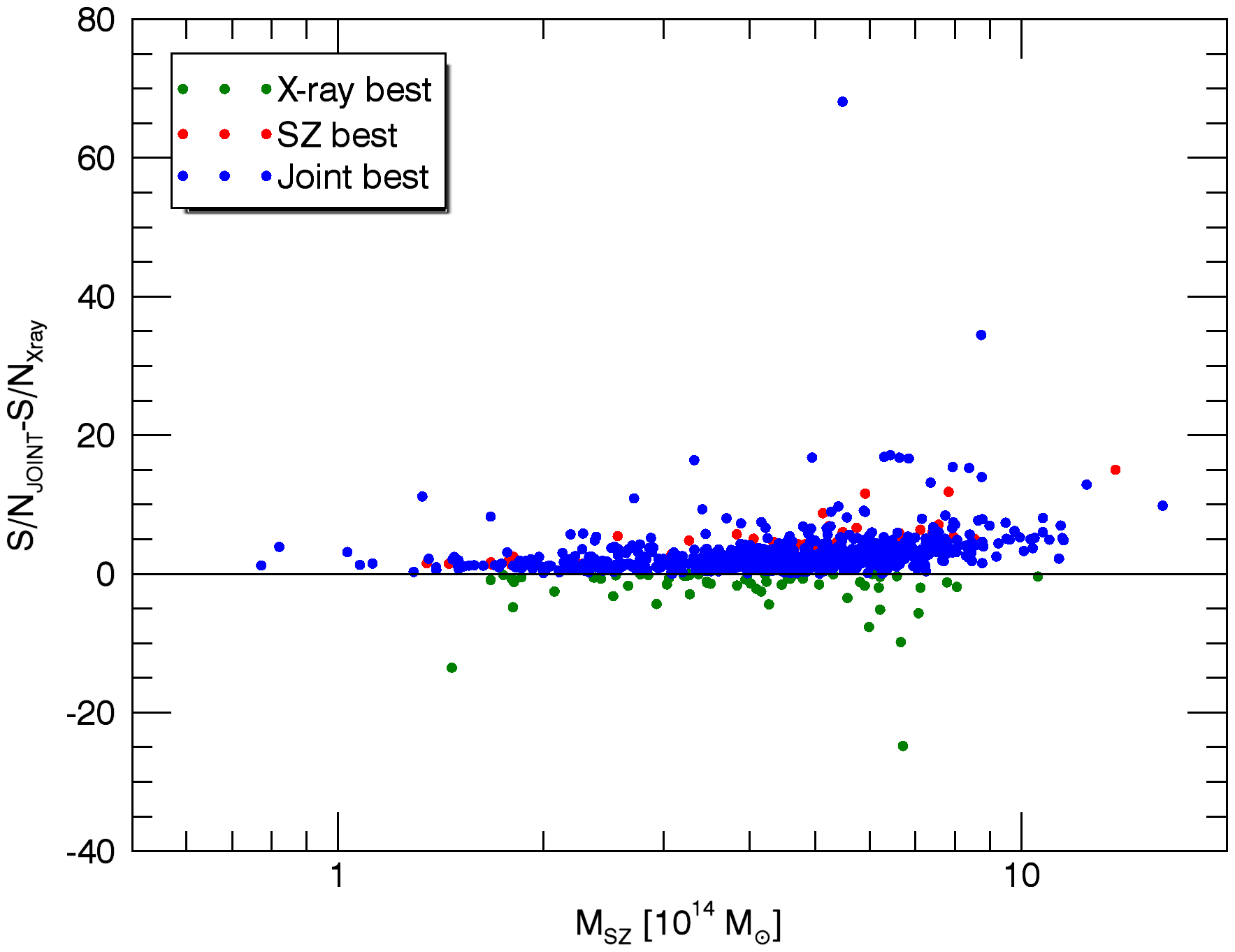}\label{fig:realmmf3_snrgainxray_vs_l500}}
	\caption{S/N gain of the joint extraction of the MCXC clusters (panels a-d) and the PSZ2 clusters (panels e-h) with respect to the individual SZ-only and X-ray-only extractions. The upper panels show the difference between the S/N of the joint extraction and the S/N of the SZ-only extraction, as a function of redshift (a) and mass (b), for the MCXC clusters. The panels in the second row show the difference between the S/N of the joint extraction and the S/N of the X-ray-only extraction, as a function of redshift (c) and mass (d), for the MCXC clusters. The four bottom panels are analogous to panels a-d, but for the PSZ2 clusters. In all the cases, the S/N is defined as the estimated flux divided by the estimated background noise. Individual measurements are color-coded according to the best extraction.}
	\label{fig:test2}
\end{figure*}

\begin{figure*}[]
	\centering
	\subfigure[]{\includegraphics[width=.74\columnwidth]{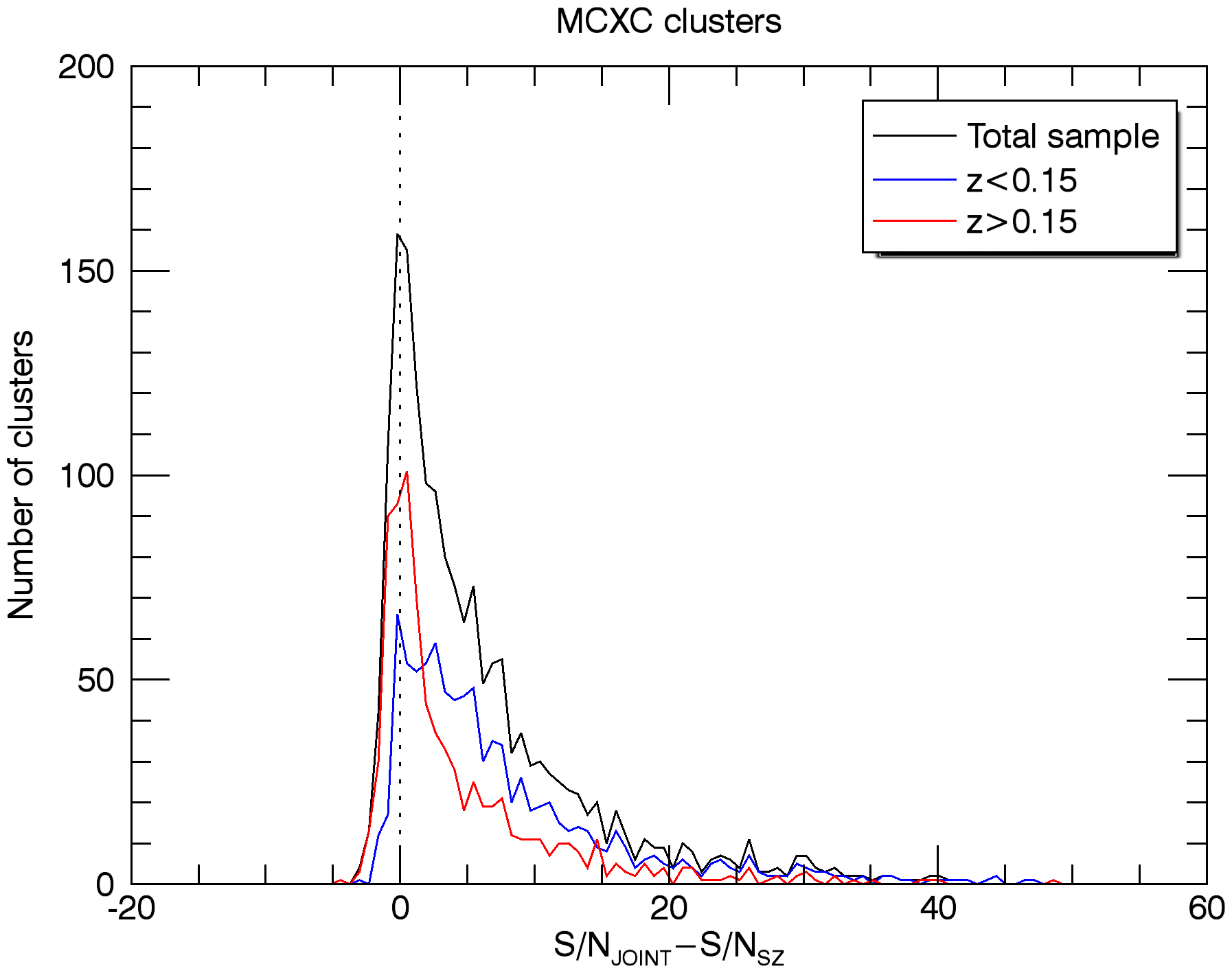}\label{fig:hist_snrgainsz_z_realmcxc}}
	\subfigure[]{\includegraphics[width=.74\columnwidth]{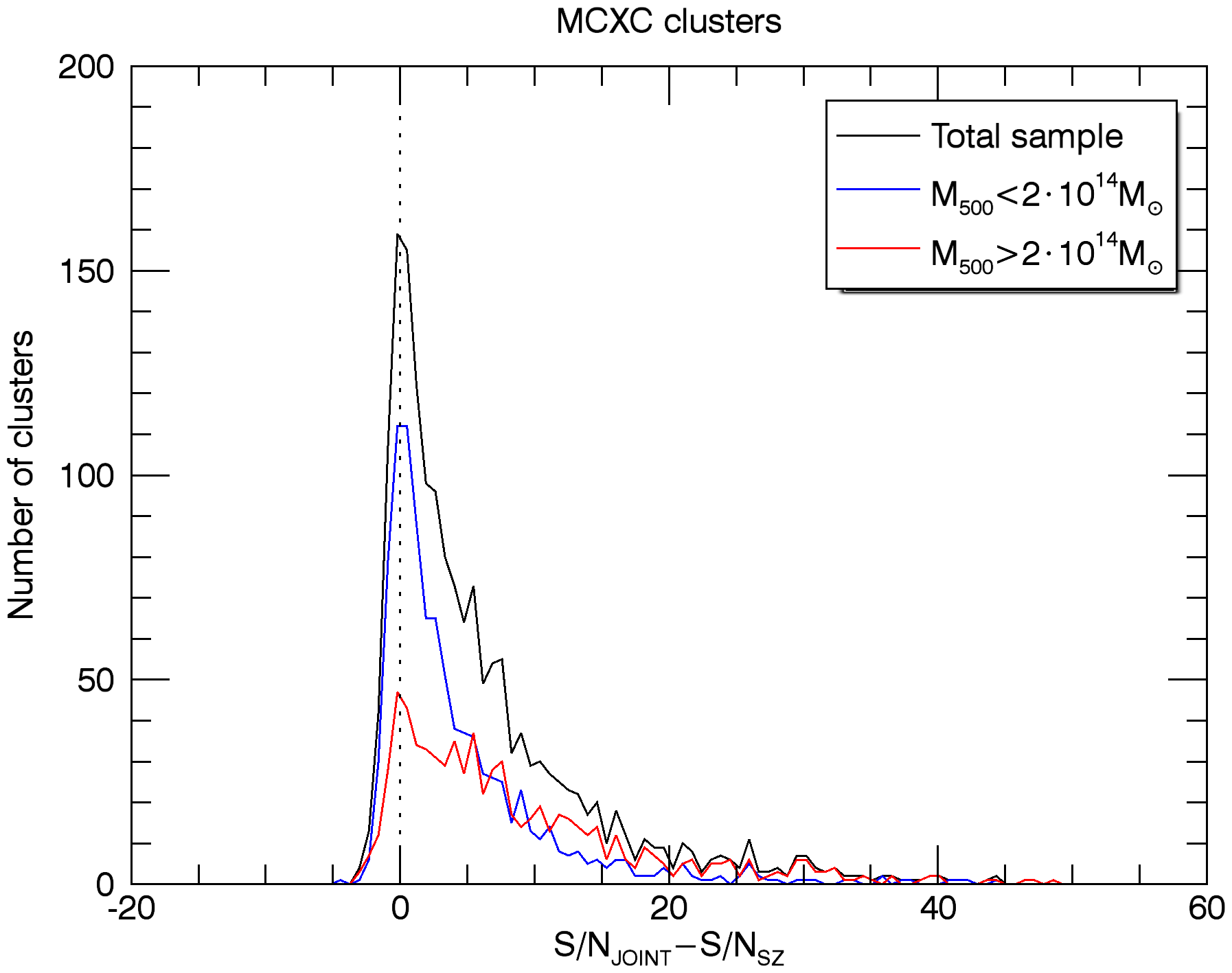}\label{fig:hist_snrgainsz_L_realmcxc}}
	\subfigure[]{\includegraphics[width=.74\columnwidth]{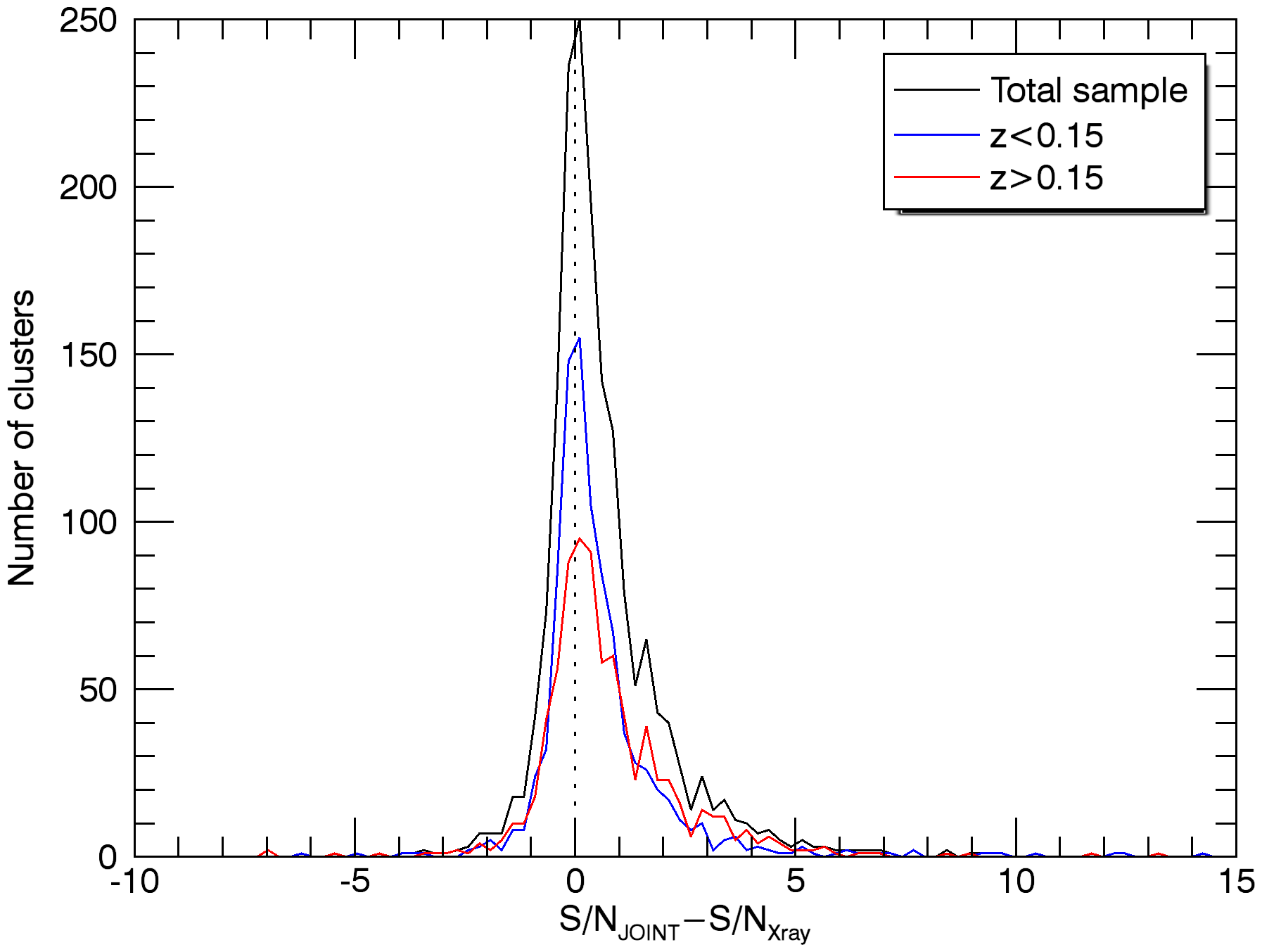}\label{fig:hist_snrgainxray_z_realmcxc}}
	\subfigure[]{\includegraphics[width=.74\columnwidth]{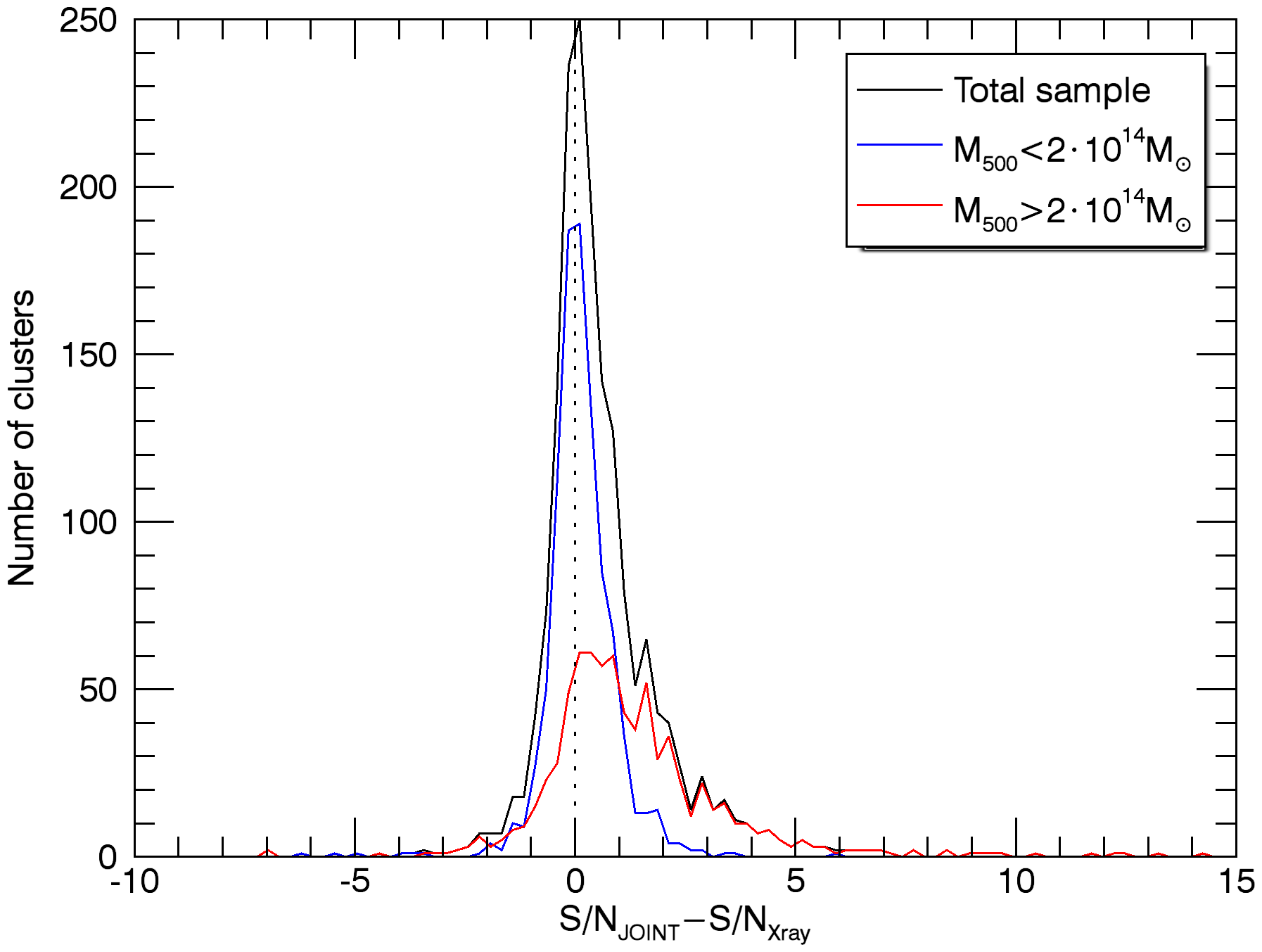}\label{fig:hist_snrgainxray_L_realmcxc}}
	\subfigure[]{\includegraphics[width=.74\columnwidth]{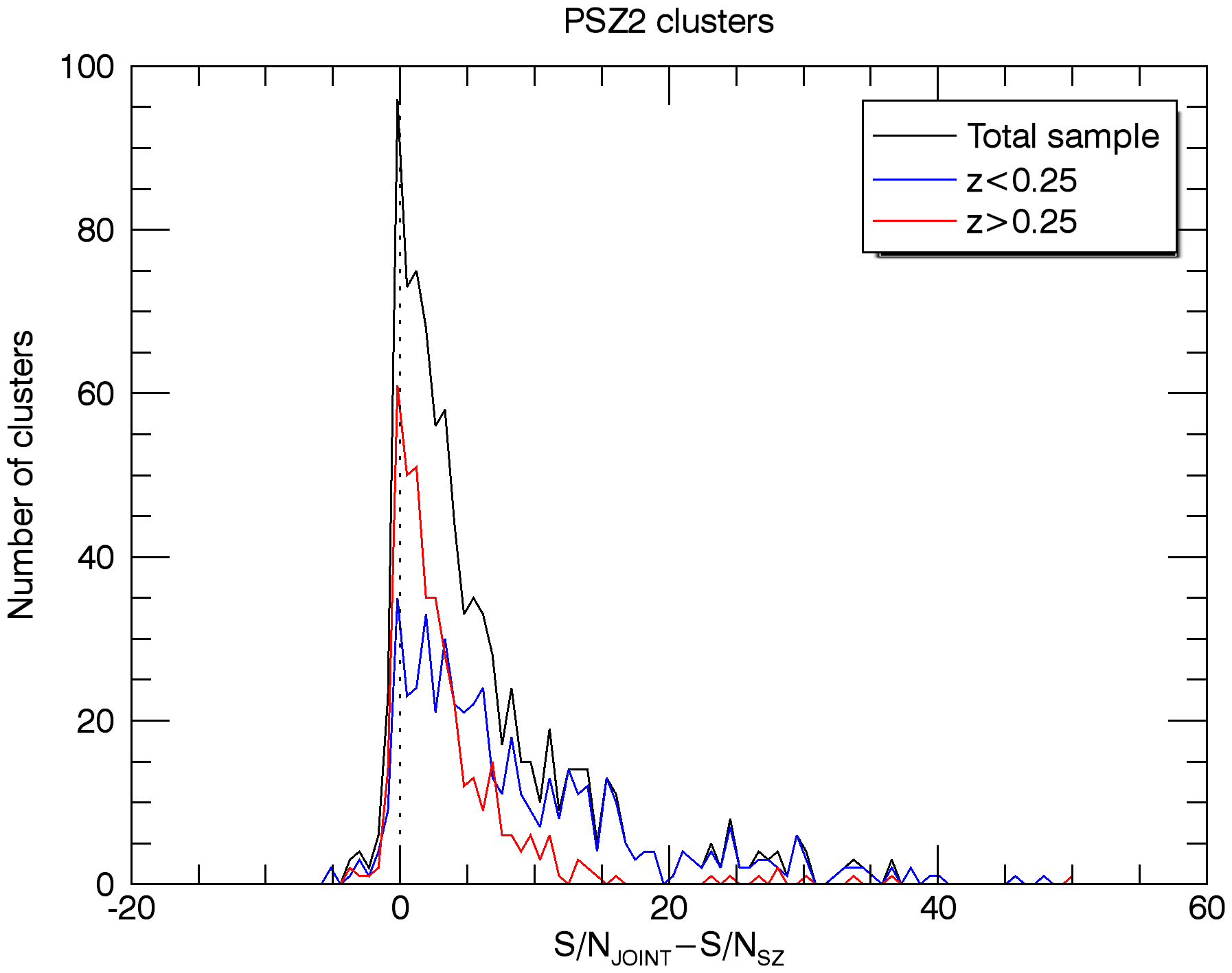}\label{fig:hist_snrgainsz_z_realmmf3}}
	\subfigure[]{\includegraphics[width=.74\columnwidth]{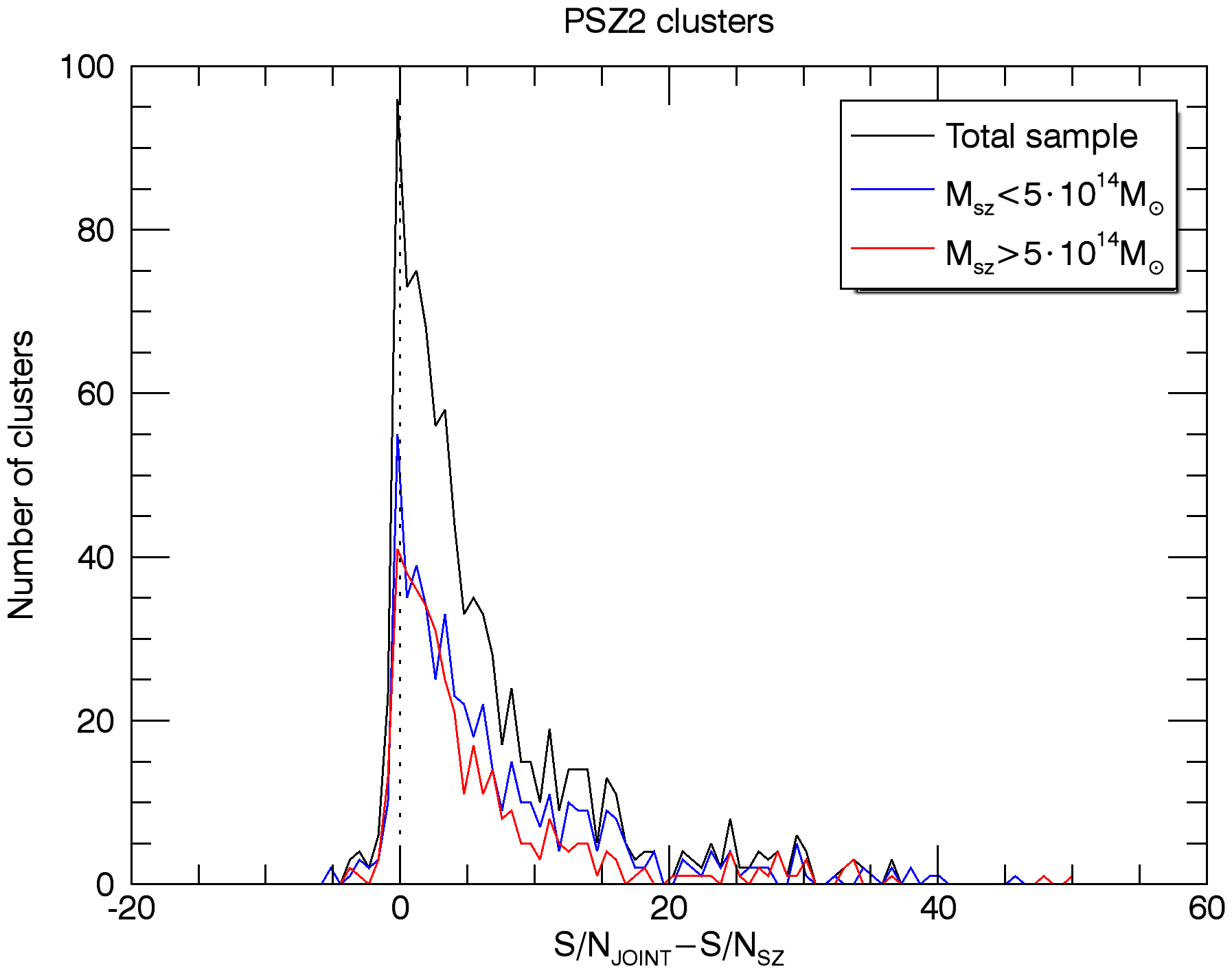}\label{fig:hist_snrgainsz_L_realmmf3}}
	\subfigure[]{\includegraphics[width=.74\columnwidth]{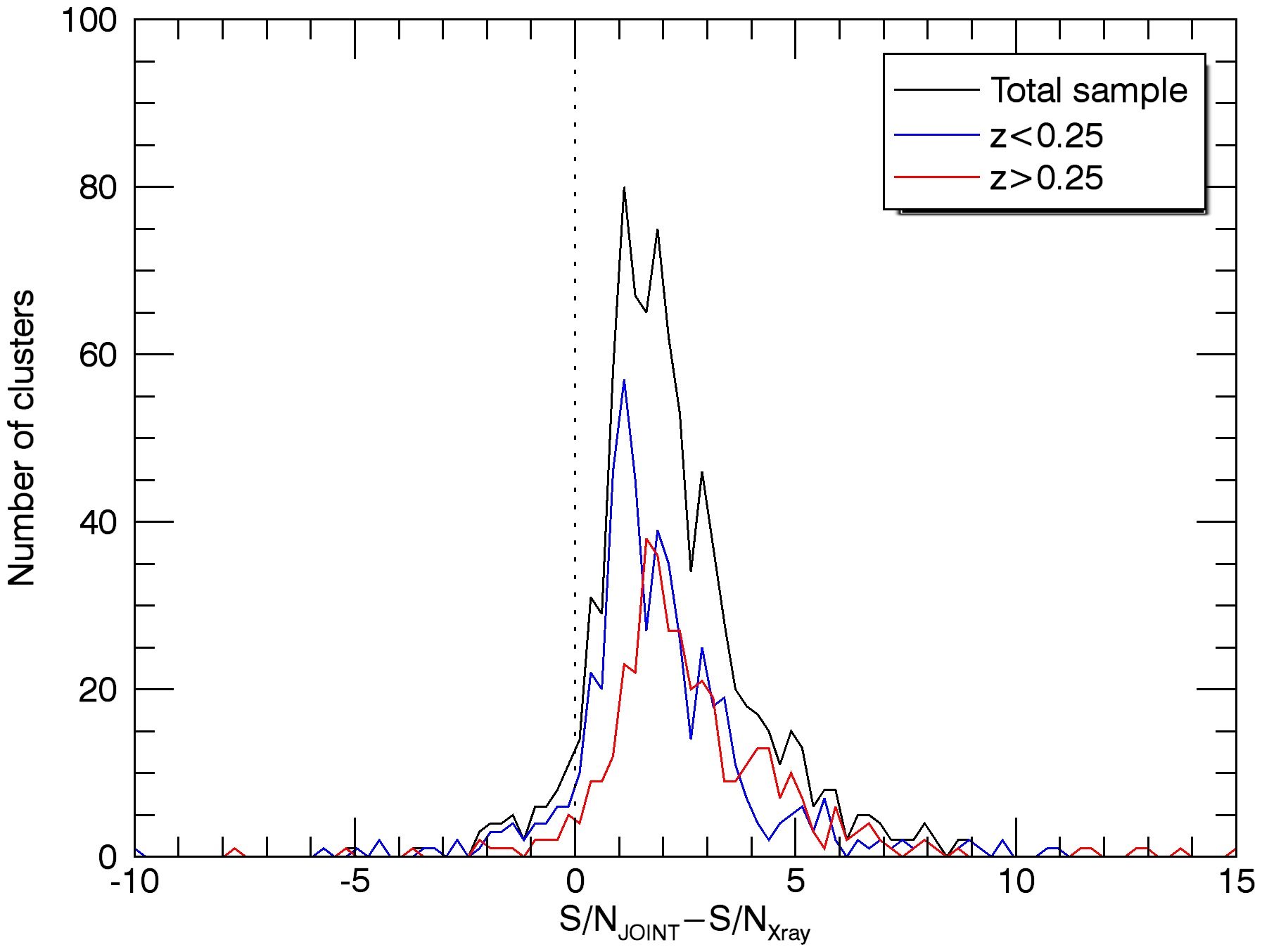}\label{fig:hist_snrgainxray_z_realmmf3}}
	\subfigure[]{\includegraphics[width=.74\columnwidth]{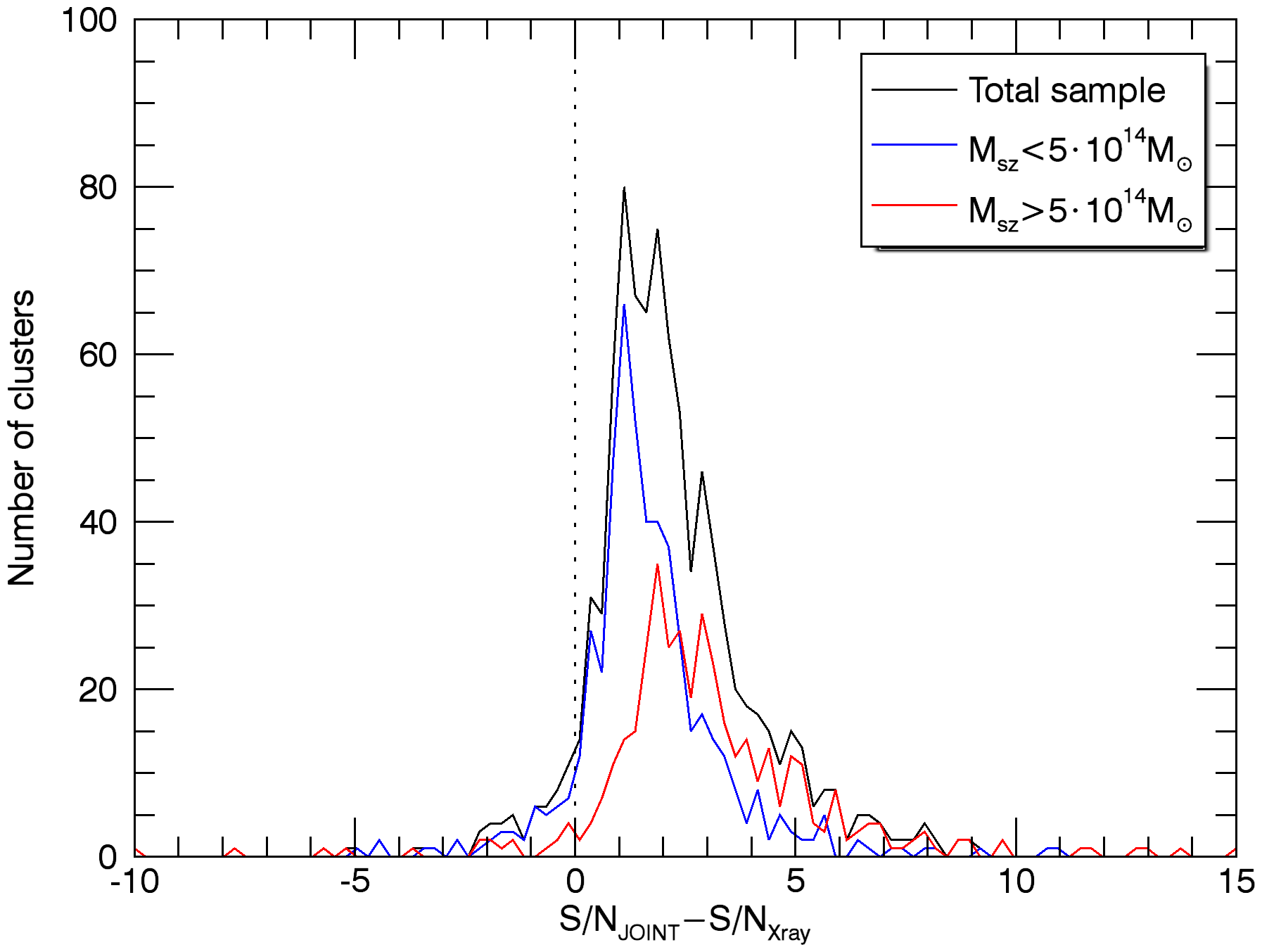}\label{fig:hist_snrgainxray_L_realmmf3}}
	\caption{Histograms of the S/N gain of the joint extraction of the MCXC clusters (panels a-d) and the PSZ2 clusters (panels e-h) with respect to the individual SZ-only and X-ray-only extractions. The upper panels show the histogram of the difference between the S/N of the joint extraction and the S/N of the SZ-only extraction, on two redshift bins (a) and two mass bins (b), for the MCXC clusters. The panels in the second row show the histogram of the difference between the S/N of the joint extraction and the S/N of the X-ray-only extraction, on two redshift bins (c) and two mass bins (d), for the MCXC clusters. The four bottom panels are analogous to panels a-d, but for the PSZ2 clusters. In all the cases, the S/N is defined as the estimated flux divided by the estimated background noise.}
	\label{fig:test}
\end{figure*}

In this section we provide some additional results related to the extraction of the MCXC and PSZ2 clusters using the three different filters: the classical SZ-MMF, the proposed X-ray matched filter and the proposed X-ray-SZ MMF. These results illustrate the advantage of the joint MMF, summarized in Fig. \ref{fig:detections_vs_snr_realrassmmf3_rpj} for the complete samples, as a function of redshift and mass. 
Figure \ref{fig:detections_vs_snr_realrassmmf3_rpj_2z} shows how the proposed joint MMF improves the detection probability of the RASS and PSZ2 clusters with respect to the single-survey filters. The samples are divided in two redshift bins to show that the gain in both regimes. 
Figures \ref{fig:test2} and \ref{fig:test} compare the estimated S/N obtained for the MCXC clusters and the PSZ2 clusters using the three detection methods as a function of redshift and mass. Figure \ref{fig:test2} shows the individual values for each cluster, while Fig. \ref{fig:test} shows the corresponding histograms. There is a S/N gain for most of the clusters, even for high redshift.

\section{Additional results on the photometry of the X-ray-SZ MMF}\label{app:simuPXCCjoint}

\begin{figure*}[]
	\centering
	\subfigure[]{\includegraphics[width=.99\columnwidth]{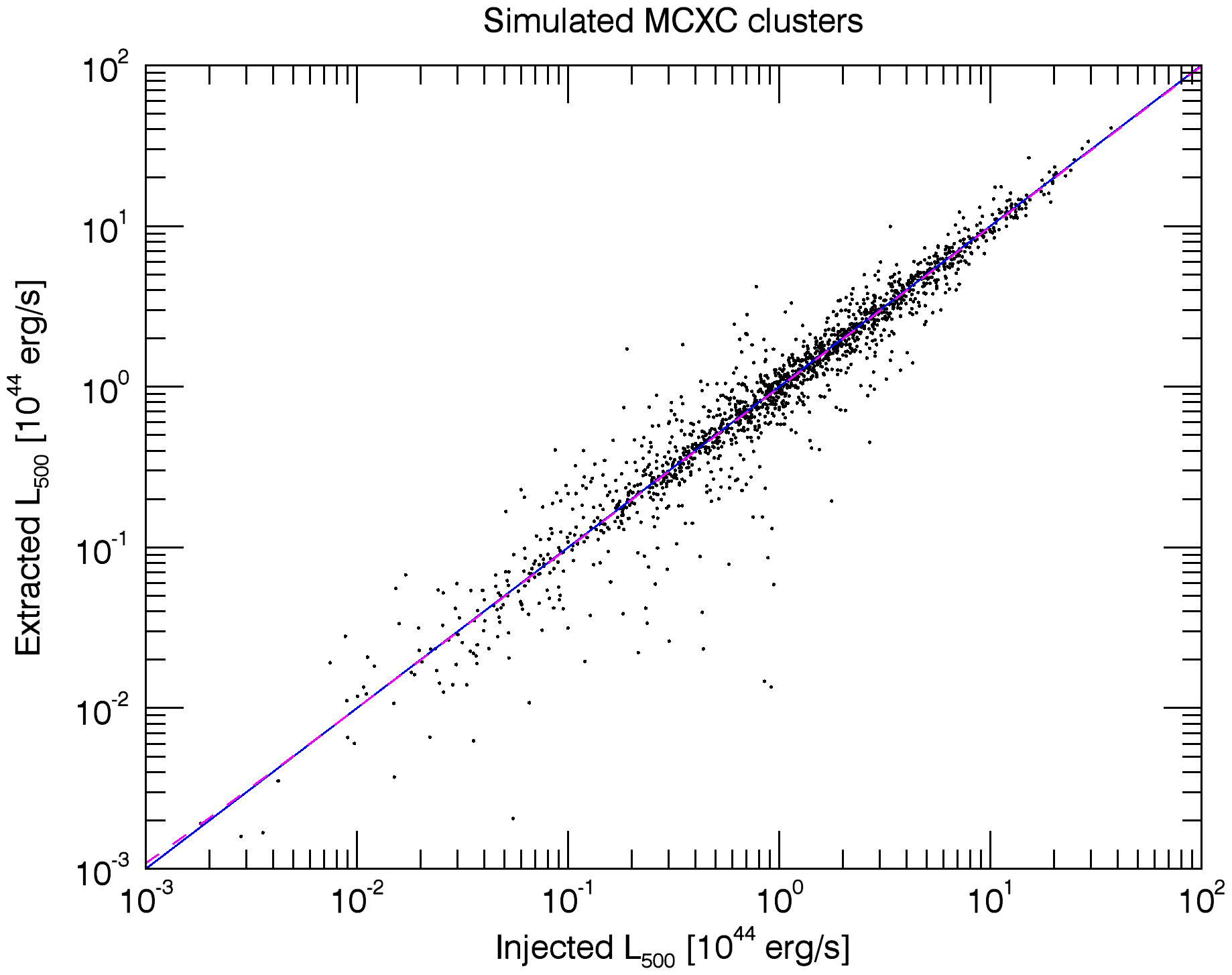}\label{fig:extraction_joint_simupxcc}}
	\subfigure[]{\includegraphics[width=.99\columnwidth]{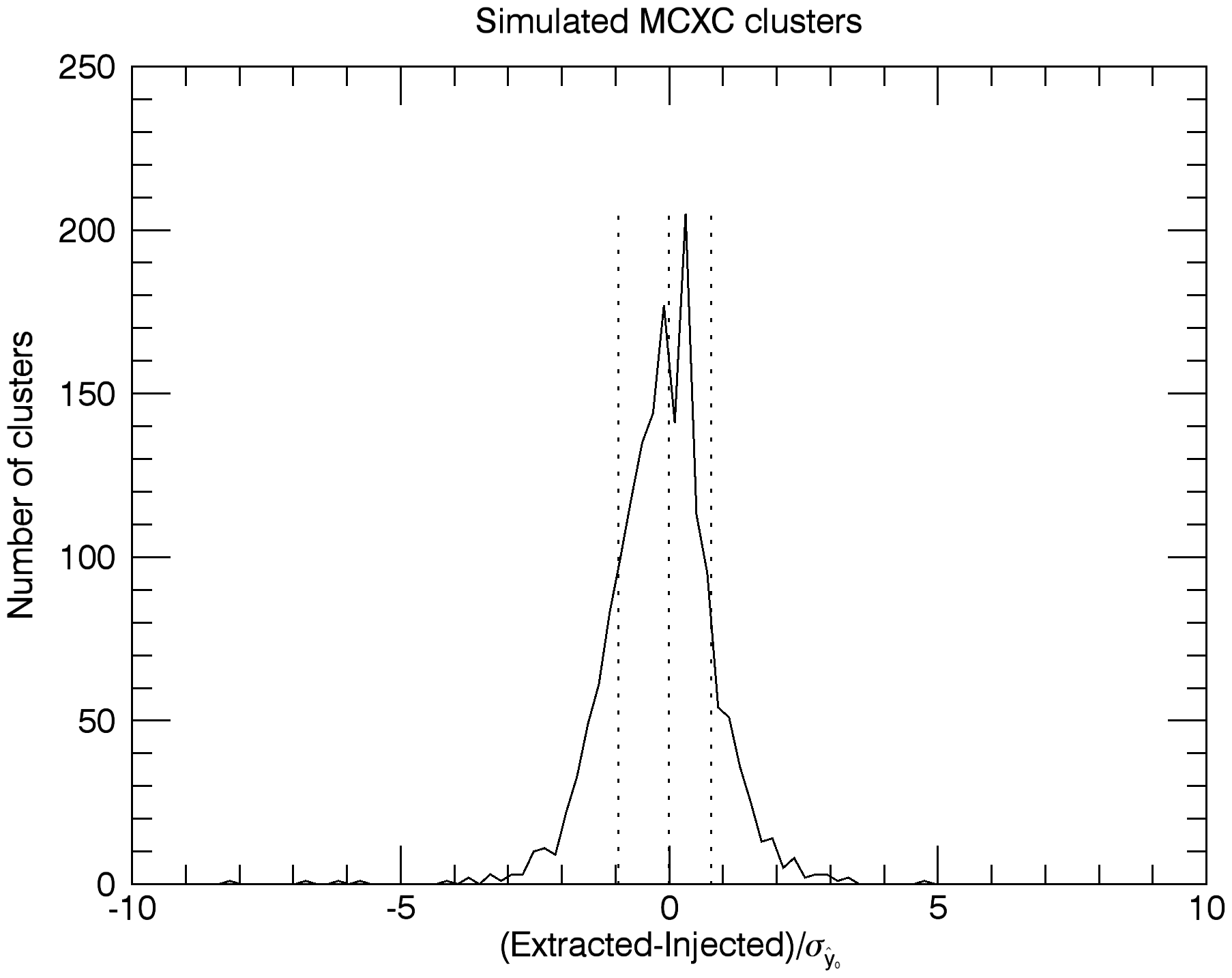}\label{fig:hist_joint_simupxcc}}
	\subfigure[]{\includegraphics[width=.99\columnwidth]{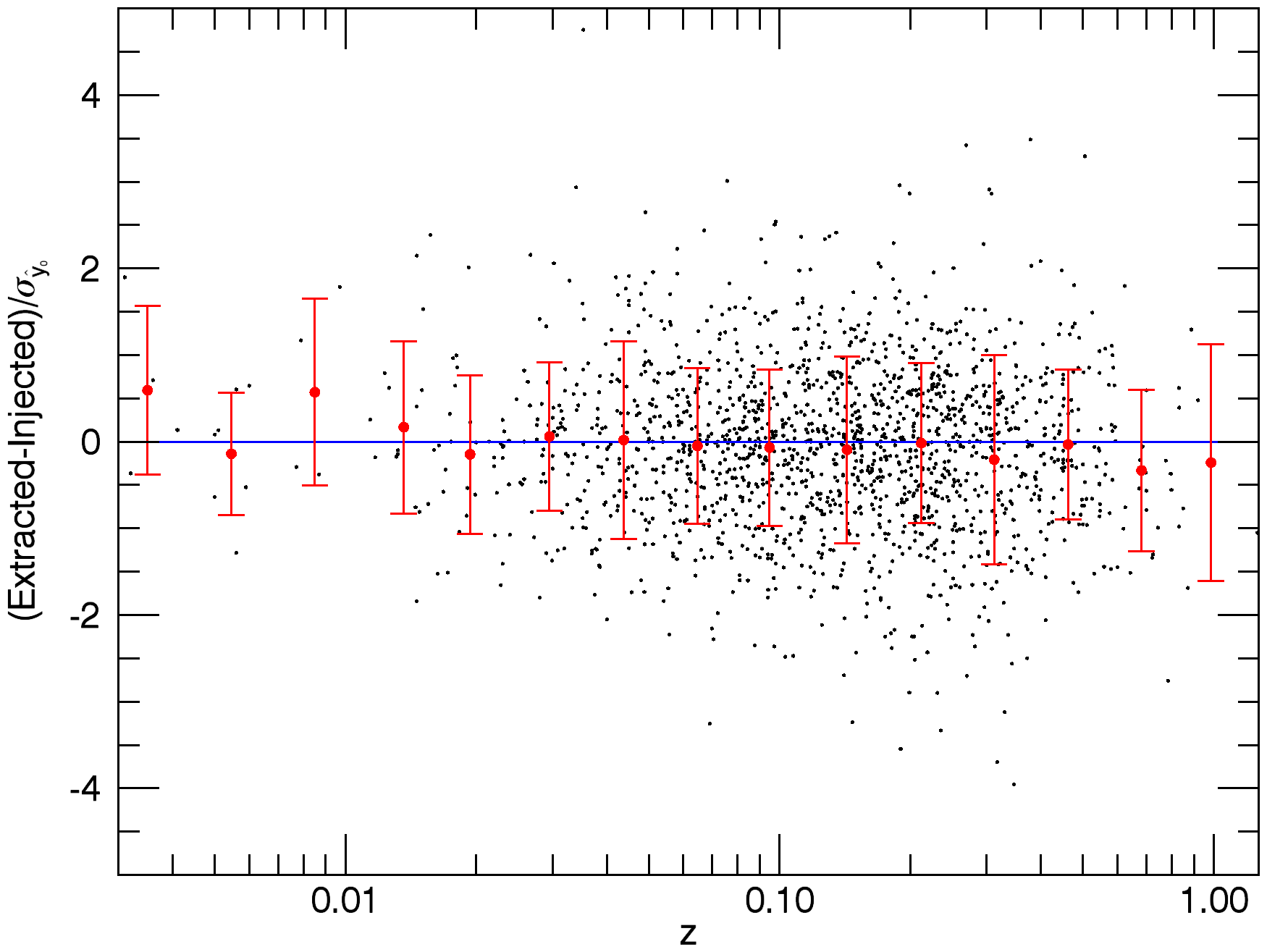}\label{fig:simupxcc_vs_z_joint}}
	\subfigure[]{\includegraphics[width=.99\columnwidth]{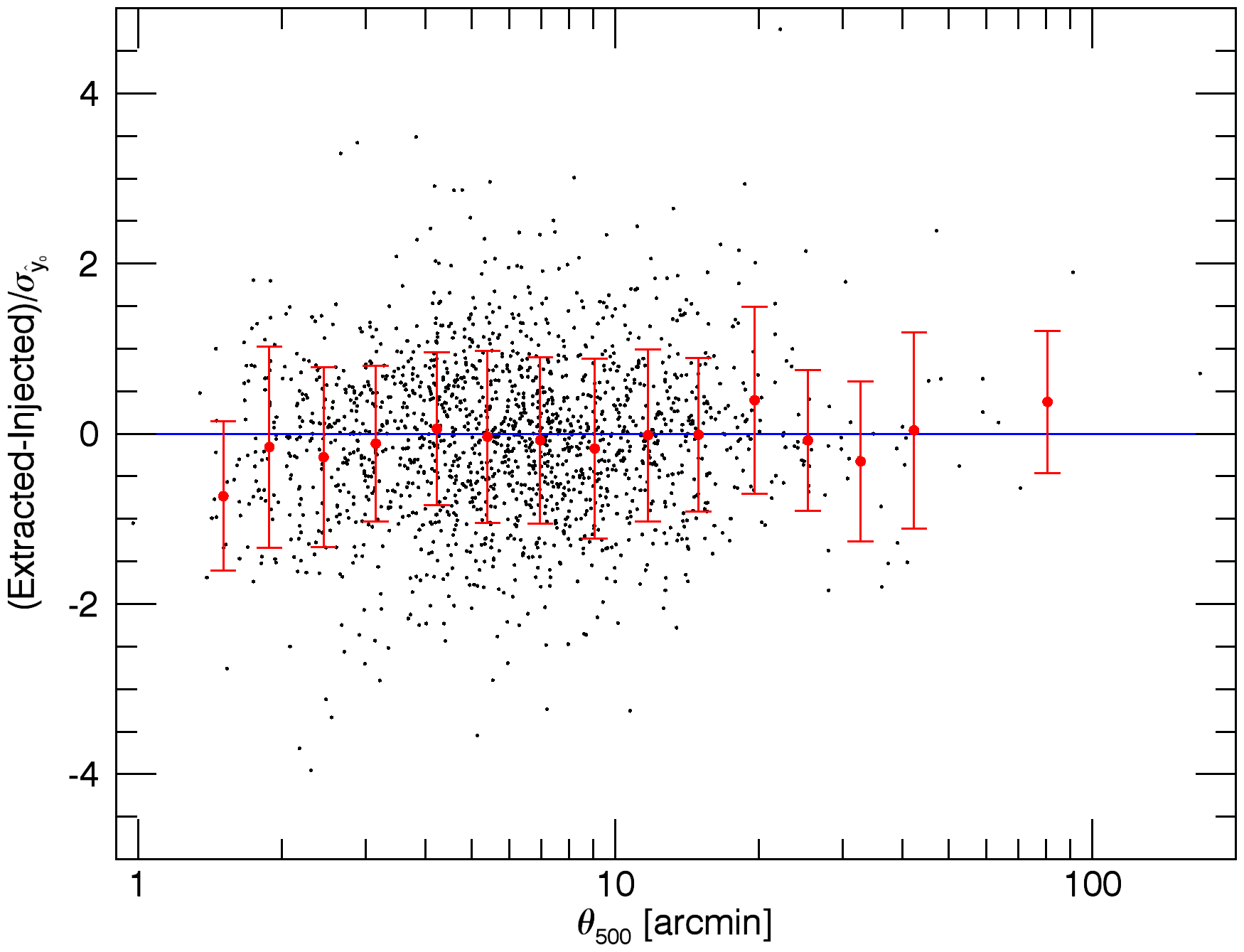}\label{fig:simupxcc_vs_theta_joint}}
	\subfigure[]{\includegraphics[width=.99\columnwidth]{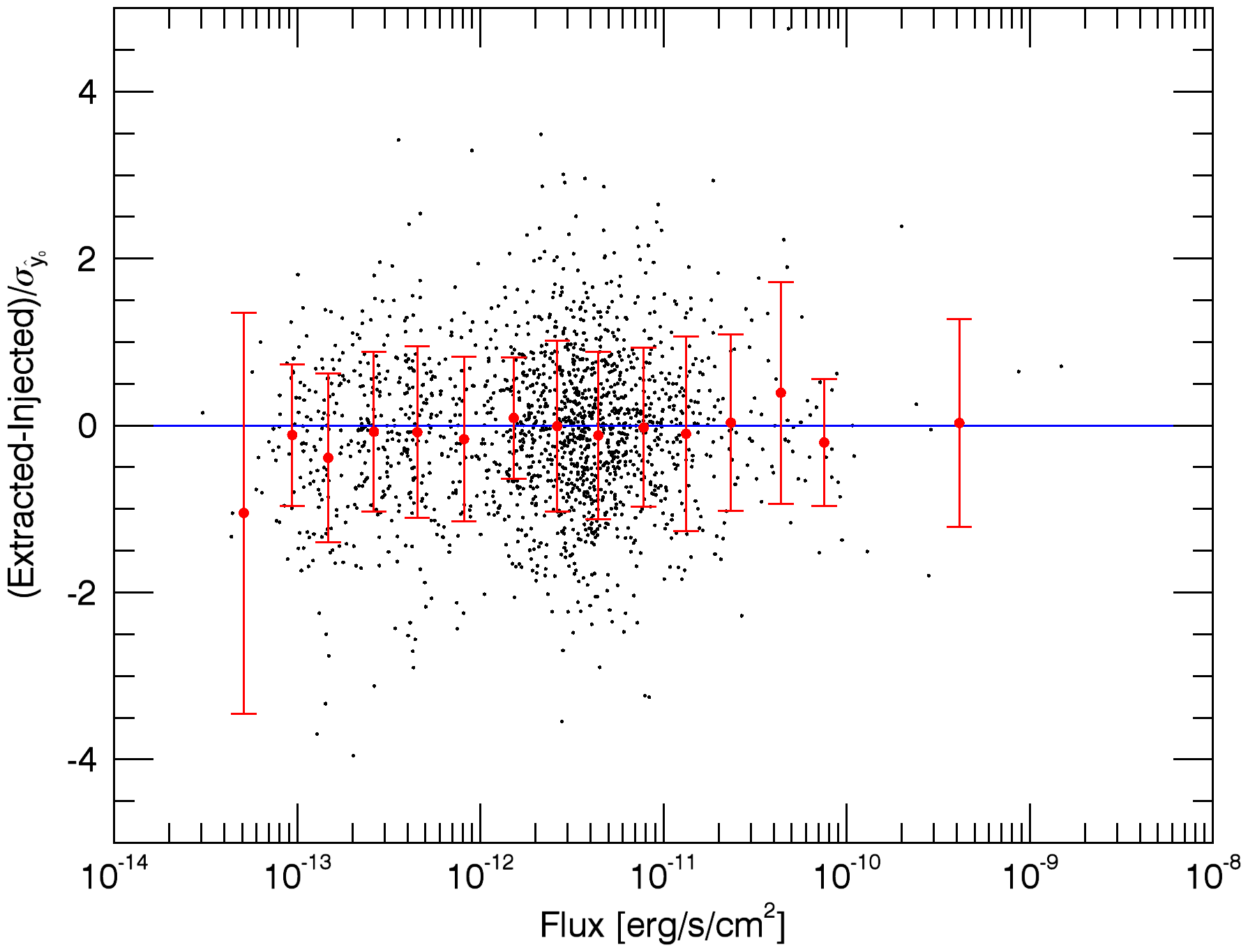}\label{fig:simupxcc_vs_flux_joint}}
	\subfigure[]{\includegraphics[width=.99\columnwidth]{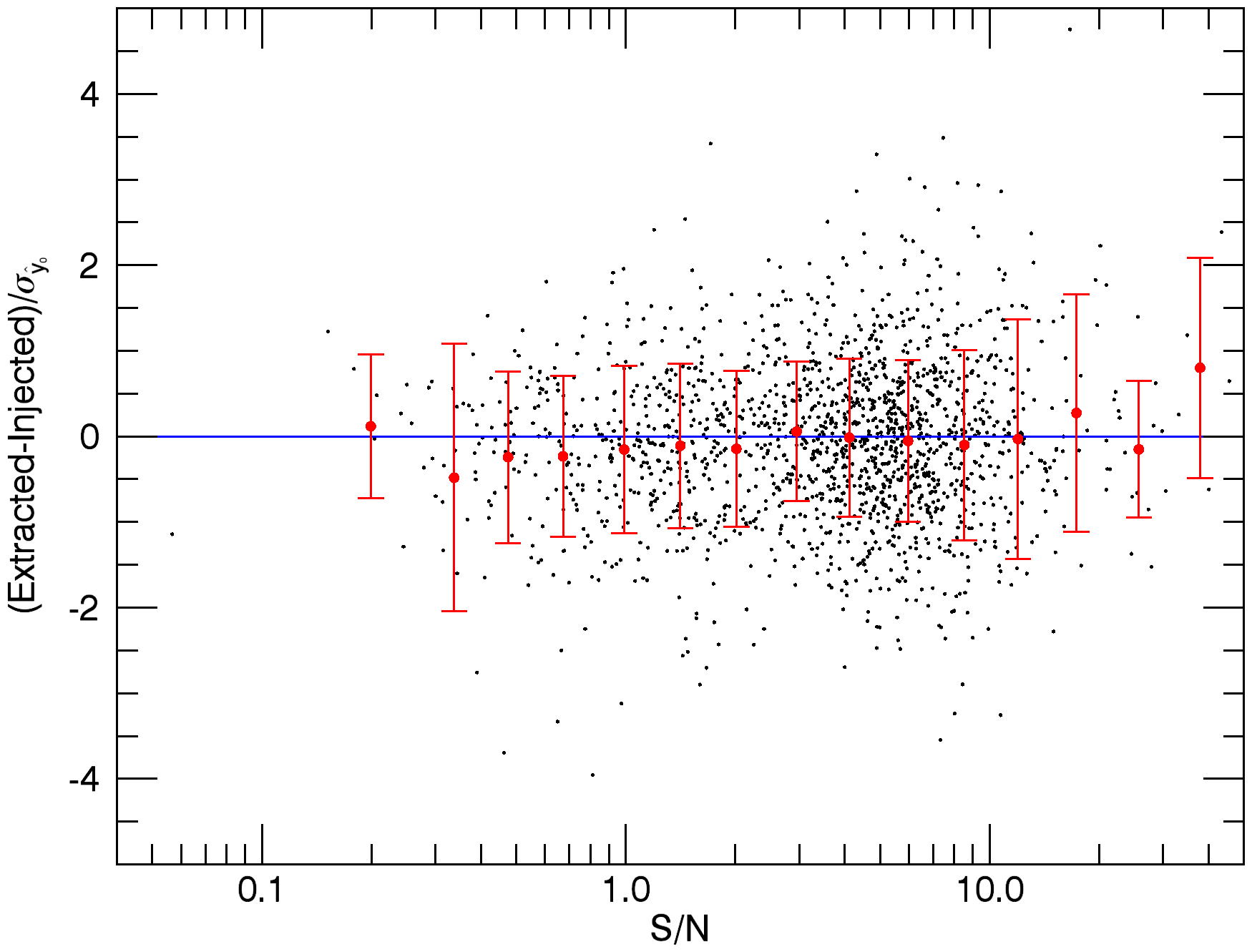}\label{fig:simupxcc_vs_snr_joint}}
	\caption{Photometry results of the extraction of simulated MCXC clusters (as described in Appendix \ref{app:simuPXCCjoint}) using the proposed X-ray-SZ MMF and assuming the position, size, and redshift of the clusters are known. The six panels are analogous to those in Fig. \ref{fig:simupxcc}. }
	\label{fig:simupxcc_joint}
\end{figure*}

In this section, we provide additional results related to the photometry of the proposed joint MMF. In particular, we present the results from an experiment in which we injected simulated clusters in real SZ and X-ray maps and extracted them using the proposed filter, assuming their positions, sizes, and redshifts are known. In particular, we injected 1743 clusters at random positions of the sky, with characteristics ($z$, $M_{500}$ and $L_{500}$) taken from the 1743 clusters in the MCXC catalogue \citep{Piffaretti2011}, to compare with the results obtained in the extraction of the real MCXC clusters. The injection was performed following the same procedure as in Sect. \ref{ssec:joint_gain}, but without scatter in the $F_{\rm X}/Y_{500}$ relation. Then, we applied the X-ray-SZ matched filter described in Sect. \ref{ssec:jointalgorithm} at each cluster position, fixing the cluster size to the true value. To convert the X-ray map into an equivalent SZ map, we assumed the same $F_{\rm X}/Y_{500}$ relation as used in the injection (Eq. \ref{eq:FxY500relation}), with the real redshift of the clusters. Figure \ref{fig:simupxcc_joint} shows the extraction results for this experiment. 

Figure \ref{fig:simupxcc_joint}a shows the extracted value of $L_{500}$ for each cluster as a function of the injected value. We used the same $L_{500}-Y_{500}$ relation as in the injection and the redshift in the catalogue to convert from the extracted $Y_{500}$ value to $L_{500}$.
The extracted flux follows the injected flux very well, with some dispersion. The best linear fit to these data is given by $y=0.990(\pm0.004)x-0.8(\pm2.5)\cdot10^{-4}$, which is very close to the unity-slope line, as shown in the figure. 

Figure \ref{fig:simupxcc_joint}b shows the histogram of the difference between the extracted and the injected value, divided by the estimated standard deviation $\sigma_{\hat{y}_0}$. Some of the properties of this histogram are summarized in Table \ref{table:simupxcc_joint_ap}. This histogram shows that there is no bias and that the estimated error bars describe the dispersion on the results well (as 68$\%$ of the extractions fall in an interval that is close to $\pm1\sigma_{\hat{y}_0}$). The small asymmetry is produced by the fact that we are approximating $y_0$ in Eq. \ref{eq:poissonvariance_joint} by its estimated value $\hat{y}_0$, which yields larger error bars for the clusters with overestimated flux and smaller error bars for the clusters with underestimated flux.

Figures \ref{fig:simupxcc_joint}c to \ref{fig:simupxcc_joint}f show the difference between the extracted and the injected value, divided by the estimated standard deviation $\sigma_{\hat{y}_0}$, as a function of the redshift, the size, the flux, and the S/N of each cluster. The extraction behaves correctly for all the values of these parameters, and they do not introduce any systematic error or bias in the results.

\begin{table}
	\caption{Main properties of the histogram in Fig. \ref{fig:simupxcc_joint}.}
	\label{table:simupxcc_joint_ap}
	\centering 
	\setlength{\tabcolsep}{4pt}
	\begin{tabular}{c c c c c}
		\hline
		\noalign{\smallskip}
		Median 				& -0.013  \\
		Mean        		& -0.064  \\
		Skewness    		& -0.657  \\
		Kurtosis    		& +5.191  \\
		Standard deviation 	& 0.994   \\
		68\% lower limit 	& -0.950  \\
		68\% upper limit 	& +0.777  \\
		\noalign{\smallskip}
		\hline
	\end{tabular}
\end{table}

\section{Effect of the scatter on the $F_X/Y_{500}$ relation}\label{app:scatter_fx_Y500}

\begin{figure}[]
	\centering
	\includegraphics[width=.99\columnwidth]{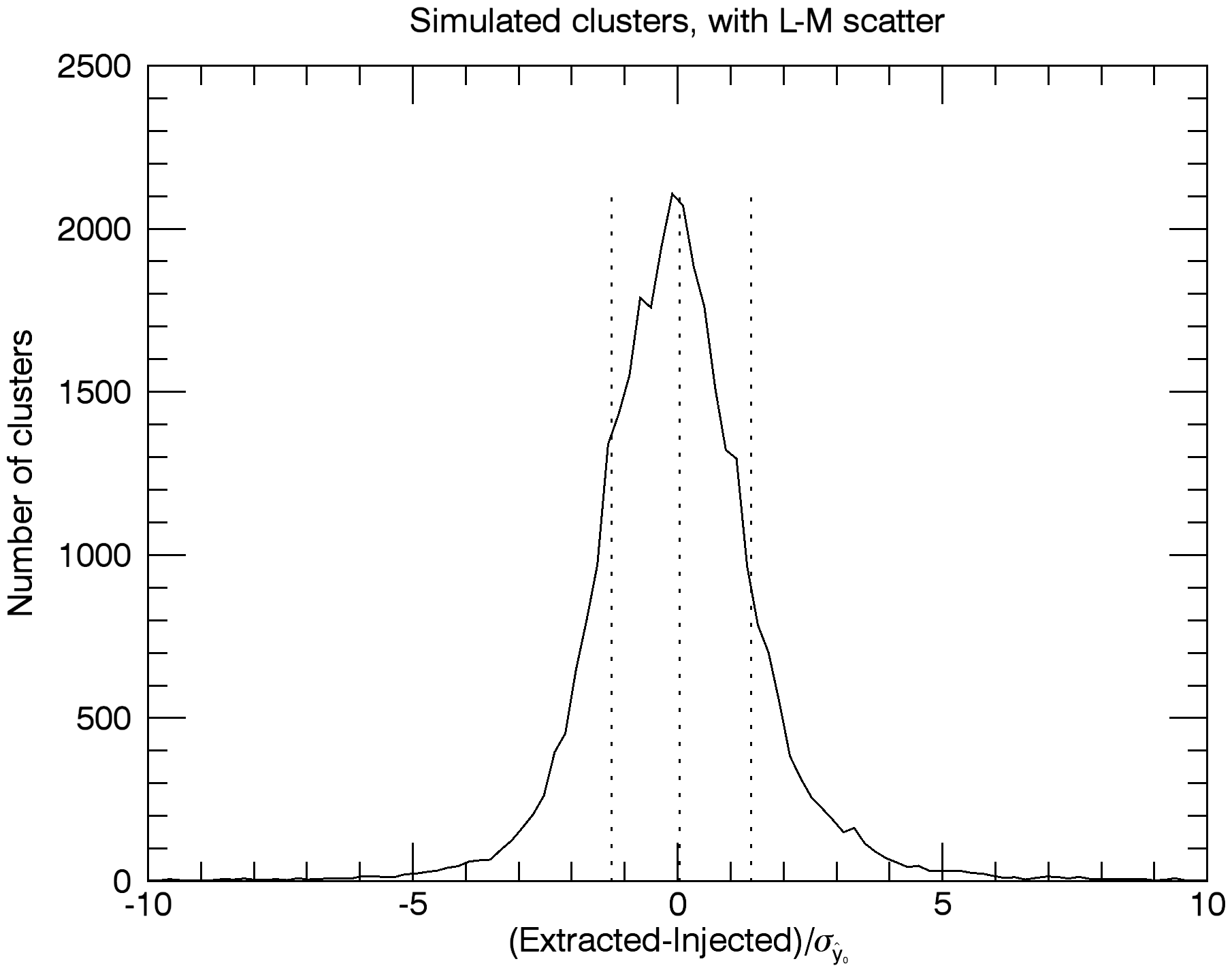}
	\caption{Histogram of the difference between the extracted and the injected $L_{500}$, divided by the estimated $\sigma_{\hat{y}_0}$ for the simulated clusters (as described in Sect. \ref{ssec:joint_gain}) extracted with the proposed X-ray-SZ MMF. The figure is analogous to Fig. \ref{fig:hist_joint_simuMZnoscatter}, but now the simulation includes scatter in the L-M relation. The central vertical line shows the median value, while the other two vertical lines indicate the region inside which 68$\% $ of the clusters lie.}
	\label{fig:hist_joint_simuMZscatter}
\end{figure}

To estimate the effect of the scatter in the $F_X/Y_{500}$ relation, we compared the results of the simulations presented in Sect. \ref{ssec:joint_photometry} (see Fig. \ref{fig:hist_joint_simuMZnoscatter}), where we simulated ideal clusters covering a large area in the mass-redshift plane, with the simulations carried out in Sect. \ref{ssec:joint_gain}, where we simulated clusters with the same positions and characteristics, but adding scatter in the L-M relation. The case with scatter shows a very good match between the extracted and the injected flux, as in the no-scatter case, with a linear fit which is very close to the unity-slope line ($y=1.010(\pm0.001)x-6.3(\pm0.9)\cdot10^{-2}$). We also checked that the extraction behaves correctly for all the values redshift, size, flux, and S/N, and that these parameters do not introduce any systematic error or bias in the results. However, we noted that the dispersion was larger than in the no-scatter case. Figure \ref{fig:hist_joint_simuMZscatter} shows the histogram of the difference between the extracted and the injected value, divided by the estimated standard deviation $\sigma_{\hat{y}_0}$ for the simulations with scatter. By comparing this to Fig. \ref{fig:hist_joint_simuMZnoscatter}, we can see that the scatter in the L-M relation increases the scatter in the extracted flux.

\end{appendix}

\end{document}